\documentclass[11pt]{article}
\usepackage[left=1in,top=.8in,right=1in,bottom=.8in,head=0in,nofoot]{geometry}

\usepackage{setspace,url,bm,amsmath} 
\usepackage{scalerel}

\usepackage{xcolor}
 \usepackage{multirow}
 \usepackage{arydshln}
 \usepackage{mathtools}
\usepackage{titlesec} 
\titlelabel{\thetitle.\quad} 
\titleformat*{\section}{\bf\Large\center}

\usepackage{float}
\usepackage{graphicx} 
\usepackage{bbm}
\usepackage{latexsym}
\usepackage{caption}
\usepackage[margin=20pt]{subcaption}
\usepackage{enumerate}
\graphicspath{ {./Graphs/} }
\def\ea{E_\textup{a}}

\def\wa{\renewcommand{\arraystretch}{1.2}}
\def\fall{\text{for all} \ }
\def\mj{\mathcal J}

\def\dgp{data-generating process}
\def\rep{ReP}
\def\reps{{ReP }}
\def\repsch{ReP schemes}
\def\repschs{{ReP schemes }}

\def\mrjq{\mr^{(J+1)(Q-1)}}
\def\hses{\hse_*}
\def\mb{\mathcal B}

\def\lmlrs{logistic and {\mlr}s }
\def\hst{H^*}
\def\hsti{(H^*)^{-1}}
\def\oh{1_{|H'|}}
\def\tbb{\tilde{\bm \beta}}
\def\gci{Gaussian correlation inequality}

\def\mr{\mathbb R}

\def\mc{\mathcal C}
\def\htxj{\htau_{x,j}}
\def\fort{\qquad\text{for} \ \ }

\def\qj{{\qiqp; \ \jej}}

\def\btb{\tilde{\bm\beta}}
\def\malogitjt{\ma_{\logit, \jt}}
\def\malogitmg{\ma_{\logit, \m}}
\def\malogitcs{\ma_{\logit, \css}}

\def\malmjt{\ma_{\lm, \jt}}
\def\malmmg{\ma_{\lm, \m}}
\def\malmcs{\ma_{\lm, \css}}

\def\matjt{\ma_{\ts, \jt}}
\def\matmg{\ma_{\ts, \m}}
\def\matcs{\ma_{\ts, \css}}

\def\mafmg{\ma_{\ff, \m}}

\newcommand{\tbt}{\tb^\T}
\def\hg{\hat\gamma}

\def\tbq{\tilde{\bm \beta}_q}

\def\wiq{\mi_{iq}}
\def\wiqq{\mi_{iQ}}
\def\wio{\mi_{i1}}
\def\qp{Q_\pluss}

\def\byqp{\by(q')}

\def\hbt{\hb^\T}
\def\htxt{\htx^\T}
\def\pzmo{p_{0,\x}}
\def\pjmo{p_{j, \x}}
\def\svxinvinv{\sigma(\vx^{-1})^{-1}}
\def\svxinv{\sigma(\vx)^{-1}}
\def\shoinv{\sigma(\ho)^{-1}}
\def\shvinv{\sigma(\hv)^{-1}}
\def\stvinv{\sigma(\tv)^{-1}}

\def\mqp{\mq_\pluss}
\def\oo{o(1)}
\def\wmlr{W_\textup{\mlogit}}
\def\wmlogit{\wmlr}
\def\tbqj{\tb_{qj}}
\def\tmlogit{T_{\mlogit}}
\def\sumiq{\sum_{i:Z_i = q}}
\def\tlm{T_\lm}

\def\twhere{\qquad\text{where} \ \ }
\def\chisq{\chi^2}
\def\llrt{\lambda_{\lrt}}
\def\rp{R_\pluss}
\def\emo{e_{Q-1}}
\def\eo{e_1}

\def\ths{\theta^*}
\def\mq{\mathcal Q}
\def\eq{e_q}
\def\eqq{e_Q}
\def\thq{\theta_q}

\def\thqmo{\theta_{Q-1}}
\def\bqz{\beta_{q0}}
\def\bqzs{\beta_{q0}^*}
\def\tbqz{\tb_{q0}}
\def\sumqmo{\sum_{\qiqp}}
\def\marem{\ma_\textup{rem}}
\def\mamgt{\ma_\mgt}
\def\majtt{\ma_\jtt}
\def\macst{\ma_\cst}

\def\pluss{{\scaleto{+}{4.5pt}}}

\def\tthn{\tilde\theta_0}
\def\maf{\ma_\textup{f}}
\def\qiqp{q\in\mqp}
\def\mqp{\mathcal Q_\pluss}
\def\qiq{q\in\mq}
\def\jej{j = \ot{J}}
\def\jij{j \in\{ \ot{J}\}}
\def\meani{N^{-1}\sumi}

\def\byq{\by(q)}
\def\mamodr{\ma_{\x, \dr}}
\def\mamojt{\ma_{\x, \jt}}
\def\mamocs{\ma_{\x, \css}}
\def\mamomg{\ma_{\x, \m}}

\def\drs{\dr=\jt, \m, \css}
\def\dg{\dagger}
\def\mgt{{\textup{t}, \m}}
\def\jtt{{ \textup{t}, \jt}}
\def\cst{{\textup{t}, \css}}

\def\yiq{Y_i(q)}
\def\fl{* = \fisher, \lin}
\def\hgs{\hat\gamma_*}
\def\succi{\succeq_\infty}

\def\lrt{\textsc{lrt}}
\def\lrts{{\textsc{lrt} }}

\def\beginp{\begin{pmatrix}}
\def\endp{\end{pmatrix}}

\def\ts{{\textup{t}}}

\def\lmt{\texttt{lm}}
\def\glmt{\texttt{logit}}
 \def\tts{T_{\ts}}

\def\glm{\textup{logit}}
\def\lm{\textup{lm}}
\def\logit{\textup{logit}}
\def\logitt{\texttt{logit}}
\def\mlogit{\textup{logit}}
\def\mlogitt{\texttt{logit}}
\def\qmo{{Q-1}}

\def\mtf{\mt_\ff }
\def\temp{\delta}
\def\nf{* = \neyman, \fisher}
\def\nfl{* = \neyman, \fisher, \lin}
\def\epm{\ep_{\textup{lm}}}
\def\mtm{\mt_{\textup{lm}}}
\def\gp{\gamma_\textsc{f}}
\def\dd{\sigma(\vxinv)}
\def\pjx{p_{j,\x}}
\def\pjt{p_{j,\ts}}
\def\pzt{p_{0,\ts}}
\def\pjlm{p_{j,\lm}}
\def\pzlm{p_{0,\lm}}
\def\pjglm{p_{j,\logit}}
\def\pjlogit{p_{j,\logit}}
\def\pzglm{p_{0,\logit}}
\def\hsx{\hat S^2_x}
\def\hsxj{\hat S^2_{x,j}}

 \def\vxm{V_{x\pluss}}
 \def\hxm{\hx_\pluss}
 
  \def\vxp{V_{x\pluss}}
 \def\hxp{\hx_\pluss}
 
\def\Vx{V_x}
\def\ff{\textup{f}}
\def\ajf{a_{j,\ff}}
\def\phiinv{\Phi^{-1}}
\def\linv{\phiinv}

\def\txbi{\tx_i^\T\theta}

\def\mlr{{multinomial logistic regression}}
\def\mlrs{{multinomial logistic regression }}
\def\mmlr{{Multinomial logistic regression}}

\def\hys{\hat Y_*}
\def\by{\bar Y}
\def\sumqi{\sumq \sum_{i:Z_i = q}}
\def\sxj{\sxxj}
\def\hxj{\hx_j}

\def\hyn{\hy_\nm}
\def\hyf{\hy_\fisher}
\def\hyl{\hy_\lin}
\def\bbr{\mathbb R}
 
\def\epf{\ep_\textup{f}}

\def\hy{\hat Y}
\def\pjf{p_{j, \textup{f}}}

\def\pqjg{p_{qj, \mlogit}}
\def\pqjmlr{p_{qj, \mlogit}}
\def\tqjmlr{t_{qj, \mlogit}}
\def\pzmlr{p_{0, \mlogit}}

\def\stx{S^2_{\tx}}

\def\mhls{Mahalanobis}

\def\wt{W_\ts}
\def\wlm{W_\lm}

\def\wlogit{W_\logit}
\def\tjt{t_{j,\ts}}
\def\tjlm{t_{j,\lm}}
\def\tjglm{t_{j,\glm}}

\def\mhl{{Mahalanobis }}

\def\mm{{\scaleto{\mathcal{M}}{4pt}}}

\def\ppinv{(\pp)^{-1}}

\def\mos{\x = \textup{t}, \lm, \logit}

\def\vl{v_\lin}

\def\dr{\diamond}
\def\m{\textup{mg}}
\def\jt{\textup{jt}}
\def\css{\textup{cs}}

\def\x{\dg}
\def\vsi{V_\Psi}

\def\cs{c_{*}}

\def\csi{c_{*}}
\def\cfi{c_{\fisher}}
\def\cni{c_{\nm}}
\def\cli{c_{\lin}}

\def\rs{\rightsquigarrow}

\def\pp{e_0e_1}

\def\sqrtn{\sqrt N}
\def\hts{\htau_*}
\def\begini{\begin{itemize}}
\def\endi{\end{itemize}}
\def\begine{\begin{enumerate}}
\def\ende{\end{enumerate}}

\def\ep{\epsilon}

\def\vxi{v_{x}}
\def\ho{\hat\Omega}

\def\mi{\mathcal{I}}

\def\vxinv{\vx^{-1}}
\def\vxiinv{\vxi^{-1}}
\def\vx{v_x}
\def\pd{\partial}
 
\def\hth{\hat\theta}
\def\tth{\tilde\theta}

\def\rss{\textsc{rss}}

\def\tx{\tilde x}
\def\tb{\tilde\beta}
\def\xij{x_{ij}}

\def\rrs{{rerandomization }}
\def\rr{{rerandomization}}
\def\mle{\textsc{mle}}
\def\mles{{\textsc{mle} }}

\def\thb{T_{\lm}}

\def\hts{\htau_*}

\def\tv{\tilde V}

\def\hv{\hat V}
\def\mn{\mathcal{N}}

\def\sxz{S_{xZ}}

\def\neyman{\textsc{n}}
\def\rtn{\sqrt N}

\def\htxj{\hat\tau_{x,j}}

\def\hse{\hat{\textup{se}}}

\def\sxxinv{(\sxx)^{-1}}
\def\sxxj{S^2_{x,j}}

\def\szz{S_{Z}^2}
\def\fwlc{The Frisch--Waugh--Lovell theorem}
\def\fwl{the Frisch--Waugh--Lovell theorem}
\def\tsesqj{{(\hse'_j)}^2}
\def\hsesq{\hat{\textup{se}}^2}

\def\txi{\tx_i}
\def\txit{\tx_i^\T}

\newcommand{\hgf}{\hat\gamma_\fisher}
\newcommand{\hglo}{\hat\gamma_{\lin,1}}
\newcommand{\hglz}{\hat\gamma_{\lin,0}}
\newcommand{\hglq}{\hat\gamma_{\lin,q}}

\newcommand{\htf}{\hat\tau_\fisher}
\newcommand{\htl}{\hat\tau_\lin}
\def\hts{\hat\tau_*}

\newcommand{\htn}{\hat\tau_\nm}

\def\Zi{Z_i}
\def\Yi{Y_i}
\def\sumi{\sum_{i=1}^N}
\def\sumN{\sum_{i=1}^N}

\def\gq{\gamma_q}

\def\bq{\bm \beta_q}
\def\ml{\mathcal{L}}

\def\sm{{Supplementary Material}}
\def\rem{{ReM }}

\def\hg{\hat\gamma}

\def\ma{\mathcal{A}}

\def\sxx{S^2_x}

\def\htx{\htau_x}

\def\ms{* = \nm, \fisher,\lin}
\def\hyf{\hY_\fisher}
\def\hyl{\hY_\lin}
\def\olss{{\textsc{ols}}}
\def\ols{{\textsc{ols} }}

\def\hb{\hat\beta }

\newcommand{\htau}{ \hat\tau}
\def\htf{ \hat\tau_\fisher}

\def\hx{\hat x}

\def\mt{\mathcal{T}}

\newcommand{\mN}{\mathcal{N}}

\newcommand{\op}{o_{\mathbb{P}}(1)}

\newcommand{\lin}{\textsc{l}}
\newcommand{\nm}{\textsc{n}}

\newcommand{\bg}{  \gamma }

\newcommand{\sumn}{\sum_{i=1}^N}
\newcommand{\sumq}{\sum_{q\in\mq}}

\newcommand{\hY}{\hat Y}

\newcommand{\pr}{\mathbb P}
\newcommand{\prgn}[1]{\noindent\textbf{#1}}
\newcommand{\prg}[1]{\smallskip\noindent\textbf{#1}}
 \newcommand{\ot}[1]{1, \ldots,#1}
 \newcommand{\zt}[1]{0, 1, \ldots,#1}

\newcommand{\ehws}{\textsc{ehw} }
\newcommand{\ehw}{\textsc{ehw}} 
\DeclareMathOperator{\diag}{\textup{diag}}

\def\T{{ \mathrm{\scriptscriptstyle T} }}

\newcommand{\cov}{\textup{cov} }

\newcommand{\fisher}{\textsc{f}}

 \newcommand{\bY}{ \bar Y}

\newcommand{\bbY}{\bar{ Y}}

\newcommand{\proj}{\textup{proj}}
\newcommand{\res}{\textup{res}}

\newcommand{\asim}{\overset{\cdot}{\sim}}
\def\assim{\stackrel{\cdot}{=}}

\newcommand{\vr}{ V_\lin}

\def\begina{\begin{eqnarray*}}
\def\enda{\end{eqnarray*}}
\def\beginy{\begin{eqnarray}}
\def\endy{\end{eqnarray}}
\def\begine{\begin{enumerate}}
\def\ende{\end{enumerate}}

\usepackage{amsthm}
\usepackage{amssymb}
\usepackage{amsmath}
\usepackage{color}

\usepackage{comment}
\theoremstyle{definition}

\newtheorem*{theorem*}{Theorem}
\newtheorem{theorem}{Theorem}
\newtheorem*{rmk*}{remark}
\newtheorem{proposition}{Proposition}
\newtheorem{lemma}{Lemma}

\newtheorem{condition}{Condition}

\newtheorem{definition}{Definition}
\newtheorem{remark}{Remark}
\newtheorem{corollary}{Corollary}
\newtheorem*{corollary*}{Corollary}

\usepackage{natbib} 
\bibpunct{(}{)}{;}{a}{}{,} 

\renewcommand{\refname}{References} 
\usepackage{etoolbox} 
\apptocmd{\sloppy}{\hbadness 10000\relax}{}{} 

\usepackage{multibib}
\newcites{sec}{References}

\usepackage{color}
\usepackage{listings}
\usepackage{hyperref}
\usepackage{booktabs}

\usepackage{lscape}

\def\ea{\mathbb E_\textup{a}}

\RequirePackage[normalem]{ulem}

\allowdisplaybreaks

\begin{document} 
\def\spacingset#1{\renewcommand{\baselinestretch}%
{#1}\small\normalsize} \spacingset{1.1}


\date{}
  \title{\bf \Large  No star is good news: a unified look at rerandomization based on $p$-values from covariate balance tests}
  \author{Anqi Zhao\thanks{Department of Statistics and Data Science, National University of Singapore (E-mail: staza@nus.edu.sg). } \hspace{.2mm} and Peng Ding\thanks{Department of Statistics, University of California, Berkeley (E-mail: pengdingpku@berkeley.edu). Ding was partially funded by the U.S. National Science Foundation (grant 1945136). We thank David Broockman, Christopher Mann, Avi Feller, and Thomas Royen for inspiring discussions.}}
  \maketitle

\begin{abstract}
Randomized experiments balance all covariates on average and provide the gold standard for estimating treatment effects. Chance imbalances nevertheless exist more or less in realized treatment allocations, complicating the interpretation of experimental results.
To inform readers of the comparability of treatment groups at baseline, modern scientific publications often report  covariate balance tables with not only covariate means by treatment group but also the associated $p$-values  from significance tests of their differences.  
The practical need  to avoid small $p$-values as indicators of poor balance motivates balance check and \rrs based on these $p$-values   from covariate balance tests (ReP) as an attractive tool for improving covariate balance in randomized experiments. 
Despite the intuitiveness of such strategy and its possibly already widespread use in practice,  the existing literature lacks results about its implications on subsequent inference,  subjecting many effectively rerandomized experiments to possibly inefficient analyses. 
To fill this gap, we examine a variety of potentially useful schemes for  \reps and quantify  their impact on subsequent inference.  Specifically, we focus on three estimators of the average treatment effect from the unadjusted, additive, and fully interacted linear regressions of the outcome on treatment, respectively, and derive their asymptotic sampling properties under \rep.  
The main findings are threefold. 
First, \reps improves covariate balance between treatment groups, thereby strengthening the causal conclusions that can be drawn from experimental data. In addition to increasing comparability of treatment groups, the improved balance also reduces the asymptotic conditional biases of the estimators and ensures more coherent inferences between covariate-adjusted and unadjusted analyses.
Second, the estimator from the fully interacted regression is asymptotically the most efficient under all \repschs examined, and permits convenient regression-assisted inference identical to that under complete randomization. 
Third, \reps improves the asymptotic efficiency of the estimators from the unadjusted and additive regressions. 
The corresponding standard regression analyses are accordingly still valid but  in general overconservative.
As a result, the combination of ReP for treatment allocation and fully interacted regression for analysis ensures both covariate balance and convenient and efficient inference. 
Importantly, our theory is design-based and holds regardless of how well the models involved in both the rerandomization and analysis stages represent the true data-generating processes.
\end{abstract}

\noindent%
\vfill

\newpage
\spacingset{1.5}

\section{Introduction}
\subsection{Rerandomization based on $p$-values}
Covariate balance increases comparability of units under different treatment conditions, strengthening the causal conclusions that can be drawn from data.
Randomized experiments balance all observed and unobserved covariates on average, providing the gold standard for estimating treatment effects. 
Chance imbalances nevertheless exist more or less in realized  allocations,  complicating the interpretation of experimental results.
Rerandomization, termed by \citet{cox:1982} and \citet{morgan2012rerandomization}, 
enforces covariate balance by discarding randomizations that do not satisfy a prespecified balance criterion.
\cite{bm} conducted a survey of leading experimental researchers 
in development economics, and suggested that rerandomization is commonly used yet often poorly documented. 

To inform readers of the comparability of treatment groups at baseline,  modern social and biomedical scientific publications are often encouraged to report baseline covariate balance tables with not only covariate means by treatment group but also the associated $p$-values from significance tests of their differences.
The practical need to avoid small $p$-values as indicators of poor balance motivates conducting rerandomization directly based on these $p$-values from balance tests \citep{bm, abs}.
Formally, \emph{rerandomization based on $p$-values} (\rep) runs one or more statistical tests to check the covariate balance of a realized randomization, and accepts the allocation if and only if the $p$-values of interest all exceed some prespecified thresholds. In their popular textbook on modern field experiments, \citet[Chapter 4.5]{gerber2012field} made this recommendation as a way to ``quickly approximate blocking".
 
Despite its decade-long existence and close relevance to randomized experiments in practice, 
the theory of \reps has not been addressed in the literature. 
\cite{hansen2008} discussed two hypothesis testing-based techniques for balance check in randomized experiments but left the issue of rerandomization and corresponding inference untouched. 
\cite{gerber2012field} gave the practical recommendation yet did not discuss its theoretical implications.
The existing discussion on rerandomization, on the other hand, focused mostly on balance criteria based on the {\mhls} distance between covariate means by treatment group
\citep{morgan2012rerandomization, morgan2015,  branson, LD2018, AOS, LD20, zach, ZDfrt, zdca, jrs, js}. 
The resulting procedure, though convenient in theory, is in general not a straightforward choice in practice. 
Another related literature is that on restricted randomization, also known as constrained randomization, which improves covariate balance by blocking, stratification, matched pairing, covariate-adaptive adjustment, etc. See, e.g., \cite{bailey}, \cite{imai}, \cite{bm}, \cite{PostStratYu}, \cite{higgins},  \cite{bugni2018, bugni2019}, \cite{colin}, \cite{hanzhong}, \cite{hanzhong2}, \cite{luke}, \cite{bai}, and \cite{ye}.
See also \cite{js} for a discussion on the connection and comparison between stratification and rerandomization.
The existing work in this literature however concerns restrictions distinct from those based on covariate balance tests.
The gap between theory and practice results in many effectively \rep-based experiments 
 being analyzed as if they were completely randomized, risking overconservative inferences that hinder the detection of statistically significant findings \citep{bm}.
The goal of this paper is to fill this gap and clarify the theoretical implications of ReP. 

\subsection{Our contributions}
First, 
we formalize {\rep} as a powerful tool for improving covariate balance in randomized treatment-control experiments, and propose a variety of potentially useful schemes based on standard statistical tests.
The proposed \repschs use $p$-values from two-sample $t$-tests, linear regression, and logistic regression to form the covariate balance criteria, respectively, allowing for easy implementation via standard software packages. 

Next, we quantify for the first time the impact of the proposed \repschs on subsequent inference, filling the gap in the literature.
Specifically, we focus on three estimators of the average treatment effect from the ordinary least squares (\olss) fits of the unadjusted, additive, and fully interacted linear regressions of the outcome on treatment, respectively, and evaluate their sampling properties under the proposed \repschs  from the {\it design-based} perspective. 
In short, the design-based perspective takes the physical act of randomization as the sole source of randomness in evaluating the sampling properties of quantities of interest, yielding results that are robust to model misspecification and unverifiable assumptions \citep{Neyman23, freedman2008randomization, Lin13, CausalImbens}.
The main findings are threefold.
First, ReP improves the covariate balance between treatment groups, which in turn (i) simplifies the interpretation of experimental results, (ii) reduces the conditional biases of the estimators \citep[c.f.][Section 4.3]{vif}, and (iii) promotes more coherent inferences between covariate-adjusted and unadjusted analyses \citep[c.f.][Section S5.3]{zdca}. 
Second, 
the estimator from the fully interacted regression is asymptotically the most efficient under all \repschs considered, with the asymptotic sampling distribution unaffected by the rerandomization.
It is thus our recommendation for subsequent  analysis under the proposed \repsch, allowing for convenient regression-based inferences identical to that under complete randomization. 
Specifically, we can use the coefficient of the treatment in the \ols fit of the outcome on the treatment, covariates, and their interactions as the point estimator of the average treatment effect, and adopt the associated Eicker--Huber--White ({\ehw}) standard error as a convenient approximation to the true standard error. 
Third, \reps improves the asymptotic efficiency of the estimators from the unadjusted and additive regressions, rendering inference based on the usual normal approximation overconservative. 
This highlights the importance of rerandomization-specific inference under \reps when the unadjusted or additive regression is used and, by contrast, demonstrates the advantage of the fully interacted regression for efficient and straightforward inference by normal approximation. 
We thus recommend the combination of ReP and fully interacted adjustment to ensure both covariate balance and efficient inference.

Lastly, 
we extend the theory of \reps  to experiments with more than two treatment arms and establish analogous results for \reps  based on $F$-tests and \mlr. 

As a novel technical contribution, 
we establish the asymptotic equivalence of the likelihood ratio test $(\lrt)$ and the Wald test for logistic and {\mlr}s from the design-based perspective, complementing the recent discussions in \citet{freedman2008randomization}, \cite{hansen2008}, and \citet{guo} on design-based inference from nonlinear regressions. 
See Theorem \ref{thm:mle_Q} in the Supplementary Material.
%
%
Without invoking any assumption of the logistic or multinomial logistic model, we view logistic and {\mlr}s as purely numeric procedures based on maximum likelihood estimation (\mle), and evaluate the sampling properties of the test statistics over the distribution of the treatment assignments.

\subsection{Notation and definitions}
For a set of tuples $\{(u_i, v_i): u_i \in \mathbb R,  \ v_i \in \mathbb R^K, \ i = \ot{N}\}$, denote by $\lmt(u_i \sim v_i)$ the \ols fit of the linear regression of $u_i$ on $v_i$,
and by $\glmt(u_i\sim v_i)$ 
the \mles fit of the logistic or multinomial logistic regression of $u_i$ on $v_i$.
We will clarify the regression specification in the context,  and focus on the numeric outputs of  \ols and \mles without invoking any of the underlying modeling assumptions. 
Assume default tests and $p$-values from
standard software packages throughout unless specified otherwise. 

For two $K\times 1$ vectors $T = (t_1, \dots, t_K)^\T$ and $a = (a_1, \dots, a_K)^\T$, denote by $|T|  \leq  a$ if $|t_k| \leq a_k$  for all  $k=\ot{K}$. 
Let $\diag(u_k)_{k=1}^K = \diag(u_1, \dots, u_K)$ denote the $K\times K$ diagonal matrix with $u_k$'s on the diagonal. 
For a $K\times K$ symmetric matrix $V = (V_{kk'})_{k,k'=\ot{K}}$ with $V_{kk} > 0$ for all $k = \ot{K}$, let $\sigma(V) = \diag(V_{kk}^{1/2})_{k=1}^K$ and
$D(V) = \{\sigma(V)\}^{-1} V \{\sigma(V)\}^{-1}$. 
Intuitively, $D(V)$ gives the corresponding correlation matrix when $V$ is a covariance matrix.
Let $\|\ep\|_\mm = \ep^\T\{\cov(\ep)\}^{-1} \ep$ denote the \mhl distance of a random vector $\ep$ from the origin. 
Let $\rs$ denote convergence in distribution, and let $\ea $ denote asymptotic expectation.

Lastly, we use the following notion of peakedness \citep{sherman} to quantify the relative efficiency between different estimators \citep{AOS}. 

\begin{definition}\label{def:peakedness}
For two symmetric random vectors $A$ and $B$ in $\mathbb R^K$, we say $A$ is {\it more peaked} than $B$ if $\pr(A\in \mc) \geq \pr(B \in \mc)$ for all symmetric convex sets $\mc$ in $\mathbb R^K$, denoted by $A \succeq B$. 
\end{definition}

For $K = 1$, a more peaked random variable has narrower central quantile ranges.
For $A$ and $B$ with finite second moments, $A \succeq B$ implies $\cov(A) \leq \cov(B)$ \citep[Proposition 4]{AOS}.
For $A$ and $B$ that are both normal with zero means, $A \succeq B$ is equivalent to $\cov(A) \leq \cov(B)$.
This suggests peakedness as a more refined measure for comparing relative efficiency of estimators than the commonly used covariance.
We formalize the intuition in Definition \ref{def:eff} below.

\begin{definition}\label{def:eff}
Assume that $\hth_1$ and $\hth_2$ are two consistent estimators for parameter $\theta \in \mathbb R^K$ as the sample size $N$ tends to infinity, 
with $\sqrtn(\hth_1 - \theta) \rs A_1$ and $\sqrtn(\hth_2 - \theta) \rs A_2$ for some symmetric random vectors $A_1$ and $A_2$.  
We say 
\begine[(i)]
\item $\hth_1$ and $\hth_2$ are {\it asymptotically equally efficient} if $A_1$ and $A_2$ have the same distribution, denoted by $\hth_1 \asim \hth_2$; 
\item $\hth_1$ is  {\it asymptotically more efficient} than $\hth_2$ if $A_1 \succeq A_2$, denoted by $\hth_1 \succi \hth_2$. 
\ende 
\end{definition}

By Definition \ref{def:eff},  an asymptotically more efficient estimator has not only a smaller asymptotic covariance but also narrower central quantile ranges.

\section{Basic setting of the treatment-control experiment}\label{sec:setup}
\subsection{Regression-based inference under complete randomization}\label{sec:setup_1}
Consider an intervention of two levels, indexed by $q = 0, 1$,  and a finite population of $N$ units, indexed by $i =1,\ldots, N$.  Let $\yiq \in \mathbb R$ be the potential outcome of unit $i$ under treatment level $q \in \{0,1\}$ \citep{Neyman23, CausalImbens}. 
The individual treatment effect is $\tau_i  = Y_i(1) - Y_i(0)$ for unit $i$, and the finite-population average treatment effect is $\tau = N^{-1} \sumN  \tau_i  = \by(1) - \by(0)$, where $\by(q)=\meani Y_i(q)$.

For some prespecified, fixed integer $N_1 >0$, 
complete randomization draws completely at random $N_1$ units to receive level $1$ of the intervention  and then assigns the remaining $N_0 = N - N_1 > 0$ units to receive level $0$. 
Let $e_q = N_q / N$ denote the proportion of units under treatment level $q \in \{0,1\}$. 

Let $Z_i \in \{0,1\}$ indicate the treatment level received by unit $i$. 
The observed outcome equals $ Y_i  = \Zi\Yi(1) + (1-\Zi)\Yi(0) $.
Let $\hY(q) = N_q ^{-1}\sumiq   Y_i $ denote the average observed outcome under treatment level $q \in \{0,1\}$. 
The difference in means $\hat{\tau}_\nm  = \hY(1) - \hY(0)$ is unbiased for $\tau$ under complete randomization \citep{Neyman23}, and can be computed as the coefficient of $Z_i$ from the simple,  unadjusted  linear regression $\lmt(Y_i \sim 1 + Z_i)$ over $i = \ot{N}$. 

The presence of covariates promises the opportunity to improve estimation efficiency. 
Let $x_i = (x_{i1}, \ldots, x_{iJ})^\T$ denote the $J$ pretreatment covariates for unit $i$, centered  at  $\bar x = N^{-1}\sumi x_i = 0_J$. 
\citet{Fisher35} suggested an estimator $\htau_\fisher$ for $\tau$, which equals  the coefficient of $Z_i$ from the additive linear regression  $\lmt(Y_i \sim 1 + Z_i + x_i)$ over $i = \ot{N}$.  
\cite{Freedman08a} criticized the possible efficiency loss by $\htf$ compared to $\htn$. 
\citet{Lin13} recommended an improved estimator, denoted by $\htau_\lin$, as the coefficient of $Z_i$ from the fully interacted linear regression  $\lmt(Y_i \sim   1 + Z_i + x_i + Z_i x_i)$ over $i = \ot{N}$, and showed its asymptotic efficiency over $\htn$ and $\htf$. 
In addition, \citet{Lin13} also showed that the corresponding {\ehw} standard errors are asymptotically appropriate for estimating the true standard errors of  $\htf$ and $\htl$.
This justifies large-sample Wald-type inference of $\tau$ based on \olss.


We adopt the {\it design-based} perspective for all theoretical statements in this article, which views the physical act of randomization, as represented by $(Z_i)_{i=1}^N$, as the sole source of randomness in evaluating the sampling properties of quantities of interest.  
Accordingly, despite the estimators $\htn$, $\htf$, and $\htl$ are all outputs from the unadjusted, additive, and fully interacted linear regressions, respectively,
we invoke no assumption of the underlying linear models but evaluate the sampling properties of $\hts$'s  over the distribution of $(Z_i)_{i=1}^N$ conditioning on the potential outcomes  and covariates.
All theoretical guarantees of the $\hts$'s are therefore design-based and hold even when the linear models are misspecified.


\subsection{Covariate balance and \rr}\label{sec:gamma}
The regression adjustment by \cite{Fisher35} and \cite{Lin13}, as it turns out, can be viewed as adjusting for imbalances in covariate means.
Let $\htx= \hx(1) - \hx(0)$ denote the difference in covariate means between treatment groups, with $\hx(q) = N_q ^{-1} \sumiq x_i$ representing the sample covariate mean under treatment level $q \in \{0,1\}$.
Let $\hgf$ be the coefficient vector of $x_i$ from the additive regression $\lmt(Y_i \sim 1+Z_i +x_i)$  over $i = \ot{N}$, and let $\hg_\lin =e_0\hglo+ e_1\hglz$, where $\hglq$ denotes the coefficient vector of $x_i$ from the treatment group-specific regression $\lmt(Y_i \sim 1 + x_i)$ over $\{i: Z_i = q\}$. 
\citet[][Proposition 1]{ZDfrt} showed that 
\begina
\hts =  \htn -  \htx^\T \hgs\quad \text{for} \ \ \fl.
\enda 
The two covariate-adjusted estimators are thus variants of $\hat{\tau}_\nm $ with corrections for the difference in covariate means.
A small $\htx$ intuitively promotes 
more coherent inferences between $\hts \ (\nf)$ and the asymptotically most efficient $\htl$, thereby enhancing the asymptotic efficiency of $\htn$ and $\htf$.

Rerandomization, on the other hand, enforces covariate balance in the design stage \citep{cox:1982, morgan2012rerandomization}, and accepts an allocation if and only if it satisfies some prespecified covariate balance criterion.
Assume complete randomization for the initial allocations throughout this article.
\cite{morgan2012rerandomization} and \cite{LD2018} studied a special type of rerandomization, known as ReM, that uses the Mahalanobis distance of $\htx$ as the balance criterion, and accepts a randomization if and only if $\|\htx\|_\mm \leq a_0$ for some prespecified threshold $a_0$. 
The practical need to avoid small $p$-values in baseline covariate balance tables instead motivates \reps that  accepts a randomization if and only if the $p$-values  from relevant balance tests all exceed some prespecified thresholds. 
To fill the gap in the literature regarding the theoretical properties of ReP, we examine nine hypothesis testing-based covariate balance criteria for conducting \reps  under the treatment-control experiment, and quantify their respective impact on subsequent inference from the design-based perspective. 
We start with three two-sample $t$-test-based criteria in Section \ref{sec:ttest} given their direct connections with the balance tables in practice, and extend the discussion to six regression-based alternatives in Section \ref{sec:lm_glm}. 
The results provide the basis for generalizations to experiments with more than two treatment arms, which we formalize in Section \ref{sec:Q_main}.

\section{ReP based on two-sample $t$-tests}\label{sec:ttest}
\subsection{Marginal, joint, and consensus rules}
The difference in covariate means provides an intuitive measure of covariate balance under the treatment-control experiment.
Depending on whether we examine the $J$ covariates separately or together, this motivates three two-sample $t$-test-based criteria for \rep. 

To begin with, recall $x_{ij}$ as the $j$th covariate of unit $i$. 
A common approach to balance check is to run one two-sample $t$-test for each covariate $j \in \{1, \dots, J\}$ based on $(x_{ij}, Z_i)_{i=1}^N$, and use the resulting two-sided $p$-value, denoted by $\pjt$,  to measure  its balance between treatment groups. 
This yields $J$ marginal $p$-values, $\{\pjt: \jej \}$, that occupy the last column of the covariate balance tables. 
An intuitive, and possibly already widely-used, criterion for ReP is then to accept a randomization if and only if $\pjt \geq \alpha_j$ for all $\jej $ for some prespecified thresholds $\alpha_j \in (0,1)$ \citep{bm}. 
We call this the \emph{marginal rule} based on $J$ marginal tests of individual covariates. 

Alternatively, we can test the difference in means of all $J$ covariates together by a multivariate version of the two-sample $t$-test, and accept a randomization if and only if the $p$-value from this joint test exceeds some prespecified threshold.  
Let  $\ho$ be the pooled estimated covariance of $\htx$.
The two-sample Hotelling's $T^2$ test takes  $\wt = \htx^\T \ho^{-1} \htx$ as the test statistic, and computes a one-sided $p$-value, denoted by $\pzt$, by comparing $\wt$ against the Hotelling's $T^2$ distribution.   
Alternatively, we can replace the Hotelling's $T^2$ distribution with the asymptotically equivalent $\chi^2_J$ distribution and compute the $p$-value based on the joint Wald test.
A \emph{joint rule}  then accepts a randomization if and only if $\pzt \geq \alpha_0$ for some prespecified threshold $\alpha_0\in (0,1)$. 

In situations where both marginal and joint balances are desired, we can adopt a \emph{consensus rule} that accepts a randomization if and only if it is acceptable under both the marginal and joint rules with $\pjt \geq \alpha_j$ for all $j = \zt{J}$.

Index  by ``\m'', ``\jt'', and ``\css'' the marginal, joint, and consensus rules, respectively.  
This defines three \repschs by two-sample $t$-tests,  summarized in Definition  \ref{def:t} below.
Of interest are their implications on the subsequent inference based on $\hts \ (\ms)$. 
We address this question in Sections \ref{sec:asym_t} and \ref{sec:infer} below.


\begin{definition}\label{def:t} 
Assume \reps by two-sample $t$-tests. 
Let 
$\ma_{\mgt} = \{  \pjt \geq \alpha_j \ \fall \jej\}$, $\ma_\jtt = \{\pzt \geq \alpha_0\}$,  and $\matcs =\ma_{\mgt} \cap\ma_\jtt =  \{\pjt \geq \alpha_j \ \fall j = \zt{J}\}$
denote the acceptance criteria under the marginal, joint, and consensus rules, respectively. 
\end{definition}

\subsection{Asymptotic theory}\label{sec:asym_t}
We derive in this subsection the asymptotic sampling properties of  $\hts \ (\ms)$ under the three \repschs in Definition \ref{def:t}.
The results demonstrate the multiple benefits of \reps in strengthening causal conclusions from experimental data, and elucidate the advantage of \cite{Lin13}'s method for convenient and efficient inference under \rep.

Let $\sxx = (N-1)^{-1}\sumi x_ix_i^\T$ denote the finite-population  covariance of the centered $(x_i)_{i=1}^N$.
Condition \ref{asym} below gives the standard regularity conditions for design-based {\it finite-population} asymptotic analysis,
which embeds the study population into an infinite sequence of finite populations with increasing sizes and defines the asymptotic distribution of any sample quantity as its limiting distribution along this hypothetical sequence. See \cite{DingCLT} for a review.
We assume this framework  for all asymptotic statements in this article. 

\begin{condition}\label{asym}
As $N \to \infty$, 
(i) $ e_q  = N_q /N$ has a limit in $(0,1)$ for $q = 0,1$, 
(ii) the first two finite-population moments of $\{Y_i(0), Y_i(1), x_i \}_{i=1}^N $  have finite limits; $\sxx$ and its limit are both nonsingular,  
and 
(iii) $N^{-1}\sumi Y_i^4(q) = O(1)$ for $q = 0,1$; $N^{-1}\sumi \|x_i\|_4^4 =O(1)$. 
\end{condition} 

Let $\gq $ be the coefficient vector of $x_i$ from $\lmt\{\yiq  \sim 1 + x_i\}$ over $i = \ot{N}$. 
This is a theoretical  fit with $\{Y_i(q)\}_{i=1}^N$ only partially observable depending on the treatment assignment.
Condition \ref{asym} ensures that $e_q $, $\gamma_q$, and $S_x^2$ all have finite limits as $N$ tends to infinity. 
For notational simplicity, we will use  the same symbols  to denote their respective limits when no confusion would arise. 
Lemma \ref{lem:cre} below follows from \cite{ZDfrt} and specifies the asymptotic distributions of $\hts \ (\nfl)$ under complete randomization. 
This provides the baseline for evaluating the efficiency gains by ReP.

\begin{lemma}\label{lem:cre}
Under complete randomization and Condition \ref{asym}, we have
\begina
\rtn   \left(\begin{array}{cc}
\hts  - \tau \\
\htx
\end{array}
\right)
&  \rightsquigarrow &
\mN \left\{ 
0_{J+1},  
  \left(\begin{array}{cc}
v_*  &  \csi^\T \\
 \csi & \vxi \end{array}\right) \right\} \qquad (\nfl) 
 \enda
with $v_x = \ppinv\sxx$, 
\begina
\cni =\sxx(e_0^{-1}\gamma_0 + e_1^{-1} \gamma_1), \quad \cfi =   \sxx(e_1^{-1}-e_0^{-1})(\gamma_1-\gamma_0 ), \quad \cli = 0_J,
\enda
 and $v_* - v_\lin = \cs^\T \vx^{-1} \cs \geq 0$ for $\nfl$. We give the explicit expressions of $v_* \ (\nfl)$ in the {\sm}.
\end{lemma}

Recall $\alpha_j \ (j = \ot{J})$ and $\alpha_0$ as the thresholds for the marginal and joint rules. 
Let $a_0$ be the $(1-\alpha_0)$th quantile of the $\chi^2_J$ distribution.
Let $a_{j}$ be the $(1-\alpha_j/2)$th quantile of the  standard normal distribution, vectorized as $a = (a_{1}, \dots, a_{J})^\T$.
Let $\ep \sim \mN(0,1)$ be a standard normal random variable. 
Let
\beginy\label{eq:a}
\ml  \sim \ep_0  \mid \{\|\ep_0\|_2^2\leq a_0\},\quad 
\mt_\ts   \sim  \ep_\ts    \mid   \{ |\ep_\ts   |  \leq  a \},\quad 
\mt'_\ts   \sim   \ep_\ts     \mid  \{ |\ep_\ts    |  \leq  a, \  \|\ep_\ts   \|_\mm  \leq a_0 \} 
\endy
be three truncated normal random vectors independent of $\epsilon$, with $\ep_0 \sim \mN(0_J,I_J)$ and
$\ep_\ts      \sim \mn\{0_J, D( \vx)\}$. 
Proposition \ref{prop:t} below gives the asymptotic sampling distributions of $\hts \ (\nfl)$ under the three \reps schemes in Definition \ref{def:t}. 
For comparison, we also include the results under \rem to highlight the connection \citep{ZDfrt}.
Let  
$
\ma_\textrm{rem} = \{\|\htx\|_\mm \leq a_0\}
$
denote the acceptance criterion under ReM with threshold $a_0$. 
Let $\hth \mid \ma$ represent the distribution of $\hth$ under rerandomization with acceptance criterion $ \ma$.

\begin{proposition}\label{prop:t}
Assume Condition \ref{asym}. Then 
\begina
 \wa
\begin{array}{lll}
\rtn (\hts   - \tau)  \mid \marem  &\ \rs \ & 
\vl^{1/2} \epsilon + c_*^\T \vx^{-1/2}  \ml,\\
\rtn (\hts   - \tau) \mid \majtt  &\ \rs \ & 
\vl^{1/2} \epsilon + c_*^\T \vx^{-1/2} \ml,\\
 \rtn (\hts   - \tau) \mid \mamgt &\ \rs \ & \vl^{1/2} \epsilon +  c_*^\T \vxinv \sigma(\vx)    \mt_\ts ,\\
 \rtn (\hts   - \tau) \mid \macst &\ \rs \ & \vl^{1/2} \epsilon +  c_*^\T \vxinv \sigma(\vx)  \mt'_\ts 
 \end{array} 
\enda 
for $\nf$, whereas 
$ \rtn (\htl  - \tau)  \mid \ma  \rs \mathcal{N}(0, v_\lin)$
 for all $\ma \in \{\marem, \matjt, \matmg, \matcs \}$.
\end{proposition}

Proposition \ref{prop:t} has two implications. 
First, 
all three estimators remain consistent under all four rerandomization schemes, with the joint rule being asymptotically equivalent to ReM.
Second, 
the asymptotic distributions of $\htl$ remain the same as that under complete randomization in Lemma \ref{lem:cre}, whereas 
  those of $\htn$ and $\htf$ change to  convolutions of normal and truncated normal when their respective $c_* \neq 0_J$.
We show in Lemma \ref{lem:peak} in the {\sm} that $\ml \succeq \ep_0$ and $\mt_\ts, \mt'_\ts \succeq \ep_\ts$ by the Gaussian correlation inequality  due to \cite{royen}.
This provides the basis for quantifying the impact of ReP on the asymptotic efficiency of each $\hts$, as well as the asymptotic relative efficiency of $\hts$ across $\nfl$
We formalize the intuition in Theorem \ref{thm:t} below. 
Let $\rho(J, a_0) =  \pr(\chi_{J+2} \leq a_0)/ \pr(\chi_{J} \leq a_0)< 1$.

\begin{theorem} \label{thm:t}
Assume Condition \ref{asym}. For all $\ma \in \{\marem, \matjt, \matmg, \matcs \}$, we have 
\begine[(i)]
\item\label{item:x} 
\begin{equation}
\label{eq::improve-tre}
(\htx \mid \ma) \succi \htx 
\end{equation}
with $\ \ \dfrac{\ea (\|\htx\|_2^2 \mid \matjt)}{\ea (\|\htx\|_2^2)} =\dfrac{\ea (\|\htx\|_2^2 \mid \marem)}{\ea (\|\htx\|_2^2)} = \rho(J, a_0)$;
\item\label{item:eff}
\renewcommand{\arraystretch}{1.2}
\beginy \label{eq::improve-tre2}
\begin{array}{l}
(\htn \mid \ma ) \succi \htn, \quad(\htf \mid \ma ) \succi \htf, \quad (\htl \mid \ma) \asim \htl, \\
 (\htl \mid \ma) \succi (\hts \mid \ma )  \quad\text{for}\quad \nf, 
 \end{array}
\endy
with $ (\htn \mid \ma ) \asim \htn \asim \htl$  if and only if $c_\neyman = 0_J$ and $ (\htf \mid \ma ) \asim \htf \asim \htl$  if and only if $c_\fisher = 0_J$; 
\item\label{item:cb} for $* \in \{\neyman, \fisher, \lin\}$,  the asymptotic conditional bias of $\hts$ given $\htx$ satisfies
\beginy\label{eq:improve-cb}
\dfrac{\ea \left[\left\{ \ea (\htau_{*} - \tau  \mid \htx, \ma)\right\} ^2\right]}
{\ea \left[\left\{ \ea (\htau_{*} - \tau  \mid \htx)\right\} ^2\right]}  \leq 1 
\endy
with $\ \ \dfrac{\ea \left[\left\{ \ea (\htau_{*} - \tau  \mid \htx, \matjt)\right\} ^2\right]}
{\ea \left[\left\{ \ea (\htau_{*} - \tau  \mid \htx)\right\} ^2\right]} 
= 
\dfrac{\ea \left[\left\{ \ea (\htau_{*} - \tau  \mid \htx, \marem)\right\} ^2\right]}
{\ea \left[\left\{ \ea (\htau_{*} - \tau  \mid \htx)\right\} ^2\right]} 
= \rho(J, a_0)$; 
\item\label{item:dis} for $* \neq ** \in \{\neyman, \fisher, \lin\}$, 
\beginy\label{eq:improve-dis}
\frac{\ea \left\{(\htau_{*} - \htau_{**})^2 \mid \ma\right\}}
{\ea\left\{(\htau_{*} - \htau_{**})^2\right\}} \leq 1 
\endy
with $\ \dfrac{\ea \left\{(\htau_{*} - \htau_{**})^2 \mid \matjt\right\}}
{\ea \left\{(\htau_{*} - \htau_{**})^2 \right\}}
= 
\dfrac{\ea \left\{(\htau_{*} - \htau_{**})^2 \mid \marem\right\}}
{\ea \left\{(\htau_{*} - \htau_{**})^2 \right\}} = \rho(J, a_0)$.
\ende
\end{theorem}

\bigskip 

For a random quantity $\hat\theta$, $(\hth\mid\ma) \succi \hth$ implies that rerandomization increases the asymptotic peakedness of $\hth$, whereas $(\hth \mid \ma) \asim \hth $ implies that rerandomization has no effect asymptotically. 
The implications of Theorem \ref{thm:t} are hence threefold. 
First, Theorem \ref{thm:t}\eqref{item:x} establishes  the utility of  the two-sample $t$-test-based \reps to improve covariate balance in terms of the asymptotic distribution of $\htx$. This gives another measure of improved covariate balance in addition to the self-evident improvement in the realized allocation.
Second,  Theorem \ref{thm:t}\eqref{item:eff} shows the utility of ReP in improving the asymptotic efficiency of $\htn$ and $\htf$, and ensures the asymptotic efficiency of $\htl$ over $\htn$ and $\htf$ under all three {\repsch} with the asymptotic efficiency unaffected by the rerandomization. 
We thus recommend the combination of ReP and $\htl$ for subsequent inference with details given in Section \ref{sec:infer}.
Third, Theorem \ref{thm:t}\eqref{item:cb}--\eqref{item:dis} are direct consequences of Theorem \ref{thm:t}\eqref{item:x} and illustrate the utility of ReP in reducing conditional biases and improving coherence across $\hts \ (\nfl)$.
In particular, we use $\ea (\htau_{*} - \tau  \mid \htx, \ma)$ and $\ea (\htau_{*} - \tau  \mid \htx)$ to measure the asymptotic conditional biases of $\hts$ given $\htx$ under rerandomization and complete randomization, respectively, in Theorem \ref{thm:t}\eqref{item:cb}, and use $\ea \{(\htau_{*} - \htau_{**})^2 \mid \ma\}$ and $\ea\{(\htau_{*} - \htau_{**})^2\}$ to measure the coherence between estimators $\hts$ and $\htau_{**}$ in Theorem \ref{thm:t}\eqref{item:dis}.

These implications together illustrate the value of ReP: despite having no effect on the asymptotic efficiency of $\htl$, ReP promotes not only covariate balance between treatment groups but also more coherent inferences across different estimators. We thus recommend the combination of ReP and fully interacted adjustment to ensure both covariate balance and efficient inference.
The theory of most previous papers on rerandomization focused on the efficiency gain by ReM analogous to Theorem \ref{thm:t}\eqref{item:eff}. Here we give more results on the multiple benefits of ReP. 

The asymptotic equivalence of $(\htl \mid \ma)$ and $\htl$ for $\ma \in \{\marem, \matjt, \matmg, \matcs \}$ in Proposition \ref{prop:t} and Theorem \ref{thm:t}\eqref{item:eff} is no coincidence but the consequence of $\htl$ being asymptotically independent of $\htx$ under complete randomization from Lemma \ref{lem:cre}. 
Balance criteria based on $\htx$ thus have no effect on $\htl$ asymptotically, with $\marem, \matjt, \matmg$, and $\matcs$ all being special cases. 
The same argument underpins the asymptotic equivalence of $\hts$ and $(\hts  \mid \ma)$ for $* \in \{\nm, \fisher\}$ when $c_* = 0_J$ under special configurations of the potential outcomes; an example is that $c_\fisher = 0_J$ when the individual treatment effects $\tau_i$ are constant across all $i$'s.
The resulting $\hts \ (\nf)$ is asymptotically identically distributed as $\htl$ under complete randomization by Lemma \ref{lem:cre}, with the asymptotic sampling distribution unaffected by rerandomization. 
 
More generally, the linear projection of $\hts$ on $\htx$ equals $
\proj( \hts \mid \htx) = \tau + c_*^\T v_x^{-1} \htx
$
with regard to the asymptotic distribution under complete randomization in Lemma \ref{lem:cre}, and is asymptotically independent of the corresponding residual, denoted by $
\res(\hts \mid \htx) = \hts - \proj(\hts \mid \htx)  = \hts - \tau - c_*^\T v_x^{-1} \htx$. This ensures
\beginy\label{eq:hts}
\hts = \proj(\hts\mid \htx) + \res(\hts \mid \htx) = \tau + c_*^\T v_x^{-1} \htx + \res(\hts \mid \htx), 
\endy
where the residual  $\res(\hts \mid \htx)$ satisfies  $\sqrtn \res(\hts \mid \htx) \rs \mn(0, v_\lin)$ and is asymptotically independent of $\htx$. 
Balance criteria based on $\htx$ can thus only affect the $c_*^\T v_x^{-1} \htx$ part in \eqref{eq:hts} asymptotically, and turn it into a truncated normal with greater peakedness when $c_* \neq 0_J$. This gives the intuition behind Proposition \ref{prop:t} and Theorem \ref{thm:t}.

 \subsection{Wald-type inference}\label{sec:infer}
Proposition \ref{prop:t} and Theorem \ref{thm:t} together establish the asymptotic distributions and relative efficiency of $\hts \ (\nfl)$ under the two-sample $t$-test-based \rep.  
The results provide two guidelines on subsequent Wald-type inference of the average treatment effect.

First, the Wald-type inference based on $\htl$ is asymptotically the most efficient and can be conducted using the same normal approximation as under complete randomization. 
Specifically, let $\hse_\lin$ denote the \ehws standard error of $\htl$ from the same \ols fit. 
\cite{Lin13} and \citet[][Lemma A16]{LD2018} ensured that it is asymptotically appropriate for estimating the true standard error of $\htl$ under both complete randomization and  the three \repschs in Definition \ref{def:t},  justifying the Wald-type inference based on $(\htl, \hse_\lin)$ and normal approximation. This illustrates the advantage of ReP for allowing for convenient regression-assisted analysis by the fully interacted regression.

Second, \reps narrows the asymptotic sampling distributions of $\htn$ and $\htf$ in general, rendering the usual normal approximation  based on $\hts \ (\nf )$ and their associated {\ehw} standard errors overconservative for estimating $\tau$. 
Rerandomization-specific sampling distributions are thus necessary for better-calibrated inference based on $\hts \ (\nf)$.
Recall the definition of $(\hat\gamma_{\lin, 0},  \hat\gamma_{\lin,1})$ from Section \ref{sec:gamma}, as the sample analogs of $(\gamma_0, \gamma_1)$. 
With $v_\lin$ and $c_*$ being the only unknowns in the asymptotic distributions of $\hts \ (\nf)$ in Proposition \ref{prop:t}, we can estimate them using $N\hse_\lin^2$ and the sample analogs
$
\hat c_\nm =\sxx(e_0^{-1}\hg_{\lin,0} + e_1^{-1} \hg_{\lin,1})$ and $\hat c_\fisher  =   \sxx(e_1^{-1}-e_0^{-1})(\hg_{\lin,1}-\hg_{\lin,0})$,
respectively,
and conduct inference based on the resulting plug-in distributions \citep{LD2018}. 
This modification mitigates the overconservativeness of the Wald-type inference based on $\hts \ (\nf)$ under ReP and, by contrast, illustrates the convenience of \cite{Lin13}'s method for efficient and well-calibrated inference.

 

\section{ReP based on linear and logistic regressions }
\label{sec:lm_glm}
\subsection{Linear and logistic regressions for assessing covariate balance}\label{sec:4.1}
The two-sample $t$-tests measure covariate balance by the difference in covariate means and are numerically equivalent to a component-wise regression of $x_i$ on $(1,Z_i)$, assessing how $x_i$ varies with different values of $Z_i$.
The idea of the propensity score \citep{rosenbaum1983central}, on the other hand, motivates an alternative measure of covariate balance by assessing how $Z_i$ varies with $x_i$. 

Consider the linear regression of  $Z_i$ on $(1, x_i)$, denoted by $\lmt(Z_i \sim 1+x_{i1} + \cdots + x_{iJ})$. 
Let $\hb_j$ denote the coefficient of the $j$th covariate $x_{ij}$ for $j = \ot{J}$.
The magnitude of $\hb_j$ gives an intuitive measure of the influence of covariate $j$ on the final assignment, with a well-balanced assignment expected to have all $\hb_j$'s close to zero; 
see, e.g., \citet[Table 1]{mmw} and \citet[Table 3]{knss} for balance tables based on these regression outputs. 
This motivates three linear regression-based {\repsch} under the marginal, joint, and consensus rules, respectively. 

To begin with, denote by $\pjlm$ the $p$-value associated with $\hb_j$ from standard software packages. The marginal rule accepts a randomization if and only if $\pjlm \geq \alpha_j$ for all $\jej $ for some prespecified thresholds $\alpha_j \in (0,1)$. 

Alternatively, let $\pzlm$ be the $p$-value from the $F$-test of $\lmt(Z_i \sim 1+x_{i1} + \cdots + x_{iJ})$ against the empty model $\lmt(Z_i \sim 1)$. 
It is a standard output of linear regression by most software packages and provides a summary measure of the magnitudes of all $\hb_j$'s. 
The joint rule then accepts a randomization if and only if $\pzlm \geq \alpha_0$ for some prespecified threshold $\alpha_0 \in (0,1)$. 
This is the recommendation by \cite{gerber2012field}.

The consensus rule, accordingly, accepts a randomization if and only if it is acceptable under both marginal and joint rules with $\pjlm\geq \alpha_j$ for all $j = \zt{J}$. 
This extends the three two-sample $t$-test-based criteria in Definition \ref{def:t} to the linear regression of $Z_i$ on $(1,x_i)$. 

One concern with the above approach based on $\lmt(Z_i \sim 1+x_{i1} + \cdots + x_{iJ})$ is that linear regression is in general not intended for  binary responses like $Z_i$. 
An immediate alternative is to consider  the logistic regression of $Z_i$ on $(1,x_i)$, denoted by $\glmt(Z_i \sim 1+x_{i1} + \cdots + x_{iJ})$, instead and form the acceptance criteria based on $p$-values from its \mles fit \citep{hansen2008}. 

Specifically, let $\pjglm$ be the $p$-value associated with the coefficient of $x_{ij}$ from $\glmt(Z_i \sim 1+x_{i1} + \cdots + x_{iJ})$ for $\jej$, and let $\pzglm$ be the $p$-value from the likelihood ratio test (\lrt) of $\glmt(Z_i \sim 1+x_{i1} + \cdots + x_{iJ})$ against the empty model $\mlogitt(Z_i \sim 1)$.
They are all standard outputs of logistic regression by most software packages, and allow us to form the marginal, joint, and consensus criteria in identical ways as those based on $\{\pjx: j = \zt{J}\}$ for $\x= \ts, \lm$.

This defines in total nine \repsch, as the combinations of three \emph{model options}--- the two-sample $t$-tests of $x_i$ (``t''), the linear regression of $Z_i$ on $(1, x_i)$ (``\lm''), and the logistic regression of $Z_i$ on $(1, x_i)$ (``\logit'')---and the marginal (``\m''), joint (``\jt''), and consensus (``\css'') rules, summarized in  
Table \ref{tb:criteria}.  
We extend below the results under the two-sample $t$-test-based schemes to the six regression-based variants, respectively.

\begin{table}[t]\caption{\label{tb:criteria}Nine \repschs under the treatment-control experiment}
\begin{center}

\renewcommand{\arraystretch}{1.5}
\begin{tabular}{c c }
\hline
Rule  &  Model option: $\mos$ \\\hline
marginal (\m)  &  {$p_{j,\x} \geq \alpha_j \ (\jej )$}   \\ 
joint (\jt)  &  {$p_{0,\x} \geq \alpha_0  $}  \\ 
consensus (\css)  &  {$p_{j,\x}\geq \alpha_j \ (j = \zt{J})$}\\
\hline
\end{tabular}

\end{center}
\end{table}

\subsection{Asymptotic theory}

We derive in this subsection the asymptotic sampling properties of $\hts\ (\ms)$ under the six linear or logistic regression-based \repschs in Table \ref{tb:criteria}. 
Echoing the comments after Theorem \ref{thm:t}, the result illustrates the utility of the six regression-based \reps schemes in improving covariate balance, reducing conditional biases, and  promoting coherence across different estimators, and establishes the asymptotic efficiency of  $\htl$ under all six criteria with the asymptotic sampling distribution unaffected by \rep. 

To this end, we first introduce an additional regularity condition that underpins the design-based properties of logistic regression.  

\begin{condition}\label{cond:glm}
Let 
\begina
\pi(\txi , \theta) = \frac{\exp(\txit \theta)}{1+\exp(\txit \theta)}  \fort \text{$\txi  = (1, x_i^\T)^\T$ and $\theta  \in \mathbb{R}^{J+1}$.}
\enda
As $N\to \infty$, 
$
H (\theta)  = -N^{-1} \sumi \pi(\txi , \theta)\{1-\pi(\txi ,\theta)\} \txi  \txit  
$ 
converges to a negative-definite matrix $H_\infty(\theta)< 0$ for all $\theta \in \mathbb{R}^{J+1}$, and the convergence is uniform on any compact set $\Theta \subset \mr ^{J+1}$.
\end{condition}

The $H(\theta)$ in Condition \ref{cond:glm} gives the Hessian matrix of the log-likelihood function under the logistic model scaled by $N^{-1}$.
To gain intuition about the uniform convergence in Condition \ref{cond:glm}, consider a superpopulation working model where the $x_i$'s are independent and identically distributed with finite second moment.
The uniform law of large numbers holds for $H(\theta)$ and ensures that it converges to some $H_\infty(\theta)$ almost surely on $\mr^{J+1}$, with the convergence being uniform on any compact set $\Theta\subset \mathbb{R}^{J+1}$ \citep[][Chapter 16]{Newey1994, Ferguson1996}.
This suggests the mildness of the requirement on uniform convergence. 
In addition, we can show that $H(\theta) \leq 0$ for all $\theta \in \bbr^{J+1}$. 
This suggests the mildness of the requirement on the negative definiteness of $H_\infty(\theta)$.  

Let 
$
\mamojt = \{p_{0,\x} \geq \alpha_0\}$, 
$\mamomg = \{p_{j,\x}\geq \alpha_j \ \fall \jej \}$, and $ \mamocs =\mamojt  \cap\mamomg  =\{p_{j,\x}\geq \alpha_j \ \fall j = \zt{J}\}$
denote the acceptance criteria under the six regression-based \reps schemes for $ \x = \lm, \logit$. 
Recall the definitions of $v_x$, $\ep$, $\ml$, $a_0$, and $a = (a_{1}, \dots, a_{J})^\T$ from Section \ref{sec:asym_t}.
Let 
$$
\mtm \ \sim \ \epm   \mid   \{ |\epm  |  \leq  a\}, \qquad
\mtm' \     \sim \   \epm    \mid  \{ |\epm|  \leq  a, \  \|\epm  \|_\mm  \leq a_0 \}$$
be two truncated normal random vectors independent of $\epsilon$, with  $ \epm    \sim \mn\{0_J, D(\vxinv)\}$. The {\gci} ensures that $\mtm,  \mtm' \succeq  \epm$; see Lemma \ref{lem:peak} in the {\sm}.
This underlies the improved asymptotic efficiency of $\htn$ and $\htf$ under regression-based ReP. We state the details in Proposition \ref{prop:lm_glm} and Theorem \ref{thm:lm_glm} below.

\begin{proposition} \label{prop:lm_glm}
Assume Condition \ref{asym} for $\x = \lm$ and Conditions \ref{asym}--\ref{cond:glm} for $\x = \logit$. For $\x \in \{\lm, \logit\}$, we have
\begina
\wa
\begin{array}{lll}
\rtn (\hts   - \tau) \mid \mamojt  &\ \rs \ & 
\vl^{1/2} \epsilon + c_*^\T \vx^{-1/2}  \ml ,\\
 \rtn (\hts   - \tau) \mid \mamomg &\ \rs \ & \vl^{1/2} \epsilon +  \cs^\T \dd  \mtm  ,\\
 \rtn (\hts   - \tau) \mid \mamocs &\ \rs \ & \vl^{1/2} \epsilon +   \cs^\T \dd  \mtm' ,
 \end{array} 
\enda 
for $\nf$, 
whereas 
$ \rtn (\htl  - \tau)  \mid \ma  \rs \mathcal{N}(0, v_\lin)$
 for all $\ma \in \{\mamodr: \x = \lm, \logit; \ \drs \}$. 
 \end{proposition}

\begin{theorem} \label{thm:lm_glm}
Theorem \ref{thm:t} holds for all $\ma \in \{\mamodr: \x = \lm, \logit; \ \drs \}$. 
\end{theorem}

All comments after Proposition \ref{prop:t} and Theorem \ref{thm:t} extend here with no need of modification. 
Proposition \ref{prop:lm_glm} gives the asymptotic sampling distributions of $\hts \ (\nfl)$ under the six regression-based \repsch, and establishes the asymptotic equivalence of the linear and logistic regression model options under all three rules.
As a direct implication of Proposition \ref{prop:lm_glm}, Theorem \ref{thm:lm_glm} highlights the utility of the regression-based \reps in improving covariate balance, reducing conditional biases, and promoting coherence between adjusted and unadjusted analyses, and 
ensures the asymptotic efficiency of $\htl$ under all six criteria. 
The fully interacted regression is thus our recommendation for subsequent inference under the linear and logistic regression-based \reps as well, with all discussion in Section \ref{sec:infer} extending here verbatim.

Juxtapose Proposition \ref{prop:lm_glm} with Proposition \ref{prop:t}.
The three joint criteria are asymptotically equivalent to not only each other but also \rem with threshold $a_0$.
This is no coincidence but the consequence of the test statistics used by these criteria all being asymptotically equivalent to $\|\htx\|_\mm$; see Remark \ref{rmk:proof} in the {\sm} for details. 
The marginal criteria based on the linear and logistic regressions, on the other hand, differ from that based on the two-sample $t$-tests even asymptotically. 
The difference is nevertheless immaterial based on simulation evidence. 

Echoing the comments at the end of Section \ref{sec:setup_1}, we view the linear and logistic regressions as purely numeric procedures based on \ols or \mles for computing the $p$-values and estimators, and invoke none of the underlying modeling assumptions in evaluating their outputs.
The results in Proposition \ref{prop:lm_glm} and Theorem \ref{thm:lm_glm} are therefore all design-based, and hold regardless of how well the underlying  linear or logistic models represent the true data-generating processes.
This concludes our discussion on \reps under the treatment-control experiment. 
The criteria based on two-sample $t$-tests are arguably the most straightforward, making the discussion on the regression-based variants  seem to be of theoretical interest only.
The logistic regression nevertheless provides a key stepping stone for extending the current results to experiments with more than two treatment arms. 
We give the details in the next section. 

\section{ReP in multi-armed experiments }\label{sec:Q_main}
 \subsection{Basic setting and covariate balance criteria}
Multi-armed experiments enable comparisons of more than two treatment levels simultaneously, and are intrinsic to applications with multiple factors of interest.
To conduct \rrs  in such settings, 
a straightforward option is to check balance for all pairs of treatment arms, and accept a randomization if and only if all pairwise comparisons pass the balance check \citep{morgan2011}. 
Depending on the number of treatment arms in question, however,  this may result in a large number of pairwise comparisons and become unwieldy in practice. 
A more practical alternative is to use a test that directly measures the balance across all treatment arms. 

To this end, the covariate-wise $F$-test provides  a natural way of extending the marginal two-sample $t$-test to more than two treatment arms,
measuring the balance of individual covariates across all treatment arms simultaneously. See, e.g., 
\cite{mmw2} and \cite{dr2} 
for balance tables based on such $F$-tests.
The \mlr, on the other hand,  is a straightforward extension of the logistic regression and provides a way to measure both covariate-wise and overall balances across all treatment arms by the idea of the propensity score. 
See \cite{gkb} for an example of  balance check based on the \mlr.
We formalize below their extensions to ReP.

Consider a multi-armed experiment with $Q > 2$ treatment levels, indexed by $\qiq   = \{\ot{Q}\}$, and a study population of $N$ units, indexed by $i = \ot{N}$. 
Renew $x_i = (x_{i1}, \ldots, x_{iJ})^\T$ as the centered covariate vector  and $Z_i \in \mq = \{1, \dots, Q\}$ as the initial treatment assignment of unit $i$.
For $j = \ot{J}$, let $\pjf$ denote the $p$-value from the marginal $F$-test on covariate $j$ based on $(x_{ij}, Z_i)_{i=1}^N$. 
The marginal $F$-test-based criterion for ReP accepts a randomization if and only if $\pjf \geq \alpha_j$ for all $\jej $ for some prespecified thresholds $\alpha_j \in (0,1)$. 
Let $\wiq = 1(Z_{i} = q)$ denote the indicator of treatment level $q$. The  $F$-test on covariate $j$ can be implemented via 
$\lmt(x_{ij} \sim 1 + \wio  + \dots + \mi_{i, Q-1})$. 

The \mlr, on the other hand, accommodates the marginal, joint, and consensus rules for ReP together via one \mles fit. 
Renew $\mlogitt(Z_i \sim 1+x_{i1}+\cdots + x_{iJ})$ as the \mlrs of $Z_i \in \mathcal Q$ on $(1, x_i)$ over $i = \ot{N}$. 
Assume without loss of generality level $Q$ as the reference level. 
The \mles fit of  $\mlogitt(Z_i \sim 1+x_{i1}+\cdots + x_{iJ})$ yields one coefficient of $x_{ij}$ for each non-reference level $\qiqp = \{\ot{Q-1}\}$, denoted by $\tb_{qj}$. 
We use the subscript $+$ to signify quantities associated with the non-reference levels. 
Let $\pqjmlr$ be the  $p$-value associated with $\tb_{qj}$ from standard software packages.
The marginal rule accepts a randomization if and only if  $\pqjg \geq \alpha_{qj}$ for all $\qiqp  $ and $\jej $ 
for some prespecified thresholds $\alpha_{qj} \in (0, 1)$. 

Alternatively, let $\pzmlr$ be the $p$-value from the \lrts of $\mlogitt(Z_i \sim 1+x_{i1}+\cdots + x_{iJ})$ against the empty model $\mlogitt(Z_i \sim 1)$.
It is a standard output of \mlrs from most software packages  and gives a summary measure of the magnitudes of $\tbqj $'s as a whole.  
The joint rule then accepts a randomization if and only if  $\pzmlr \geq \alpha_0$ for some prespecified threshold $\alpha_0 \in (0, 1)$. 
The consensus rule, accordingly, accepts a randomization if and only if it is acceptable under both the marginal and joint rules. 
This defines three additional criteria for conducting \reps  under multi-armed experiments. We summarize the definitions in Table \ref{tb:criteria_g}.

Other criteria  can be formed based on tests for multivariate analysis of variance \citep{morgan2011} or linear regression of $\wiq = 1(Z_i = q)$ on $(1, x_i)$.
These alternatives nevertheless involve more technical subtleties and can be unwieldy in practice. 
To save space, we give in this section the theory of ReP for multi-armed experiments based on marginal $F$-tests and \mlrs due to their practical convenience, 
and relegate details on the alternative criteria to the {\sm}.

\begin{table}[t]\caption{\label{tb:criteria_g} Four R\lowercase{e}P schemes under multi-armed experiments}

\renewcommand{\arraystretch}{1.5}
\begin{center}\begin{tabular}{ccc c}
\hline
  {Rule}    & $F$-test & Multinomial logistic regression \\\hline
  marginal & $\pjf \geq \alpha_j \ \fall \jej$ & $\pqjmlr \geq \alpha_{qj} \ \fall qj$  \\
   joint  & n.a. & $\pzmlr \geq \alpha_0  $\\
  consensus & n.a. & $\pqjmlr \geq \alpha_{qj} \ \fall qj, \ \pzmlr \geq \alpha_0  $ \\
\hline
\end{tabular}\end{center}
\end{table}

\subsection{Treatment effects and regression estimators}\label{sec:po_Q}
We next define the average treatment effect and regression estimators under  multi-armed experiments, extending the notation and definitions in Section \ref{sec:setup}. 
Renew $\yiq \in\mathbb R  $ as the potential outcome of unit $i$ if assigned to treatment level $\qiq = \{\ot{Q}\} $.
The observed outcome equals $Y_i = \sumq   \wiq \yiq  $ with $\wiq = 1(Z_{i} = q)$.
Renew $\byq   = N^{-1} \sumn \yiq  $ as the average potential outcome under treatment level $\qiq$, vectorized as 
$\bbY = (\bar Y(1),   \dots, \by(Q) )^\T \in \mathbb R^Q$. 
The goal is to estimate the finite-population average causal effect 
\beginy\label{eq:tau}
\tau = G \bY
\endy 
for some prespecified contrast matrix $G$ with rows orthogonal to $1_Q$. 
The $\tau = \bar Y(1) - \bar Y(0)$ under the treatment-control experiment 
is a special case with $\by = (\by(1), \by(0))^\T$  and $G = (1,-1)$.

Consider 
\begina
\neyman:&&\lmt(Y_i \ \sim \  \wio  + \dots + \wiqq ),   \\ 
\fisher:&&\lmt(Y_i \ \sim \  \wio  + \dots + \wiqq + x_i),  \\
\lin:&& \lmt(Y_i \ \sim \ \wio  + \dots + \wiqq  + \wio x_i + \dots + \wiqq x_i)  
\enda
 as the unadjusted, additive, and fully interacted linear regressions of $Y_i$ on $\{\wiq: q\in\mathcal Q\}$  and $x_i$, respectively, indexed by $* =$  {\neyman} (unadjusted), {\fisher} (additive), and {\lin} (fully interacted).
Let 
$$\hys =( \hys(1), \dots, \hys(Q))^\T \quad (\nfl)$$ denote the coefficient vectors of $( \wio,\dots,\wiqq)^\T$ from these  three regressions, respectively.
They are consistent for estimating $\by$ under complete randomization \citep{lu2016covariate, zdca}, and allow us to estimate $\tau = G\bar Y$ by
\begina
\hts = G\hys \qquad (\ms).
\enda
Of interest are the validity and relative efficiency of $\hts$'s under \rep.
We give the details in the next subsection.

\subsection{Asymptotic theory}\label{sec:thm_Q}
We present in this subsection the asymptotic theory of \reps under multi-armed experiments. Assume throughout that the initial allocation is obtained by complete randomization.
The experimenter assigns completely at  random $N_q > 0$ units to level $\qiq  $ with $\sumq  N_q  = N$, and accepts the allocation if and only if the assignments satisfy the prespecified covariate balance criterion.

\subsubsection{Baseline efficiency under complete randomization}
%
Recall the definitions of $\hx(q) = N_q ^{-1}\sumiq x_{i}$, $e_q = N_q/N$, $\gq$, and Condition \ref{asym} in Sections \ref{sec:setup}--\ref{sec:ttest} under the treatment-control experiment. 
Renew them  for multi-armed experiments with $q\in\mq = \{\ot{Q}\}$.
Let $\hx = (\hx(1)^\T, \dots, \hx(Q)^\T)^\T \in \mathbb R^{JQ}$ and $\gamma_\fisher = \sumq e_q \gamma_q$. 
Lemma \ref{lem:clt_Q} below follows from \cite{zdca} and states the asymptotic distributions of $\hys \ (\nfl)$ under complete randomization. 
The results ensure  $\htl \succi \htn , \htf$ under complete randomization and provide the baseline for evaluating the efficiency gains under ReP.

\begin{lemma}\label{lem:clt_Q}
Under complete randomization and  the multi-armed version of Condition \ref{asym}, we have
\begina
\sqrtn\beginp
\hys -\by\\
\hx
\endp &\rs &  \mn\left\{0_{Q+JQ}, \beginp
V_* & \Gamma_* V_x\\
V_x\Gamma_*^\T & V_x
\endp\right\} \qquad(\nfl) 
\enda
with $V_x = N\cov(\hx) = \{\diag(e_q ^{-1})_{\qiq } -1_{Q\times Q}\} \otimes \sxx$, 
\begina
\Gamma_\neyman = \diag(\gq ^\T)_{\qiq  }, \quad\Gamma_\fisher = \diag\{ (\gq  - \gp )^\T\}_{\qiq  }, \quad\Gamma_\lin = 0_{Q\times JQ},
\enda
 and
$V_* = V_\lin + \Gamma_* \Vx \Gamma^\T_* \geq V_\lin$ for $\nfl$. 
We give the explicit expressions of $V_*$'s in the  \sm.
\end{lemma}

\subsubsection{\reps based on the marginal $F$-tests}
Let 
$
\maf = \{\pjf \geq \alpha_j \ \fall \jej \}
$
denote the acceptance criterion under \reps  based on the marginal $F$-tests. 
Renew $\ep \sim \mn(0_Q, I_Q)$ as a $Q\times1$ standard normal random vector. 
Let 
$\mtf   \sim    \epf \mid \{ \sumq   e_q   \ep_{\ff ,qj} ^2 \leq a'_j  \sxj \ \fall   \jej  \}$
be a truncated normal random vector independent of $\ep$, 
where $\epf = (\ep_{\ff,qj})_{q \in \mq; \ j = \ot{J}} \sim \mn(0_{JQ},\Vx)$, $\sxj = (N-1)^{-1}  \sumi x_{ij}x_{ij}^\T$, and $a'_j $ denotes the $(1-\alpha_j)$th quantile of the $\chi _{Q-1}^{2} $ distribution.
Lemma \ref{lem:peak} in the {\sm} ensures $\mtf\succeq  \epf$ by the \gci.

\begin{proposition}\label{prop:F}
Assume  the multi-armed version of Condition \ref{asym}. Then  
\begina
\sqrtn(\hys -\bar Y) \mid \maf &\rs&   \vr^{1/2}  \epsilon + \Gamma_*\mtf  \quad(\nf ),\\
\sqrtn(\hyl -\bar Y) \mid \maf  &\rs & \mn(0_Q, \vr).
\enda 
\end{proposition}

Renew $\htx = (G_x \otimes I_J) \hx$, where $G_x$ is some prespecified contrast matrix with rows orthogonal to $1_Q$. It defines a general measure of the difference in $\{\hx(q): q\in\mq\}$, extending $\htx = \hx(1) - \hx(0)$  under the treatment-control experiment to multi-armed experiments \citep{zdca}. 

\begin{theorem}\label{thm:F}
Assume Condition \ref{asym}. 
\begine[(i)]
\item Equations \eqref{eq::improve-tre}--\eqref{eq::improve-tre2} hold for $\htx = (G_x \otimes I_J) \hx$, $\hts = G\hys \ (\nfl)$, and $\mathcal{A}= \maf$
 for  arbitrary $G$ and $G_x$.
In particular, 
$ (\htn \mid \maf ) \asim \htn \asim \htl$ if $\Gamma_\neyman= 0_{Q\times JQ}$ and $ (\htf \mid \maf ) \asim \htf \asim \htl$ if $\Gamma_\fisher = 0_{Q\times JQ}$. 
\item Analogous to equations \eqref{eq:improve-cb}--\eqref{eq:improve-dis},  for $* \neq ** \in \{\neyman, \fisher, \lin\}$, we have 
\begina
\dfrac{\ea \left[\big\| \ea (\htau_{*} - \tau  \mid \htx, \maf)\big\| ^2_2\right]}
{\ea \left[\big\|\ea (\htau_{*} - \tau  \mid \htx)\big\|_2^2\right]}  \leq 1,\qquad
\frac{\ea \left\{\big\| \htau_{*} - \htau_{**}\big\| _2^2 \mid \maf\right\}}
{\ea\left\{\big\| \htau_{*} - \htau_{**}\big\|_2^2\right\}} \leq 1.  
\enda 
\ende 
\end{theorem}

Echoing the comments after Theorems \ref{thm:t} and \ref{thm:lm_glm}, 
Theorem \ref{thm:F}  illustrates the utility of the marginal $F$-test-based \reps for improving covariate balance, reducing conditional biases, and promoting coherence across different estimators under multi-armed experiments, and ensures the asymptotic efficiency of $\htl$ with identical asymptotic sampling distribution as under complete randomization.
Subsequent inference can thus be conducted based on $\htl$ and its \ehws covariance  in full parallel with the discussion in Section \ref{sec:infer}. We relegate the details to the {\sm}.

\subsubsection{\reps based on the \mlr}
We now address the \repschs based on the \mlrs of $Z_i$ on $(1,x_i)$. 
Condition \ref{cond:glm_Q} below extends Condition \ref{cond:glm} 
to multi-armed experiments and underpins the design-based properties of  \mlr.

\begin{condition}\label{cond:glm_Q}
For $\txi  = (1, x_i^\T)^\T$ and $\theta = (\theta_1^\T, \dots, \theta_{Q-1}^\T)^\T \in \bbr^{(J+1)(Q-1)}$ with $\theta_q  \in \mathbb{R}^{J+1}$, let 
\begina
\pi_q (\txi , \theta) =  \frac{\exp(\txbi_q )}{1+\sum_{q' \in \qp} \exp(\txbi_{q'}) }\qquad (\qiqp),
\enda
and let 
$H(\theta) = (H_{qq' }(\theta))_{q,q' \in\mqp}$ with
%
$
H_{qq'}(\theta) =N^{-1} \sumi  \pi_q (\txi , \theta)\{\pi_{q'}(\txi , \theta)  - 1(q=q')\} \txi \txit.
$
As $N\to \infty$, $H(\theta)$
converges to a negative-definite matrix $H_\infty(\theta)< 0$ for all $\theta \in\mathbb{R}^{(J+1)(Q-1)}$, and the convergence is uniform on any compact set $\Theta \subset \mathbb{R}^{(J+1)(Q-1)}$.
\end{condition}

The same intuition about Condition \ref{cond:glm} extends here, with $H(\theta) \leq 0$ giving the Hessian matrix of the scaled log-likelihood function under the multinomial logistic model. 

Let 
$
\ma_{\mlogit, \jt} = \{\pzmlr \geq \alpha_0\}$, 
$\ma_{\mlogit, \m}  =  \{\pqjmlr \geq \alpha_{qj}\ \fall qj\}$, and
$ \ma_{\mlogit, \css} =\ma_{\mlogit, \jt} \cap \ma_{\mlogit, \m}= \{  \pzmlr \geq \alpha_0;  \ \pqjmlr \geq \alpha_{qj} \ \fall  qj\}
$
denote the acceptance criteria under the joint, marginal, and consensus rules, respectively. 

Let $a_{qj}$ be the $(1-\alpha_{qj}/2)$th quantile of the  standard normal distribution for $\qiqp$ and $\jej$.
Without introducing new notation, renew $a_0$ as the $(1-\alpha_0)$th quantile of the $\chi^2_{J(Q-1)}$ distribution, and renew $a = (a_{qj})_\qj$ as the vectorization of $a_{qj}$ in lexicographical order.
The definitions of $a_0$ and $a$ reduce to those in \eqref{eq:a}  
with $(\alpha_j, a_j) = (\alpha_{1j},a_{1j})$ when $Q=2$.

Recall that $\epsilon \sim \mn(0_Q, I_Q)$.
Let 
\begina
\ml \, \sim\,\ep_0  \mid \{\|\ep_0\|_2^2\leq a_0\}, \quad
\mt_\textrm{logit}  \, \sim\,    \ep_\textrm{logit}  \mid   \{ |\ep_\textrm{logit} |  \leq  a \},  \quad
\mt'_\textrm{logit} \, \sim\,  \ep_\textrm{logit}   \mid  \{ |\ep_\textrm{logit} |  \leq  a, \  \|\ep_\textrm{logit}\|_\mm  \leq a_0 \}
\enda
be three truncated normal random vectors independent of $\ep$, 
with $\ep_0 \sim \mN(0_{J(Q-1)},I_{J(Q-1)})$ and $\ep_\textrm{logit}  \sim \mn\{0_{J(Q-1)}, D (\vsi  )\}$. We have $\ml\succeq  \ep_0$ and $\mt_\mlogit,  \mt'_\mlogit\succeq  \ep_\mlogit$ by the \gci; see Lemma \ref{lem:peak} in the {\sm}.

Recall the definition of $\Gamma_* \ (\nfl)$ from Lemma \ref{lem:clt_Q}. 
Let  $\Gamma'_*  = \Gamma_* \{(I_{Q-1}, - e^{-1}_Q e_\pluss )^\T\otimes I_J\}$ for $\nfl$
with $e_\pluss  = (e_1, \dots, e_{Q-1})^\T$. 
Let $\Psi    = \{\Phi ^{-1}\diag(e_\pluss )\} \otimes  \sxxinv$ with $\diag(e_\pluss ) = \diag(e_q )_{\qiqp}$ and $
\Phi = \diag(e_\pluss )   - e_\pluss  e_\pluss ^\T$. 
We use the subscript $+$ to signify quantities associated with the non-reference levels, and give the intuition behind $\Gamma_*'$ and $\Psi$ in the \sm. 
%
Let  
\begina
\vxm = N\cov(\hxm) =  ( \rp^{-1}   -1_{(Q-1)\times (Q-1)} )\otimes \sxx, \qquad V_{\Psi  }  = N\cov(\Psi \hxm) = \Psi \vxp \Psi^\T,
\enda
with $\hxm = (\hx(1)^\T, \dots, \hx(Q-1)^\T)^\T$  and $\cov(\cdot)$ denoting covariance under complete randomization.

\begin{proposition}\label{prop:glm_Q}
Assume the multi-armed version of Condition \ref{asym} and Condition \ref{cond:glm_Q}. Then 
\begina
\wa
\begin{array}{lll}
\sqrtn(\hys -\bar Y) \mid \ma_{\mlogit, \jt} &\ \rs \ &  \vr^{1/2}  \epsilon +  \Gamma'_*  \vxm^{1/2}\ml, \\
 \sqrtn(\hys -\bar Y) \mid \ma_{\mlogit, \m} &\ \rs \ &  \vr^{1/2}  \epsilon +  \Gamma'_*  \Psi ^{-1}    \sigma(\vsi  )  \mt_\mlogit,\\
\sqrtn(\hys -\bar Y) \mid \ma_{\mlogit, \css}  &\ \rs \ &  \vr^{1/2}  \epsilon +  \Gamma'_*  \Psi ^{-1}    \sigma(\vsi  )  \mt'_\textrm{logit}
 \end{array} 
\enda 
for $\nf$, 
whereas 
$ \sqrtn(\hyl -\bar Y) \mid \ma_{\mlogit, \dr}  \rs  \mn(0_Q, \vr)$
 for $\drs$. 
\end{proposition}

\begin{theorem}\label{thm:glm_Q}
Theorem \ref{thm:F} holds if we replace all $\maf$ with $\ma_{\mlogit,\dr}$ for $\drs$. 
\end{theorem}

All comments after Theorem \ref{thm:F} extend here after changing the ``marginal $F$-test-based'' to ``\mlr-based''. 
We omit the details to avoid repetition. 
Echoing the comments after Theorem \ref{thm:lm_glm}, the results in Proposition \ref{prop:glm_Q} and Theorem \ref{thm:glm_Q} are all design-based, and hold regardless of how well the models underlying the multinomial logistic and linear regressions for rerandomization and analysis represent the true data-generating processes.
The proof of Proposition \ref{prop:glm_Q} further introduces a novel technical result on the asymptotic equivalence of the \lrts and the Wald test for logistic and multinomial logistic regressions from the design-based perspective. We relegate the details to Theorem \ref{thm:mle_Q} in the {\sm}.

\section{Numerical examples}\label{sec:simu}
We now illustrate the finite-sample properties of \reps by simulation.
Due to space limitations, we focus on the treatment-control experiment in the main paper, and relegate additional examples based on a real-world multi-armed experiment to the {\sm}.
The results are coherent with the asymptotic theory in Sections \ref{sec:ttest}--\ref{sec:Q_main}, featuring considerably improved covariate balances and the overall efficiency of $\htl$ over 
$\htn$ and $\htf$.

Consider a treatment-control experiment with $N=500$ units, indexed by $i = \ot{N}$, and treatment arm sizes $(N_0,N_1) = (400, 100)$. 
For each $i$, we draw a $J=5$ dimensional covariate vector $x_i =  (x_{i1}, \dots, x_{i5})^\T$ with $x_{ij}$ as independent Uniform$(-1,1)$, and generate the potential outcomes as  
$Y_i(0) \sim \mathcal{N}( - \sum_{j=1}^5 x^3_{ij} , 0.1^2)$ and $Y_i(1) \sim \mathcal{N}( \sum_{j=1}^5 x^3_{ij}  , 0.4^2)$. We center the $Y_i(0)$'s and $Y_i(1)$'s respectively to ensure $\tau = \by(1) - \by(0)= 0$, and fix $\{Y_i(0), Y_i(1), x_i\}_{i=1}^N$ in the simulation. 
For each iteration, we draw a random permutation of $N_1$ 1's and $N_0$ 0's  to obtain the initial allocation under complete randomization.

\begin{figure}

\centering 

 \begin{subfigure}{0.8\textwidth}
 \centering
\includegraphics[width =\textwidth]{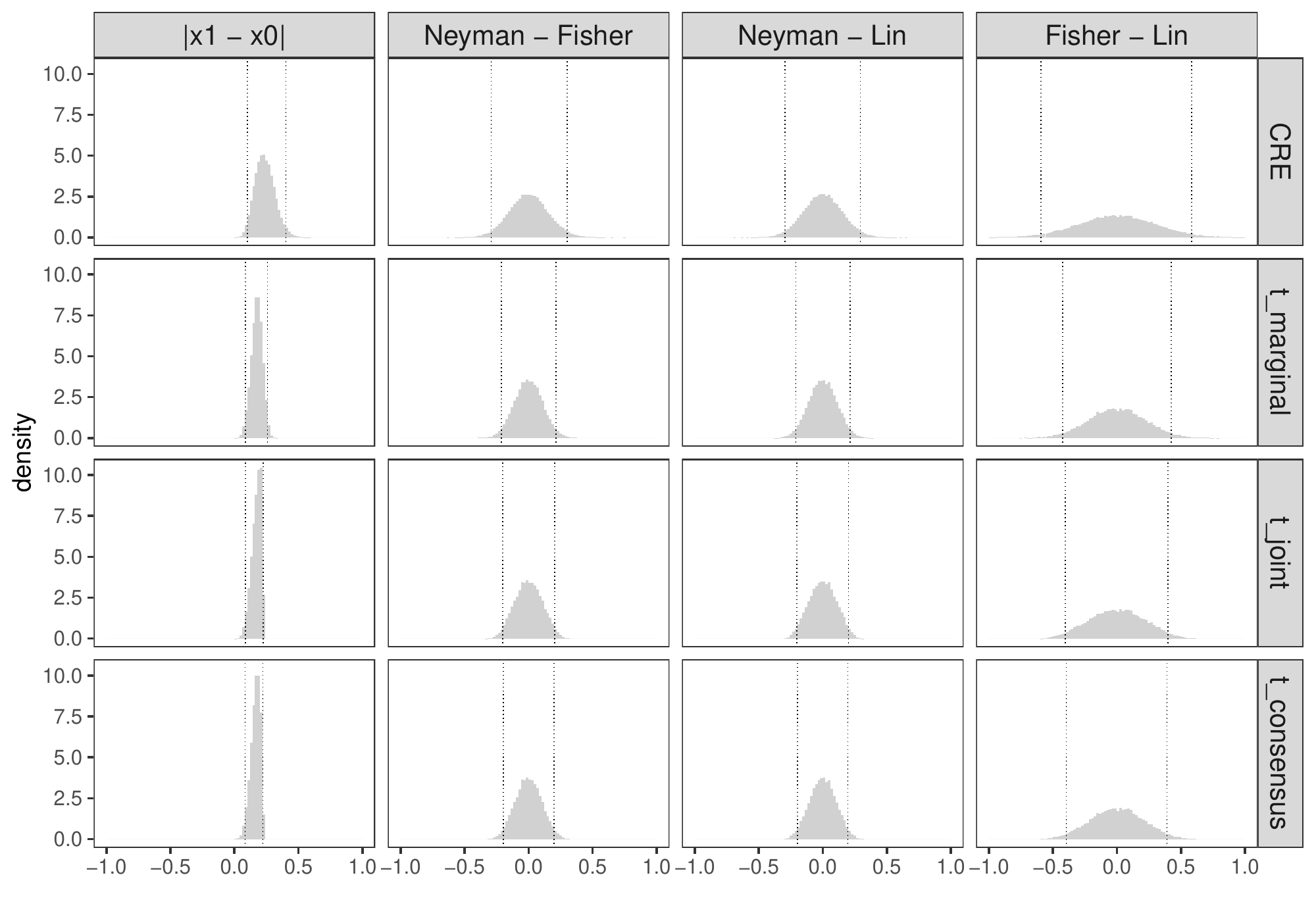}
\label{fig:a}
\caption{$\alpha_j = 0.15$ for $\jej $, and $\alpha_0 = 0.55$.}
\end{subfigure}

\bigskip
\medskip

\begin{subfigure}{0.8\textwidth}
\centering
\includegraphics[width =\textwidth]{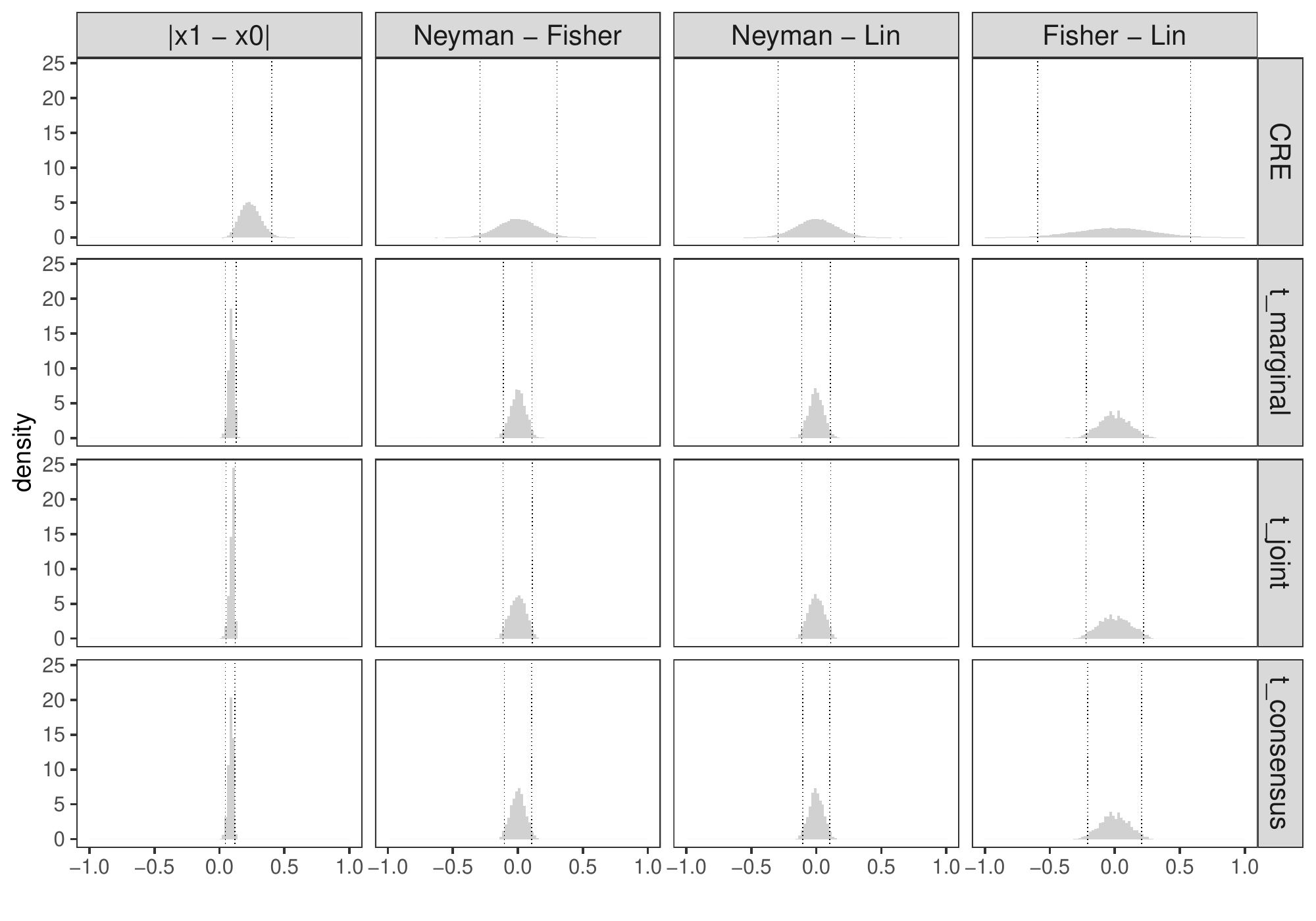}
\caption{$\alpha_j = 0.50$ for $\jej $, and $\alpha_0 = 0.95$.}
\label{fig:b}
\end{subfigure}

\caption{\label{fig:21}Distributions of $\|\htx\|_2 = \|\hat x(1) - \hat x(0)\|_2$, $\htau_\nm - \htau_\fisher$, $\htau_\nm-\htau_\lin$, and $\htau_\fisher-\htau_\lin$ under complete randomization and the three two-sample $t$-test-based {\repsch} over $50000$ independent initial allocations. 
The results under complete randomization, labeled as ``CRE",  are summarized over all $50000$ allocations, whereas those under  \rep, labeled as ``t\_marginal", ``t\_joint", and ``t\_consensus", respectively, are summarized over the subsets of allocations that satisfy the respective  balance criteria.
The vertical lines correspond to the $0.025$ and $0.975$ empirical quantiles, respectively.}
\end{figure}

\begin{figure}

\centering 

 \begin{subfigure}{0.8\textwidth}
 \centering
\includegraphics[width =.85\textwidth]{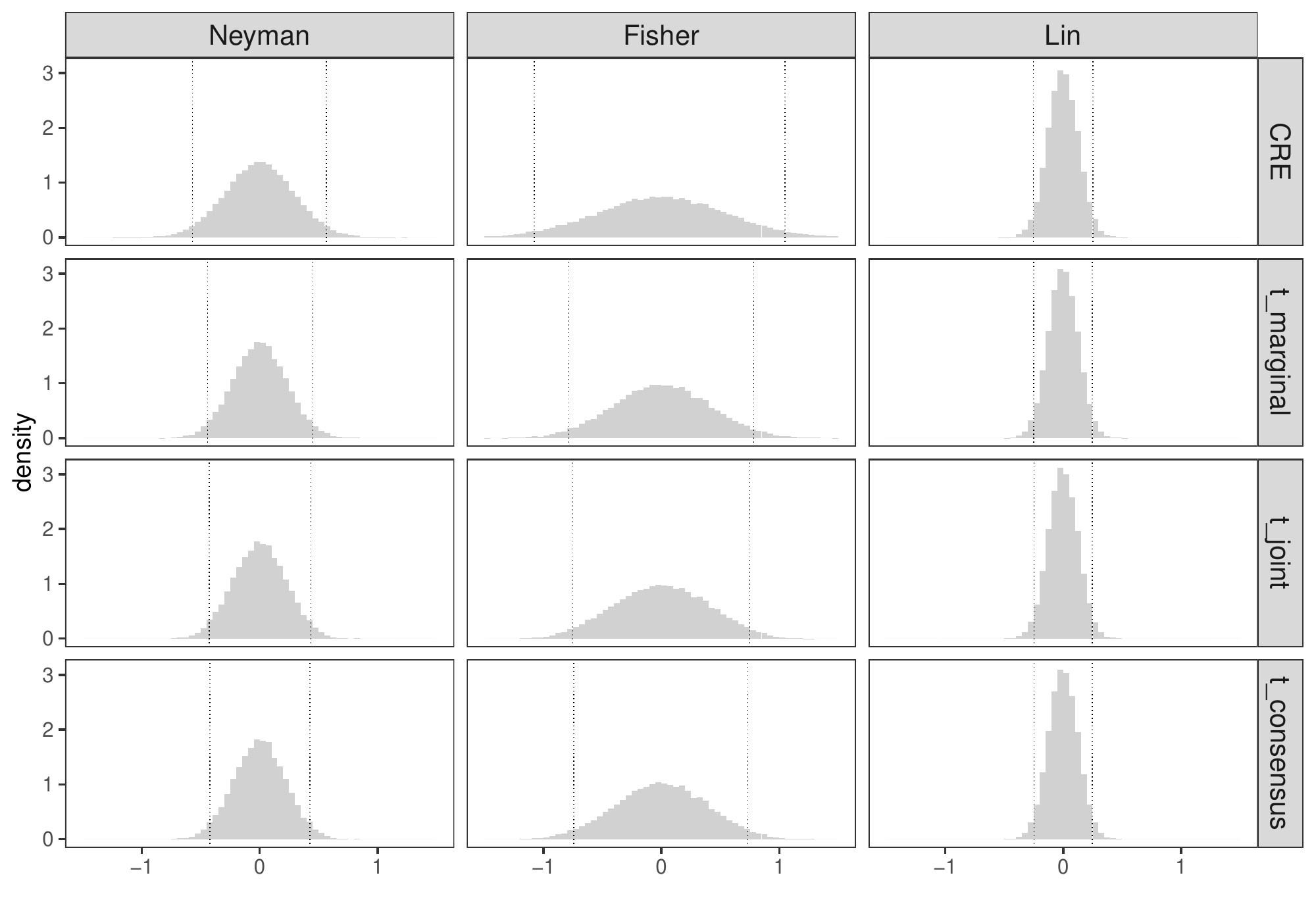}
\label{fig:a}
\caption{$\alpha_j = 0.15$ for $\jej $, and $\alpha_0 = 0.55$.}
\end{subfigure}

\bigskip
\medskip

\begin{subfigure}{0.8\textwidth}
\centering
\includegraphics[width =.85\textwidth]{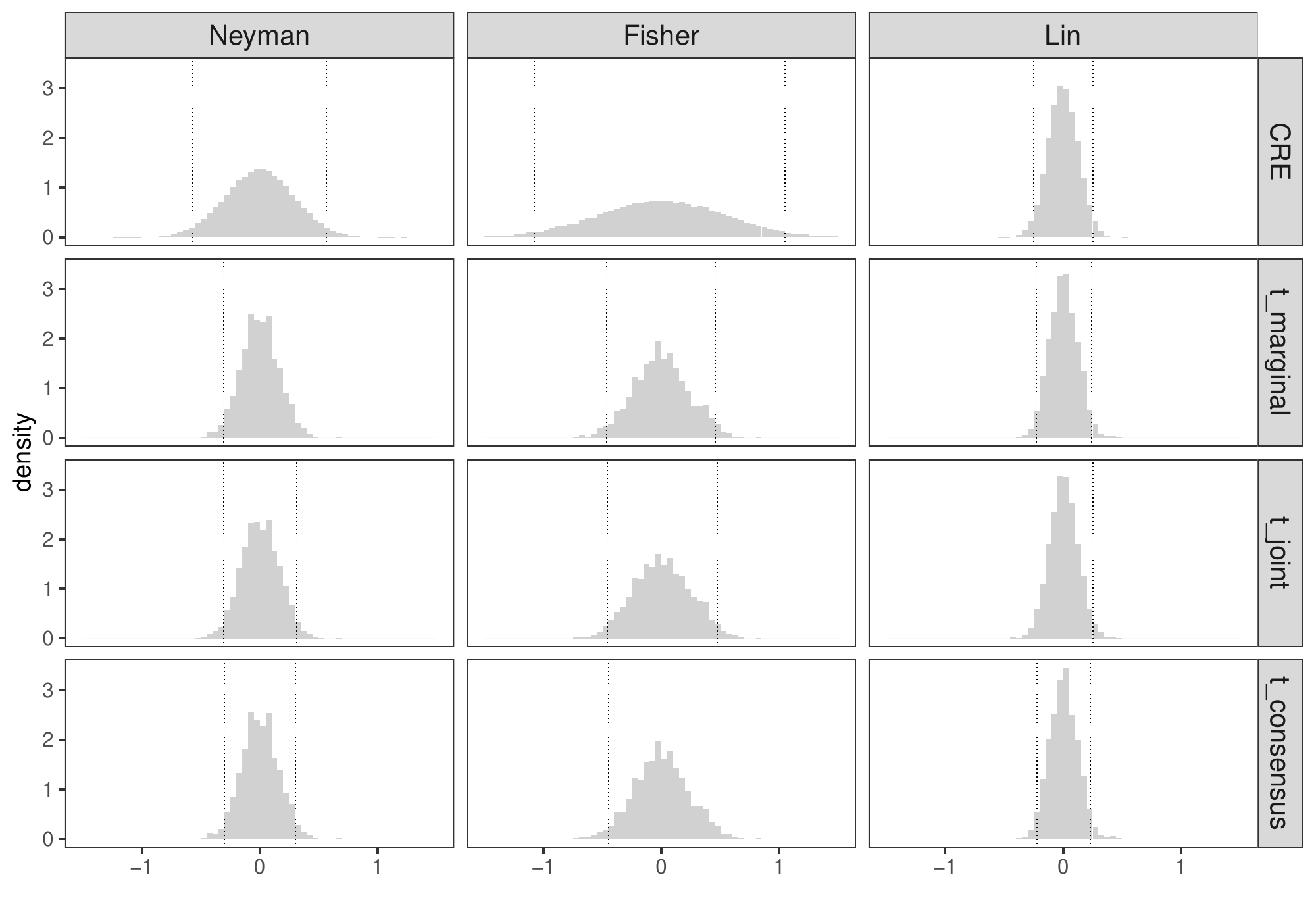}
\caption{$\alpha_j = 0.50$ for $\jej $, and $\alpha_0 = 0.95$.}
\label{fig:b}
\end{subfigure}

\caption{\label{fig:2}Distributions of $\hts \ (\ms)$ under complete randomization and the three two-sample $t$-test-based {\repsch} over $50000$ independent initial allocations. 
The results under complete randomization, labeled as ``CRE",  are summarized over all $50000$ allocations, whereas those under  \rep, labeled as ``t\_marginal", ``t\_joint", and ``t\_consensus", respectively, are summarized over the subsets of allocations that satisfy the respective  balance criteria.
The true $\tau$ is $0$. 
The vertical lines correspond to the $0.025$ and $0.975$ empirical quantiles, respectively.}
\end{figure}

Figure \ref{fig:21} shows the distributions of $\|\htx\|_2 =\|\hat x(1) - \hat x(0)\|_2$, $\htau_\nm - \htau_\fisher$, $\htau_\nm-\htau_\lin$, and $\htau_\fisher-\htau_\lin$ under complete randomization and the three two-sample $t$-test-based {\repsch} over $50000$ independent initial allocations. 
The results under complete randomization are summarized over all $50000$ allocations, whereas those under  \reps are summarized over the subsets of allocations that satisfy the respective  balance criteria.
We vary the thresholds for the marginal rule from $\alpha_j = 0.15$ to $\alpha_j = 0.5$ for $\jej$, and choose  $\alpha_0$ accordingly to ensure that the joint rule has approximately the same acceptance rate as the marginal rule.   
The message is coherent across different rules and thresholds: \reps considerably reduces the difference in covariate means and the differences across different estimators, both in line with the theoretical results in Theorem \ref{thm:t}. 
Compare sub-plots (a) and (b) under different values of $\alpha_j$'s. 
More stringent thresholds result in greater reduction in the differences when everything else stays the same. 

Figure \ref{fig:2} shows the distributions of $\hts \ (\ms)$.
The message is coherent across different rules and thresholds: \reps considerably improves the efficiency of $\htn$ and $\htf$ but leaves that of $\htl$ unchanged, both in line with Theorem \ref{thm:t}. 
Compare sub-plots (a) and (b) under different values of $\alpha_j$'s. 
Increasing the thresholds improves the efficiency of $\hts \ (\nf)$ when everything else stays the same.

\section{Discussion}
\reps provides a powerful tool for improving covariate balance in randomized experiments. 
We examined thirteen \rrs  schemes under this category and quantified their theoretical properties from the design-based perspective.
The theory clarifies three important issues regarding the impact of ReP on subsequent inference. 
First, ReP improves covariate balance, which in turns reduces conditional biases and ensures more coherent inferences across different estimators.
Second,  the estimator from the fully interacted regression is asymptotically the most efficient under all schemes examined, with the asymptotic sampling distribution remaining unchanged by \rep. 
We can thus conduct subsequent inference based on this estimator and its {\ehw} standard error or covariance via identical procedure as that under complete randomization.
Third,  \reps improves the efficiency of the estimators from the unadjusted and additive regressions asymptotically, necessitating rerandomization-specific inference to avoid overconservativeness. 
These results illustrate the value of \reps for strengthening the causal conclusions that can be drawn from experimental data, and highlight the value of the fully interacted adjustment for efficient and well-calibrated inference under \rep.

We focused on the thirteen criteria in Tables \ref{tb:criteria} and \ref{tb:criteria_g} because of their already widespread use for balance checks in practice. 
The variety of other test options for balance check promises a whole spectrum of alternative schemes for conducting ReP, catering to the needs of different studies. We give the details in the {\sm}. 


We focused on asymptotic results in establishing the theoretical properties of ReP.
Finite-sample exact unbiasedness is in general lost after rerandomization or regression adjustment. 
An exception is that the unadjusted difference in means remains unbiased under ReP when the treatment groups are of equal sizes; see \cite{morgan2012rerandomization}. 
Alternatively, the Fisher randomization test provides a way to conduct finite-sample exact inference under ReP.   \cite{ZDfrt} established the theory of the Fisher randomization test under ReM. All results therein extend to ReP with minimal modifications.

\bibliographystyle{plainnat}
\bibliography{refs_rerandomization}

\newpage
\spacingset{1.5} 

\setcounter{equation}{0}
\setcounter{section}{0}
\setcounter{figure}{0}
\setcounter{example}{0}
\setcounter{proposition}{0}
\setcounter{corollary}{0}
\setcounter{theorem}{0}
\setcounter{table}{0}
\setcounter{condition}{0}
\setcounter{lemma}{0}
\setcounter{remark}{0}
\setcounter{definition}{0}

\renewcommand {\thedefinition} {S\arabic{definition}}
\renewcommand {\theproposition} {S\arabic{proposition}}
\renewcommand {\theexample} {S\arabic{example}}
\renewcommand {\thefigure} {S\arabic{figure}}
\renewcommand {\thetable} {S\arabic{table}}
\renewcommand {\theequation} {S\arabic{equation}}
\renewcommand {\thelemma} {S\arabic{lemma}}
\renewcommand {\thesection} {S\arabic{section}}
\renewcommand {\thetheorem} {S\arabic{theorem}}
\renewcommand {\thecorollary} {S\arabic{corollary}}
\renewcommand {\thecondition} {S\arabic{condition}}
\renewcommand {\thepage} {S\arabic{page}}
\renewcommand {\theremark} {S\arabic{remark}}

\setcounter{page}{1}

\begin{center}
\bf \Large 
Supplementary Material  
\end{center}

Section \ref{sec:Q} gives the additional results for ReP in multi-armed experiments.

Section \ref{sec:ext} presents extensions to alternative covariate balance criteria for \rep.


Section \ref{sec:test_statistics_app} reviews the test statistics that underlie the $p$-values we use to form \rep. 

Section \ref{sec:lemma} states the key lemmas for proving the results in the main paper. In particular, Theorem \ref{thm:mle_Q} is a novel technical result, and formalizes the design-based properties of the \mles outputs from logistic and {\mlr}s. The result establishes the asymptotic equivalence of the \lrts and the Wald test for logistic and multinomial logistic regressions from the design-based perspective.  

Section \ref{sec:proof_main} gives the proofs of the results in the main paper. 



Section \ref{sec:Q_app} gives the proof of Theorem \ref{thm:mle_Q}. 


Assume centered covariates with $\bar x = \meani x_i = 0_J$ throughout to simplify the presentation. 
For two sequences of random vectors $\{A_N\}_{N=1}^\infty$ and $\{B_N\}_{N=1}^\infty$ with $A_N\rs A$ and $B_N\rs B$ in $\mr^m$, write $A_N \succi B_N$ if $A \succeq B$, and write $A_N \asim B_N$ if $A$ and $B$ have the same distribution.
Definition \ref{def:eff} in the main paper is a special case of this definition of $\succi$ and $\asim$, 
with $\hth_1 \asim \hth_2$ and $\hth_1 \succi \hth_2$ being abbreviations of 
$\sqrtn (\hth_1 - \theta) \asim \sqrtn (\hth_2 -\theta)$  and $\sqrtn (\hth_1 - \theta) \succi \sqrtn (\hth_2 - \theta)$, respectively, when the meaning of $\theta$ is clear from the context.

\section{Additional results for ReP with multiple arms}\label{sec:Q}

\subsection{Simulation based on a multi-armed field experiment}
We now illustrate our theory for ReP in general experiments using data from a multi-armed  experiment studying the effect of public recognition on giving \citep{karlan2014}. 

\cite{karlan2014} conducted a field experiment in collaboration with Dwight Hall, a service club at Yale University, as part of their annual fundraising campaign from October 2007 to May 2008. 
The sample frame consisted of $4,168$ alumni in Dwight Hall's prior-donor database who had a valid phone number and had not already made a donation between January 2007 and October 2007. 
Campaign staffs made calls to these potential donors and randomly assigned them to $Q = 4$ treatment scripts with equal probability:
\begine[(1)]  
\item Control: We are hoping you will continue your support to Dwight Hall with a gift of \$$100$.
\item 100 circle: We are hoping you will continue your support to Dwight Hall with a gift of \$$100$. With a donation of at least \$$100$, you will become a member of our Friend donor circle. Friends will be listed by name in the Dwight Hall Fall 2008 newsletter.
\item 500 circle: We are hoping you will continue your support to Dwight Hall with a gift of \$$100$. With a donation of at least \$$500$, you will become a member of our Benefactor donor circle. Benefactors will be listed by name in the Dwight Hall Fall 2008 newsletter.
\item 100 circle and 500 circle: We are hoping you will continue your support to Dwight Hall with a gift of \$$100$. With a donation of at least \$$100$, you will become a member of our Friends donor circle. With a donation of at least \$$500$, you will become a member of our Benefactor donor circle. Both Friends and Benefactors will be listed by name in the Dwight Hall Fall 2008 newsletter.
\ende
The three active treatment scripts, indexed by $q = 2, 3, 4$, contain information on the donor circles, whereas the control treatment script, indexed by $q = 1$, does not. 

One outcome of interest is the probability of a gift large enough to publish in the Friends circle, represented by $Y_i = 1(\text{unit $i$ made a donation of at least \$$100$})$.  
For arbitrary $\tau = G\bY$ of interest, we can compute $\hys \ (\nfl)$ by the linear regressions in Section \ref{sec:po_Q} of the main paper, and form the point estimators as $\hts = G\hys$. 
For concreteness, we consider the inference of $\tau = 3^{-1}\{\by(2) + \by(3) + \by(4)\} - \by(1)$, as the average treatment effect of the three active treatment scripts relative to the control. The results for other treatment effects are similar and thus omitted. 

We include $J = 7$ pretreatment covariates: marital status, gender, age, average prior gift, and indicators of whether the largest last gift falls between $\mathcal{I}$ for $\mathcal{I} = (0, 100)$, $[100, 500)$, and $[500, +\infty)$, respectively. 
We exclude units with incomplete covariate information, resulting in $N = 2,298$ units in $Q=4$ treatment groups with $(N_1, N_2, N_3, N_4) = (526, 610, 584, 578)$. 
The sampling distributions of $\hts \ (\nfl)$ depend on all potential outcomes including the unobserved ones. To make the simulation more realistic, we impute the unobserved potential outcomes by simple model fitting. 
Specifically, we fit a logistic regression of the observed outcome $Y_i$ on the indicators of the treatment levels,
all covariates, and their interactions, and then impute
the unobserved potential outcomes, namely $\yiq  $ for $q \in\{1, 2, 3, 4\} \backslash \{Z_i\}$, based on the fitted probability of $\yiq  = 1$ with a cutoff at $0.5$. 
For the simulated data set, the true value of $\tau$ equals $0.0203$. 

\begin{figure}[ht!]
\centering
\includegraphics[width =\textwidth]
{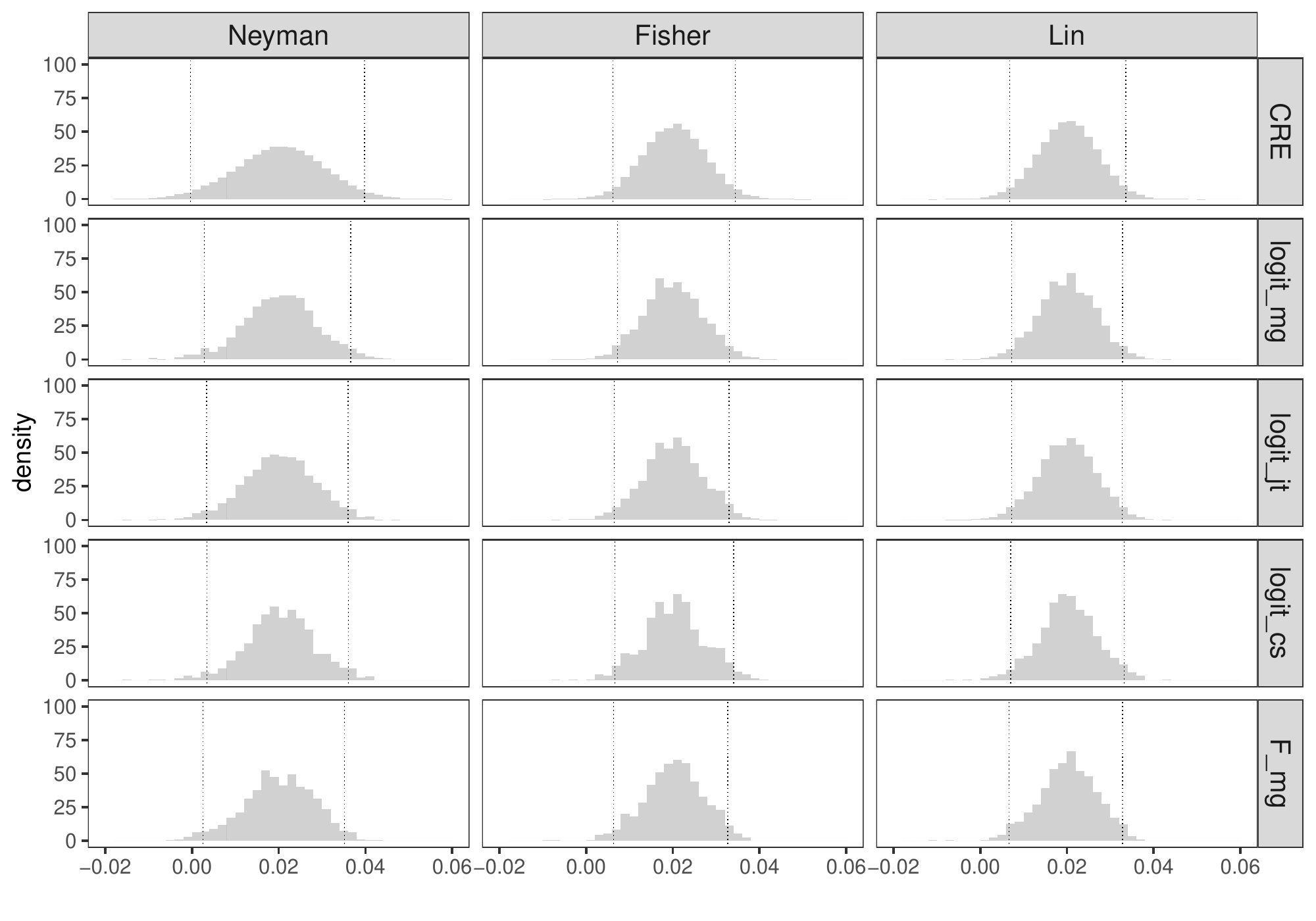}

\caption{\label{fig:hey}Distributions of $\hts \ (\ms)$ under complete randomization and the four {\repsch} in Table \ref{tb:criteria_g} over $50000$ independent initial allocations. 
The results under complete randomization, labeled as CRE,  are summarized over all $50000$ allocations, whereas those under  \rep, labeled as logit\_marginal, logit\_joint,  logit\_consensus, and F\_mg, respectively, are summarized over the subsets of allocations that satisfy the respective  balance criteria.
The true $\tau$ is $0.0203$.
The vertical lines correspond to the $0.025$ and $0.975$ empirical quantiles, respectively.}
\end{figure}

Figure \ref{fig:hey} shows the distributions of $\hts \ (\ms)$ under complete randomization and the four {\repsch} from Table \ref{tb:criteria_g} over $50000$ independent initial allocations.
We set $\alpha_{qj} = 0.2$ for the marginal rule based on the \mlr, 
$\alpha_0=0.96$  for the joint rule based on the \mlr, and $\alpha_j=0.55$ for the marginal rule based on the $F$-tests to ensure that the resulting \reps procedures have approximately the same acceptance rate.  

The results are coherent with the asymptotic theory in Theorems \ref{thm:F} and \ref{thm:glm_Q}, and illustrate the duality between \reps and regression adjustment in improving the efficiency of $\htn$. 
The similarity between $\htf$ and $\htl$, on the other hand, is likely the result of the $\gq $'s being relatively homogeneous across treatment groups such that $\Gamma_\fisher \approx 0$. 
This is not surprising given the binary nature of the outcome, which inevitably restricts the variability in the $\gq $'s. 
We have observed substantial efficiency gain of $\htl$ over $\htf$ in simulation with heterogeneous $\gamma_q$'s but omit the details to save space.
Importantly, recall that $\hts \ (\nfl)$ are based on outputs from three linear regressions that are in general considered suboptimal for fitting binary outcomes.  
The results here highlight the validity of the proposed inferential procedure regardless of how well the linear models involved approximate the true {\dgp}.

\subsection{Wald-type inference}\label{sec:infer_Q}
Recall from Theorems \ref{thm:F} and \ref{thm:glm_Q} that $\htl$ is asymptotically the most efficient under ReP with multiple arms, with $ \sqrtn(\hyl -\bar Y) \mid \ma  \rs  \mn(0_Q, \vr)$ for all $\ma \in \{\maf, \ma_{\mlogit, \dr}: \drs\}$. 
This suggests subsequent inference based on $\htl$ and its \ehws covariance by normal approximation. 
 
Specifically, let $\hv_\lin'$ be the {\ehw} covariance of $\hyl$ from the same \ols fit.  
\cite{zdca} showed that it is asymptotically appropriate for estimating the true sampling covariance under complete randomization; see Lemma \ref{lem:cre_Q} in Section \ref{sec:cre_app}.
The same reasoning as in \citet[][Lemma A16]{LD2018} ensures that the asymptotic appropriateness extends to \reps based on the marginal $F$-tests as well. 
This, together with the asymptotic normality of $\htl$ from Lemma \ref{lem:clt_Q} and Proposition \ref{prop:F} in the main paper, justifies the Wald-type inference of $\tau$ based on $(\htl, G\hv_\lin' G^\T)$ under both complete randomization and \rep. 
The Fisher randomization test can be conducted similarly using $\htl^\T(G\hv_\lin' G^\T)^{-1}\htl$ as the test statistic for both the strong and weak null hypotheses \citep{wuanding2020jasa}.
This illustrates the convenience of the fully interacted regression for inference under general experiments.

The asymptotic sampling distribution of $\hts \ (\nf)$, on the other hand, is altered by \reps into a convolution of independent normal and truncated normal when $\Gamma_* \neq 0$, resulting in greater peakedness than that under complete randomization. 
Inference based on the usual normal approximation, as a result, is overly conservative, deterring statistically significant findings.
This, again,  illustrates the value of the fully interacted regression for efficient and well-calibrated inference under \reps for general experiments. 

\subsection{Extensions of the joint $t$-test and linear regression}
The marginal $F$-tests give an immediate extension of the marginal two-sample $t$-tests to more than two treatment arms. 
The range of tests commonly used in multivariate analysis of variance, on the other hand, provide reasonable substitutes to the Hotelling's $T^2$ test under the joint rule \citep{morgan2011}. 
Common choices of test statistics include Wilks' $\Lambda$, the Lawley--Hotelling trace, the Pillai--Bartlett trace, and Roy's largest root. 
One complication is that the distributions of these test statistics are not well studied under the design-based inference. 
\cite{morgan2011} recommended using the Fisher randomization test to generate an empirical distribution for the test statistic of choice. 
This is sound in theory but can become unwieldy in practice. 

Generalization of the linear regression-based criteria can be accomplished by a dichotomization of the treatment assignment variable.
Recall that $\wiq = 1(Z_{i} = q)$ denotes the indicator for assignment to treatment arm $q\in\mq= \{1, \dots, Q\}$ in a general experiment.
We can conduct balance checks based on $Q$ separate linear regressions as $
\lmt( \wiq \sim 1 + x_{i1}+\cdots + x_{iJ})$ over $i = \ot{N}$ for $q  = \ot{Q}$.
The process yields one joint and $J$ marginal $p$-values for each $q\in\mq$, measuring the influence of $x_{ij}$'s on assignment to treatment level $q$.
We can form acceptance criteria accordingly based on whether some or all of them exceed some prespecified thresholds. 
Despite the conceptual straightforwardness of this approach, however,  it requires additional data transformations, and gives only measures of covariate balance for the $Q$ treatment levels separately.

\subsection{Covariate-wise and treatment-wise $p$-values}\label{sec:matrix}

The \mlrs of $Z_i$ on $(1, x_i)$ provides a powerful tool for conducting \reps under general experiments. 
We formed the marginal rule based on $(\pqjmlr)_\qj$, with one $p$-value for each estimated coefficient $\tb_{qj}$, corresponding to the correlation between covariate $j$ and assignment to treatment level $\qiqp$. 
Alternatively, most standard software packages also allow us to test the overall effect of a given covariate $j \in \{\ot{J}\}$ across all treatment levels; see, e.g., the \texttt{test} command in stata. 
Denote by $p_{\cdot j,\mlogit}$ the resulting $p$-value associated with the overall effect of covariate $j$. 
We can also form the balance criterion using the $p_{\cdot j, \mlogit}$'s, and accept a randomization if and only if $p_{\cdot j, \mlogit} \geq \alpha_j$ for all $\jej$.
The resulting criterion parallels the marginal rules under the treatment-control experiment, and yields analogous results with $\htl$ being our recommendation. 

More generally, we can arrange the $\tb_{qj}$'s into a matrix,  with rows corresponding to the $Q-1$ non-reference treatment levels and columns corresponding to the $J$ covariates:

\smallskip

\begin{center}
\begin{tabular}{cccccccccc}\hline 
Non-reference  & \multicolumn{5}{c}{Covariate} & Treatment-wise\\\cline{2-6}
treatment level & $1$ & $\ldots$ & $j$ & $\ldots$  & $J$ &  $p$-value\\\hline
1 & $\tb_{11}$ & $\ldots$& $\tb_{1j}$ & $\ldots$& $\tb_{1J}$ & $p_{1\cdot, \mlogit}$\\
$\vdots$ &&&&& & $\vdots$ \\
$q$ & $\tb_{q1}$ & $\ldots$& $\tb_{qj}$ & $\ldots$& $\tb_{qJ}$ & $p_{q\cdot, \mlogit}$\\
$\vdots$ &&&&& & $\vdots$\\
$Q-1$ & $\tb_{Q-1,1}$ & $\ldots$& $\tb_{Q-1,j}$ & $\ldots$& $\tb_{Q-1,J}$ & $ p_{Q-1,\cdot, \mlogit}$\\\hline
Covariate-wise & \multirow{2}{*}{$p_{\cdot 1, \mlogit}$} & \multirow{2}{*}{$\ldots$} & \multirow{2}{*}{$p_{\cdot j, \mlogit}$} & \multirow{2}{*}{$\ldots$} & \multirow{2}{*}{$p_{\cdot J, \mlogit}$} & \multirow{2}{*}{$p_{0, \mlogit}$}\\
$p$-value&&&&&&\\\hline
\end{tabular}
\smallskip
\end{center}
The $\pqjmlr$'s, $p_{\cdot j, \mlogit}$'s, and $\pzmlr$ then correspond to the cells, columns, and the entire matrix, respectively, 
measuring the deviations of the corresponding $\tb_{qj}$'s from $0$.  

By symmetry, we can also conduct one Wald test for each row of the matrix, $\tbb_{q} = (\tb_{q1}, \ldots, \tb_{qJ})^\T$, 
 and accept a randomization if and only if the resulting treatment-wise $p$-values, denoted by $p_{q\cdot, \mlogit}$ for $\qiqp$, satisfy some prespecified criterion. 
The magnitude of $p_{q\cdot , \mlogit}$ intuitively reflects the covariate balance between the treatment level $\qiqp$ and the reference level $Q$. 
The acceptance rule based on $\{p_{q\cdot, \mlogit}: \qiqp\}$ hence
involves $Q-1$ pairwise comparisons of the non-reference levels to the reference level, simplifying the approach that consults all pairwise comparisons. 

This yields four types of $p$-values, namely $\pqjmlr$, $p_{\cdot j, \mlogit}$, $p_{q \cdot, \mlogit}$, and $\pzmlr$, summarized in the first row of Table \ref{tb:matrix}. 
They measure the covariate balance at the treatment-covariate pair, covariate, treatment, and overall levels, respectively, and provide the ingredients for defining a whole spectrum of balance criteria under \rep. 
The marginal, joint, and consensus rules in Table \ref{tb:criteria_g} use $\{p_{qj, \mlogit}: \qiqp; \ \jej\}$, $p_{0, \mlogit}$, and their union to form the acceptance criteria, respectively, 
but the choice can be general.
A key consideration is that the joint $p$-value $p_{0, \mlogit}$ is invariant to non-degenerate transformation of the covariate vector, in the sense of $x_i' = Ax_i$ for some nonsingular $J\times J$ matrix $A$, whereas the treatment-covariate-wise, covariate-wise, and treatment-wise $p$-values  in general are not unless $A$ is diagonal.
Emphases on specific covariates or treatment levels, on the other hand, justify the use of covariate- or treatment-wise $p$-values, respectively.

The same discussion extends to the $p$-values from
(i) the treatment-wise regressions $\lmt(\wiq \sim 1 + x_{i1}+\cdots + x_{iJ})$ over $i = \ot{N}$ for $\qiq$; 
(ii) the covariate-wise regressions $\lmt(x_{ij}\sim 1+\wio + \dots + \mi_{i, Q-1})$ over $i = \ot{N}$ for $\jij$; and (iii) the two-sample $t$-test of $\{x_{ij}: Z_i = q\}$ and $\{x_{ij}: Z_i = Q\}$ for each pair of $(q, j)\in \mqp \times \{1, \ldots, J\}$, respectively. 

Specifically, recall that the treatment-wise regression $\lmt(\wiq \sim 1 + x_{i1}+\cdots + x_{iJ})$ extends the linear regression of $Z_i$ on $(1, x_i)$ under the treatment-control experiment to general experiments,  measuring the influence of covariates on assignment to treatment level $\qiq$. 
Denote by $\hat {\bm \beta}_{q} = (\hb_{q1}, \ldots, \hb_{qJ})^\T$ the coefficient vector of $(x_{i1}, \ldots, x_{iJ})^\T$ from the \ols fit. 
It yields two types of $p$-values, namely 
\begine[(i)]
\item the marginal $p$-value associated with each individual $\hb_{qj}$, denoted by $p_{qj, \lm}$; and
\item the treatment-wise $p$-value from the $F$-test of $\lmt(\wiq \sim 1 + x_{i1}+\cdots + x_{iJ})$ against $\lmt(\wiq \sim 1)$, denoted by  $p_{q\cdot,\lm}$. 
\ende
They are analogous to the $p_{qj, \mlogit}$'s and $p_{q\cdot, \mlogit}$'s from the \mlr, respectively,  and allow us to form balance criteria like (i) $p_{qj, \lm} \geq \alpha_{qj}$ for all $qj$, or (ii) $p_{q\cdot , \lm} \geq \alpha_q$ for all $\qiq$. See the second row of Table \ref{tb:matrix}.

Next, recall $\lmt(x_{ij}\sim 1+\wio + \dots + \mi_{i, Q-1})$ as a regression formulation for computing the $\pjf$ from the marginal $F$-test of $(x_{ij}, Z_i)_{i=1}^N$. 
The resulting fit, 
in addition to yielding $\pjf$ as the $p$-value from the $F$-test of $\lmt(x_{ij}\sim 1+\wio + \dots + \mi_{i, Q-1})$ against $\lmt(x_{ij}\sim 1)$, also yields one marginal $p$-value for the coefficient of each $\wiq \ (\qiqp)$, denoted by $p_{qj,\ff}$. 
The two types of $p$-values are analogous to the $p_{qj, \mlogit}$'s and $p_{\cdot j, \mlogit}$'s, respectively, and allow us to form balance criteria accordingly.  
 The $\pjt$'s under the treatment-control experiment are a special case with $Q =2$ and $\wio = Z_i$. 
See also \citet[][Table 2]{bblp} and \citet[][Tables 1 and A1]{dr2} for applications that use the $p_{qj,\ff}$'s for balance check.

Last but not least, we can conduct one two-sample $t$-test of $\{x_{ij}: Z_i = q\}$ and $\{x_{ij}: Z_i = Q\}$ for each pair of $(q, j) \in \mqp\times \{\ot{J}\}$, comparing the balance of covariate $j$ between treatment groups $\qiqp$ and $Q$. 
Denote by $p_{qj, \ts}$ the resulting two-sided $p$-value. 
It can be implemented by fitting $\lmt(x_{ij} \sim 1+\wiq)$ over $\{i: Z_i \in \{q,Q\}\}$, and allows us to form balance criteria like $p_{qj, \ts} \geq \alpha_{qj}$ for all $qj$ for some prespecified thresholds $\alpha_{qj}\in (0,1)$.

This yields four strategies for forming treatment-covariate-wise,  covariate-wise, treatment-wise, and joint $p$-values  under general experiments, summarized in Table \ref{tb:matrix}.
The \mlrs accommodates
 all four types of $p$-values via one \mles fit, and is hence our recommendation in general.

\begin{table}[t]\caption{\label{tb:matrix}Four strategies for forming the treatment-covariate-wise,  covariate-wise, treatment-wise, and joint $p$-values under general experiments. Let $\mathcal J= \{\ot{J}\}$}
\centering
\resizebox{\columnwidth}{!}{\begin{tabular}{cccccc}\hline
  &  Treatment-covariate-wise  & Covariate-wise &  Treatment-wise& \\
& ($q,j$) & ($j$) &  ($q$) &  Joint  \\\hline\\

$\glmt (Z_i \sim 1+x_{i1}+\cdots + x_{iJ})$  &  \multirow{2}{*}{$p_{qj, \mlogit}$} &  \multirow{2}{*}{$p_{\cdot j, \mlogit}$} &  \multirow{2}{*}{$p_{q\cdot, \mlogit}$} &  \multirow{2}{*}{$p_{0, \mlogit}$}\\
over $i = \ot{N}$   &  &  & &   \\\\
$
\lmt(\wiq \sim 1 +x_{i1}+\cdots + x_{iJ})$   &  \multirow{2}{*}{$p_{qj,\lm}$} &  \multirow{2}{*}{n.a.} &  \multirow{2}{*}{$p_{q\cdot,\lm}$} &  \multirow{2}{*}{n.a.}\\
over $i = \ot{N}$ for $\qiq$ &&&&\\\\
$\lmt(x_{ij}\sim 1+\wio + \dots + \mi_{i, Q-1})$ 
& \multirow{2}{*}{$p_{qj,\ff}$} &  \multirow{2}{*}{$\pjf$} &  \multirow{2}{*}{n.a.} & \multirow{2}{*}{n.a.} \\
 over $i = \ot{N}$ for $j \in \mj$ &&&&\\ \\
Two-sample $t$-test of \\
$\{x_{ij}: Z_i = q\}$ and $\{x_{ij}: Z_i = Q\}$
& $p_{qj, \ts}$ & n.a. & n.a. & n.a.\\
  for $j \in \mj$ and $\qiqp$ &&&&\\
\hline
\end{tabular}}
\end{table}

\section{Alternative covariate balance criteria}\label{sec:ext}

\subsection{Rerandomization with tiers of covariates}
When covariates vary in a priori importance, 
\cite{morgan2015} proposed rerandomizing based on Mahalanobis distance within tiers of covariate importance, imposing more stringent criteria for covariates that are thought
to be more important. 
Extension of \reps to such settings is straightforward under the marginal rules by setting the covariate-wise thresholds according to the importance. 
To construct joint rules for \reps with tiers of covariates, we can conduct one joint test for each tier of covariates, and set the tier-wise thresholds according to the importance. 
The consensus rules then follow as the  intersections of the corresponding marginal and joint rules. 
These {\rep} schemes complicate the asymptotic distributions of $\hts \ (\nf)$ but keep that of $\htl$ unchanged. We recommend the same analysis based on $\htl$.

\subsection{Alternative joint tests from regression}
The discussion so far assumed default tests from standard software packages. 
As a result, we conducted an $F$-test to construct the joint criterion under the linear regression model option, namely $\malmjt = \{\pzlm \geq \alpha_0\}$, and conducted an \lrts to construct the joint criterion under the logistic and \mlrs model options, namely $\malogitjt = \{\pzglm \geq \alpha_0\}$, respectively. 
The Wald test, on the other hand, enables definition of the joint criteria in a unified way.

Specifically, recall $\lmt(Z_i \sim 1+x_i)$ and $\glmt(Z_i \sim 1+x_i)$ as the linear and logistic regressions under the treatment-control experiment. 
Let $\hb = (\hb_1, \ldots, \hb_J)^\T$ and $\tb= (\tb_1, \ldots, \tb_J)^\T$ be the coefficient vectors of $x_i$ from the \ols and \mles fits, respectively, with $\hv$ and $\tv$ as the corresponding estimated covariances.
The Wald tests of $\hb$ and $\tb$ compare $\wlm = \hb^\T \hv^{-1}\hb$ and $\wlogit = \tb^\T\tv^{-1}\tb$ against the $\chi^2_J$ distribution, and define two alternatives to the $F$-test and \lrt, respectively, measuring the magnitudes of $(\hb_j)_{j=1}^J$ and $(\tb_j)_{j=1}^J$ as a whole. 
All results in Proposition  \ref{prop:lm_glm} and Theorem \ref{thm:lm_glm} on $\malmjt$ and $\malogitjt$ extend verbatim to the resulting {\repsch}, with $\htl$ being our recommendation. 

Likewise for all results in Proposition \ref{prop:glm_Q} and Theorem \ref{thm:glm_Q} on $\malogitjt$ to extend verbatim to the \reps based on the Wald test of $\tb = (\tbqj)_\qj$ under the general experiment, with $\htl$ being our recommendation. 

See Lemmas \ref{lem:equiv_lm}, \ref{lem:equiv}, \ref{lem:joint_Q} and Theorem \ref{thm:mle_Q} in Section \ref{sec:lemma} for the proof of the asymptotic equivalence of the Wald test to the $F$-test and \lrt, respectively. 
See also \citet[Table 1]{mmw} for an application of the Wald test to balance check.

 \subsection{EHW standard errors for balance test}\label{sec:ext_ehw}
We assumed the default $p$-values from standard software packages for forming the balance criteria. 
The standard error and covariance involved in their computation are hence in general the classic standard errors and covariances derived under homoskedasticity. 
Alternatively, we can form the test statistics using the {\ehw} robust standard errors and covariances as in the analysis stage, and compute the $p$-values accordingly.  
 We give the explicit forms of the resulting test statistics in Remark \ref{rmk:1} in Section \ref{sec:unif_app}, and show in Remark \ref{rmk:proof} in Section \ref{sec:wc_app} their equivalence with the classic counterparts as $N$ tends to infinity.  
All results thus extend to \reps based on the robustly studentized test statistics, with $\htl$ being our recommendation.

\subsection{$p$-values from other standard statistical tests}
The discussion so far concerned $p$-values from the linear, logistic, and {\mlr}s in standard software packages. 
Alternatively, the probit and multinomial probit regressions provide two intuitive variants to the logistic and multinomial logistic regressions of $Z_i$ on $(1,x_i)$, respectively, accommodating marginal, joint, and consensus rules via one \mles fit. 
We conjecture that the results are analogous, and leave the technical details to future work.

In addition, the two-sample Kolmogorov--Smirnov test and the chi-square test of independence exemplify  alternative test choices for forming covariate balance criteria under the treatment-control and general experiments, respectively. 
See \cite{gkb} and \cite{chkl} for their applications to covariate balance check.  
We leave the theory on their properties for rerandomization to future work.

\section{Test statistics and equivalent forms for acceptance criteria in Tables \ref{tb:criteria} and \ref{tb:criteria_g}}\label{sec:test_statistics_app}

We review in this section the test statistics that underlie the $p$-values we use to form the {\repsch} in Tables \ref{tb:criteria} and \ref{tb:criteria_g}, respectively.  
To avoid repetition, we treat the logistic regression for treatment-control experiments as a special case of the \mlrs with $Q=2$ and level 2 relabeled as level 0. 

Assume $\alpha_j \ (\jej )$ and $\alpha_0$  as the thresholds for the marginal and joint criteria under the treatment-control experiment, respectively. 

Assume $\alpha_j \ (\jej )$, $\alpha_{qj} \ (\qiqp; \ \jej)$, and $\alpha_0$ as the thresholds for the marginal $F$-tests and the marginal and joint tests  based on the \mlr, respectively, under the general experiment.

\subsection{Two-sample $t$-tests}
\prgn{Marginal tests.}
Let $
\htxj =  \hx_j(1)- \hx_j(0)$, where $\hx_j(q) = N_q ^{-1}\sumiq x_{ij}$, be the difference in means of the $j$th covariate, equaling the $j$th component of $\htx$.
The pooled standard error for $\htxj$ equals
\begina
\hse_j = \sqrt{\frac{(N_1-1)\hsxj(1) + (N_0-1)\hsxj(0)}{N-2} \left( \frac{1}{N_1} + \frac{1}{N_0} \right)},
\enda
  with $\hsxj(q) = (N_q -1)^{-1}\sumiq  \{x_{ij} - \hx_j(q)\}^2$ for $q = 0, 1$.   
The two-sample $t$-test uses  \begina
\tjt  = \htxj/\hse_j  
\enda 
as the test statistic,  and computes $\pjt$ based on the $t_{N-2}$ distribution as
\begina
\pjt = \pr(|A| \geq |\tjt|),\twhere A\sim t_{N-2}. 
\enda
Let $\tts = (t_{1, \ts}, \ldots, t_{J, \ts})^\T$.
The marginal criterion equals
\beginy\label{eq:mg_t}
\matmg &\ =\ & \{\pjt \geq \alpha_j \ \fall \jej  \}\\ &=& \{|\tjt| \leq a_{j, \ts} \ \fall \jej  \}\nonumber\\
&=& \{ |\tts|  \leq  a_\ts  \},\nonumber
\endy
with $a_\ts = (a_{1, \ts}, \ldots, a_{ J, \ts})^\T$ and $a_{j, \ts}$ denoting the $(1-\alpha_j/2)$th quantile of the $t_{N-2}$ distribution. 
Numerically, $\htxj$, $\hse_j$, $\tjt$, and $\pjt$ equal the coefficient, classic standard error, $t$-value, and $p$-value associated with $Z_i$ from the \ols fit of $\lmt(x_{ij} \sim 1+Z_i)$ over $i = \ot{N}$, respectively.
This gives an alternative implementation of the marginal $t$-tests via \olss.

\prg{Joint  test.}
Recall $W_\ts = \htx^\T\ho^{-1}\htx$ as the test statistic for the joint two-sample $t$-test. 
The pooled estimated covariance equals
\beginy\label{eq:ho}
\ho =  \frac{(N_1-1)\hsx(1) + (N_0-1)\hsx(0)}{N-2}   \left(\frac{1}{N_1} + \frac{1}{N_0}\right) , 
\endy
with  $\hsx(q) = (N_q -1)^{-1}\sumiq  \{x_i  - \hx(q)\}\{x_i  - \hx(q)\}^\T$ for $q = 0,1$. 
Assume the joint Wald test for computing $\pzt$, with $\chisq_J$ as the reference distribution. The acceptance criterion equals 
\begina
\matjt = \{\pzt \geq \alpha_0\} = \{\wt \leq a_0 \},
\enda
where $a_0$ denotes the $(1-\alpha_0)$th quantile of the $\chisq_J$ distribution.

\subsection{Linear regression}

\prgn{Marginal tests.}
Let $\hb = (\hb_1, \ldots, \hb_J)^\T$ denote the coefficient vector of $x_i = (x_{i1}, \ldots, x_{iJ})^\T$ from $\lmt(Z_i \sim 1+x_{i1}+\cdots + x_{iJ})$.  
Let $\hv$ be the associated estimated covariance, with $\hv_{jj}$ as the $(j,j)$th element for $\jej$. 
The marginal test of $\hb_j$ takes 
\begina
\tjlm  = \hb_j /\hv_{jj}^{1/2}
\enda
as the test-statistic, and computes $\pjlm$ based on the $t_{N-1-J}$ distribution as 
\begina
\pjlm = \pr( |A| \geq |\tjlm|), \twhere A \sim t_{N-1-J}. 
\enda
Let $  T_\lm  = (t_{1, \lm}, \ldots, t_{J, \lm})^\T$.
The marginal criterion equals
\beginy\label{eq:mg_lm}
\malmmg &\ =\ & \{\pjlm \geq \alpha_j \ \fall \jej  \}\\&=& \{|\tjlm| \leq a_{j, \lm} \ \fall \jej  \}\nonumber\\
&=& \{ |\tlm|  \leq  a_\lm \},\nonumber
\endy
with $a_\lm = (a_{1, \lm}, \ldots, a_{ J, \lm})^\T$ and $a_{j, \lm}$ denoting the $(1-\alpha_j/2)$th quantile of the $t_{N-1-J}$ distribution. 

\prg{Joint test.}
 The $F$-test for linear regression compares $\lmt(Z_i \sim 1+x_{i1}+\cdots+x_{iJ})$ against the empty model $\lmt(Z_i \sim 1)$. Let $\rss_0$ and $\rss_1$ denote the residual sums of squares from the null and full regressions, respectively.
The test statistic equals
\begina
F = \frac{(\rss_0-\rss_1)/J}{\rss_1/(N-1-J)},
\enda
and is compared against the $F_{J, N-1-J}$ distribution to compute $\pzlm$ as 
\begina
\pzlm = \pr( A \geq F), \twhere A \sim F_{J, N-1-J}.
\enda
The acceptance criterion equals 
\begina
\malmjt = \{\pzlm \geq \alpha_0\} = \{F \leq f_{J, N-1-J} \},
\enda
where $ f_{J, N-1-J}$ denotes the $(1-\alpha_0)$th quantile of  the $F_{J, N-1-J}$ distribution.

Whereas the $F$-test is the default joint test for linear regression returned by most software packages, we can also compute the joint $p$-value by a Wald test with test statistic $W_\lm = \hb^\T \hv^{-1} \hb$. 
The resulting $p$-value equals 
\begina
\pzlm' = \pr( A \geq \wlm), \twhere A \sim \chisq_J,
\enda
with $
\{\pzlm' \geq \alpha_0\} = \{\wlm \leq a_0 \}$. 

Lemma \ref{lem:equiv_lm} below gives the numeric correspondence between $F$ and $\wlm$, and underpins the asymptotic equivalence between $\pzlm$ and $\pzlm'$ for constructing the joint criterion from linear regression; see Lemma \ref{lem:equiv}. 
For  two sequences of events $(\ma_N)_{N=1}^\infty$ and $(\mb_N)_{N=1}^\infty$, write $\ma_N \assim \mb_N$  if $\pr(\ma_N \backslash \mb_N) = \oo$ and $\pr(\mb_N \backslash \ma_N) = \oo$.

\begin{lemma}\label{lem:equiv_lm}
$F = J^{-1} \wlm$ and $f_{J, N-1-J} = J^{-1} a_0 + \oo$ such that 
\begina
\malmjt = \{\pzlm \geq \alpha_0\}  = \{F \leq f_{J, N-1-J}\} = \{\wlm \leq a_\ff \}   \assim   \{\wlm \leq a_0\} = \{\pzlm' \geq \alpha_0\} ,
\enda
where $a_\ff = J \cdot f_{J, N-1-J} = a_0 + \oo$. 
\end{lemma}

\subsection{\mmlr}
Recall level $Q$ as the reference level when fitting $\glmt(Z_i \sim 1+x_{i1}+\cdots+x_{iJ})$ by \mle. 
The fitting algorithm assumes that $(x_i, Z_i)_{i=1}^N$ are independent samples from 
$$Z_i \mid x_i \ \sim\ \text{multinomial}\{ 1; \left( \pi_1(x_i), \dots, \pi_Q(x_i) \right)\},
$$ with 
\beginy
\label{eq:mlr} \qquad
\log\frac{\pi_q (x_i)}{\pi_Q(x_i)} = \bqz  + x_i^\T \bq \quad \text{for}  \ \qiqp  = \{1, \ldots, Q-1\}  \ \text{and}  \ \bq  = (\beta_{q1}, \ldots, \beta_{qJ})^\T. 
\endy

Let $\tbq   = (\tb_{q1}, \dots, \tb_{qJ})^\T$ be the \mles  of $\bq$ for $\qiqp$.  
Let $\tb=(\btb_1^\T, \dots, \btb_{Q-1}^\T)^\T = (\tbqj )_{\qiqp; \ \jej} \in \mathbb{R}^{J(Q-1)}$, with $\tv = (\tv_{qj, q'j'})$ as the  estimated covariance from the same \mles fit. 
The notation simplifies to $\tb = (\tb_1, \ldots, \tb_J)^\T$ and $\tv = (\tv_{jj'})_{j, j' = \ot{J}}$ under the treatment-control experiment with $Q = 2$, $\mqp = \{1\}$, $\tb = \btb_1$, and $(\tb_j, \tv_{jj'}) = (\tb_{1j},  \tv_{1j, 1j'})$. 

\prg{Marginal tests.}
The marginal test of $\tb_{qj}$ 
uses 
\begina
\tqjmlr = \tbqj /\tv_{qj,qj}^{1/2}
\enda
as the test statistic, and computes $\pqjmlr$ as 
\begina
\pqjmlr = \pr( |A| \geq |\tqjmlr|), \twhere A \sim \mn(0, 1). 
\enda
Let $\tmlogit = (\tqjmlr)_\qj \in \mr^{J(Q-1)}$ in lexicographical order of $qj$. The marginal criterion equals
\begina
\malogitmg &\ =\ & \{\pqjg \geq \alpha_{qj} \ \fall \qiqp \ \text{and} \ \jej\} \nonumber\\
& =&\{ |\tqjmlr| \leq a_{qj} \ \fall \qiqp \ \text{and} \ \jej\} \nonumber \\
&=& \{|\tmlogit|  \leq  a\}, 
\enda
with 
$
a  = (a_{qj})_\qj
\in \mr^{J(Q-1)}$ and $a_{qj}$ denoting the $(1-\alpha_{qj}/2)$th quantile of the  standard normal distribution.

When $Q=2$, we have
$(\tb_j, \tv_{jj}, \pjlogit, \alpha_j) = (\tb_{1j},  \tv_{1j, 1j}, p_{1j, \logit}, \alpha_{1j})$ for $\jej$.
The definition of $a$ simplifies to $a   = (a_{1 }, \ldots, a_{J } )^\T \in \mr^J$, with $a_{j}$ equaling the $(1-\alpha_j/2)$th quantile of the standard normal distribution.
The definition of $\tmlogit$ simplifies to $\tmlogit = (t_{1, \logit}, \ldots, t_{J, \logit})^\T$ with 
\begina
t_{j, \logit} = \tb_j / \tv_{jj}^{1/2}.
\enda 
The marginal criterion equals
\beginy\label{eq:mg_logit}
\malogitmg &\ =\ & \{\pjlogit \geq \alpha_{j} \ \fall  \jej\} \\
& =&\{ |\tjglm| \leq a_{j} \ \fall  \jej\} \nonumber \\
&=& \{|\tmlogit|  \leq  a\}. \nonumber
\endy

\prg{Joint test.}
The \lrts for \mlrs tests $\logitt(Z_i \sim 1+x_i)$ against the empty model $\logitt(Z_i \sim 1)$. 
Denote by $\llrt$ the resulting test statistic, with the explicit form given in \eqref{eq:lmd}. 
Then $\pzmlr$ is computed based on the $\chi^2_{J(Q-1)}$ distribution as
\begina
\pzmlr = \pr(A \geq \llrt), \qquad \text{where} \ \ A \sim \chi^2_{J(Q-1)}. 
\enda
The joint criterion equals
\begina
\malogitjt = \{\pzmlr \geq \alpha_0\} = \{\llrt \leq a_0\}, 
\enda
where $a_0$ denotes the $(1-\alpha_0)$th quantile of the $\chi^2_{J(Q-1)}$ distribution. 
The definition of $a_0$ reduces to that under the treatment-control experiment, namely the $(1-\alpha_0)$th quantile of the $\chi^2_{J}$ distribution, with $Q=2$.

Whereas the \lrts is the default joint test for \mlrs returned by most software packages, we can also compute the joint $p$-value  via a joint Wald test with test statistic $\wmlogit = \tb^\T \tv^{-1} \tb$. 
The resulting $p$-value equals 
\begina
\pzmlr' = \pr( A \geq \wmlr), \twhere A \sim \chisq_{J(Q-1)}, 
\enda
with $
\{\pzmlr' \geq \alpha_0\} = \{\wmlr \leq a_0 \}$. 
Theorem \ref{thm:mle_Q} in Section \ref{sec:lemma} ensures
$\llrt- \wmlr = \op$. 
This underpins the asymptotic equivalence between the \lrts and the Wald test for constructing the joint criterion from \mlr; see Lemma \ref{lem:equiv}.

\subsection{Marginal $F$-tests under the general experiment}
Renew $\hx_j(q) = N_q ^{-1}\sumiq x_{ij}$  for $\qiq = \{1, \ldots, Q\}$. 
The $F$-test of covariate $j$ uses
\beginy\label{eq:fj}
F_j  = \frac{\sumq   N_q  \hxj^2(q) / (Q-1) }{ \sumq  \sum_{i:Z_i = q} \{x_{ij}-\hxj(q)\}^2 / (N-Q)} 
\endy 
as the test statistic, and compares it against the $F_{Q-1, N-Q}$ distribution to compute $\pjf$ as
\begina
\pjf = \pr(A \geq F_j), \qquad \text{where} \ \ A \sim F_{Q-1, N-Q}.
\enda
The marginal criterion equals 
\beginy\label{eq:afmg}
\mafmg = \{ \pjf \geq \alpha_j \ \fall \jej \} = \{ F_j \leq \ajf \ \fall \jej\},
\endy
where $\ajf$ denotes the $(1-\alpha_j)$th quantile of the $F_{Q-1,N-Q}$ distribution.

\subsection{Unification}\label{sec:unif_app}

Table \ref{tb:stat} summarizes the regression realizations, test statistics, and reference distributions for the thirteen \repschs in Tables \ref{tb:criteria} and \ref{tb:criteria_g}. 
Table \ref{tb:criteria_stat} summarizes the alternative expressions of the nine criteria in Table \ref{tb:criteria} for treatment-control experiments in terms of the test statistics. 

In particular, recall $ T_\x =(t_{1,\x}, \dots, t_{J,\x})^\T$ as the vector of the marginal $t$-statistics for $\mos$ under the treatment-control experiment. 
Direct comparison shows that $\ho_{jj} =  \hse_j ^2$ for $\jej $, and allows us to unify the marginal criteria in \eqref{eq:mg_t}, \eqref{eq:mg_lm}, and \eqref{eq:mg_logit} as 
\begina
\ma_{\dg,\m} \ = \ \{ |T_\x | \leq  a_\x\} \qquad (\mos),
\enda 
with $a_\logit = a$ and 
\begina
\begin{array}{cllcl}
T_\ts &=\ &  \diag( \hse_j^{-1})_{j=1}^J \htx &\ =\ & \shoinv \htx, \\
\thb &=\ & \diag(\hv_{jj}^{-1/2})_{j=1}^J \hb &=&\shvinv \hb, \\ 
T_\logit &=\ & \diag(\tv_{jj}^{-1/2})_{j=1}^J \tb &=& \stvinv \tb.
\end{array}
\enda
The expressions in the last column of Table \ref{tb:criteria_stat} also extend to the criteria based on the \mlrs for general experiments with renewed definitions of $\tmlogit = (\tqjmlr)_\qj $, $a = (a_{qj})_\qj $, and $a_0$.

\begin{table}[t]\caption{\label{tb:stat}Regression realizations, test statistics, and reference distributions for the thirteen \repschs  in Tables \ref{tb:criteria} and \ref{tb:criteria_g}. 
Let $\mi_{i,\pluss}  = (\mi_{i1}, \ldots, \mi_{i,Q-1})^\T$ for $i= \ot{N}$.  }
\begin{center}
\resizebox{\columnwidth}{!}{
\begin{tabular}{ccc c c cc cc}
\hline
 &Model &Regression&\multicolumn{6}{c}{Test statistics and reference distributions}\\\cline{4-9}
$Q$  &option & realization &\multicolumn{2}{c}{Marginal} &\multicolumn{2}{c}{Joint (default)} &\multicolumn{2}{c}{Joint (Wald)}  \\\hline
\multirow{3}{*}{$2$} & \ts & $\lmt( x_{ij}  \sim 1+Z_i)$&
$\htxj/\hse_j$&$ t_{N-2} $ &
& & $ \htx^\T \ho^{-1} \htx$&$ \chisq_J$ \\
& \lm   & $\lmt(Z_i \sim 1+x_i )$ & $\hb_j / \hv_{jj}^{1/2}$ &$ t_{N-1-J}$ & $F$&$ F_{J, N-1-J}$ &$\hb^\T \hv^{-1} \hb$&$ \chisq_J$ \\
&\logit &$\logitt(Z_i  \sim 1+x_i)$ & $\tb_j / \tv_{jj}^{1/2}$&$ \mn(0,1)$ & $\llrt$&$\chisq_{J}$& $\tb^\T \tv^{-1} \tb$&$ \chisq_{J}$\\
\hline 
\multirow{2}{*}{$ \geq2$} &  \ff &$\lmt(x_{ij} \sim 1 + \mi_{i,\pluss} )$& $F_j$ & $F_{Q-1,N-Q}$ & && &  \\
&\logit &$\glmt(Z_i  \sim 1+x_i)$ & $ \tb_{qj} / \tv_{qj,qj}^{1/2}$&$ \mn(0,1)$ & $\llrt$&$\chisq_{J(Q-1)}$& $\tb^\T \tv^{-1} \tb$&$ \chisq_{J(Q-1)}$\\
\hline
\end{tabular}
}\end{center}
\end{table}

\begin{table}[ht]\caption{\label{tb:criteria_stat}Acceptance criteria for $\mos$ under the treatment-control experiment.}
\begin{center}\begin{tabular}{ccccc c}
\hline
 & by $p$-values  & t & lm & logit    \\\hline
$\ma_{\dg,\m}$   &  {$\pjmo \geq \alpha_j \ (\jej )$}  & $|T_\ts|  \leq  a_\ts$ & $|T_\lm|  \leq  a_\lm$ & $|T_\logit|  \leq  a$ \\ 
$\ma_{\dg,\jt}$  &  {$\pzmo \geq \alpha_0  $} & $W_\ts  \leq a_0  $ & $W_\lm  \leq a_\ff $ & $\llrt  \leq a_0  $ \\ 
$\ma_{\dg,\css}$   &  {$\pjmo\geq \alpha_j \ (j = \zt{J})$}& $|T_\ts|  \leq  a_\ts$, $W_\ts\leq a_0$ & $|T_\lm|  \leq  a_\lm$, $W_\lm\leq a_\ff$& $|T_\logit|  \leq  a$, $\llrt \leq a_0$\\
\hline
\end{tabular}\end{center}
\begin{flushleft}
The test statistics satisfy 
\begina
\begin{array}{lllllllll}
T_\ts &=& \shoinv \htx, &\quad\thb   &=&\shvinv \hb, &\quad T_\logit  &=& \stvinv \tb,\\
\wt &=& \htxt \ho^{-1}\htx, &\quad \wlm &=& \hbt \hv^{-1}\hb, &\quad\wlogit &=& \tbt \tv^{-1}\tb,
\end{array}
\enda
with $a_\dg = a + \oo \ (\dg = \ts, \lm)$ and $a_\ff = a_0 + \oo$. 
\end{flushleft}
\end{table}

 \begin{remark}\label{rmk:1}
Recall $\hse_j$ as the classic standard error of $\htxj$  from $\lmt(\xij  \sim 1+Z_i)$. 
Echoing the discussion in Section \ref{sec:ext_ehw}, we can replace it with the {\ehw} standard error from the same \ols fit, denoted by $\hse_j'$, and conduct the marginal two-sample $t$-test based on the robustly studentized $t$-statistic $\tjt' = \htxj / \hse'_j$ for $\jej$. 

Extensions to other criteria are straightforward by replacing the $\ho$, $\hv$, and $\tv$ in Tables \ref{tb:stat} and \ref{tb:criteria_stat}  with their respective heteroskedasticity-robust counterparts. 
In particular, the {\ehw} counterparts of $\hv$ and $\tv$ can be obtained as direct outputs from the same linear, logistic, and {\mlr}s, respectively. 
The robust counterpart of $\ho$ can be computed as 
$
\ho' =   N_1^{-1}\hsx(1) + N_0^{-1}\hsx(0) 
$, recalling $\hsx(q)$ as the sample covariance of $\{x_i: Z_i = q\}$ for $q = 0,1$. 
The resulting robust variant of the Hotelling's $T^2$ statistic, namely $W_\ts' = \htx^\T\ho' \htx$, defines the multivariate Behrens--Fisher $T^2$ statistic. 
We show in Remark \ref{rmk:proof} in Section \ref{sec:wc_app} below the asymptotic equivalence between the classic and robust test statistics for defining \rep. 
\end{remark}

\section{Lemmas}\label{sec:lemma}
We give in this section the key lemmas for quantifying the asymptotic sampling properties of $\hts \ (\nfl)$ under \rep.

\subsection{Asymptotic theory under complete randomization}\label{sec:cre_app}
We review in this subsection the theory of regression adjustment under complete randomization. 
Assume a general experiment with $\mq = \{\ot{Q}\}$ throughout. 
The treatment-control experiment  is a special case with $Q=2$ and level 2 relabeled as level 0. 

Recall $\gq $ as the coefficient vector of $x_i$ from $\lmt\{\yiq  \sim 1 + x_i\}$ over $i = \ot{N}$.
Let $\gp  = \sumq   e_q \gq $, and let $S_* = (S_{*,qq'} )_{q, q'\in\mq} \ (\nfl)$ be the finite-population covariance matrices of  
$Y_{i, \nm}(q) = \yiq$, $Y_{i, \fisher}(q) = \yiq   -  x_i^\T\gp $, and  $Y_{i, \lin}(q) =  \yiq   -  x_i^\T\bg_q$, respectively, 
with
$S_{*, qq' }= (N-1)^{-1}\sumn\{Y_{i, *}(q)  - \byq  \}\{Y_{i, *}(q')  - \byqp \}$. 
Let 
\begina
V_* =     \diag(S_{*, qq}/e_q)_{\qiq  }  -  S_*  \qquad (\nfl).
\enda 
Condition \ref{asym} ensures that $e_q $, $\gq $, $\gp$, $S_x^2$, and $V_*$ all have finite limits as $N$ tends to infinity. 
For notational simplicity, we will use  the same symbols  to denote their respective limits when no confusion would arise. 
Recall that $\hx = (\hx(1)^\T, \dots, \hx(Q)^\T)^\T$, with $\hx(q) = N_q ^{-1}\sumiq x_{i}$ for $\qiq$. 
Let $\hv'_*$ be the {\ehw} covariance of $\hys$ from the same \ols fit. 
Lemma \ref{lem:cre_Q} follows from \cite{zdca}, and clarifies the design-based properties of $\hys$ and $\hv'_*$. 
\begin{lemma}\label{lem:cre_Q}
Assume a completely randomized general experiment and Condition \ref{asym}. For $ \ms$, 
\begine[(i)]
\item\label{item:normal_Q} 
\begina
\sqrtn\beginp
\hys -\by\\
\hx
\endp &\ \rs \ & \mn\left\{0_{Q+JQ}, \beginp
V_* & \Gamma_* V_x\\
V_x\Gamma_*^\T & V_x
\endp\right\},
\enda 
with $V_x = N\cov(\hx) = \{\diag(e_q ^{-1})_{\qiq } -1_{Q\times Q}\} \otimes \sxx$, 
$V_* = V_\lin + \Gamma_* \Vx \Gamma^\T_* \geq V_\lin$
 for 
\begina
\Gamma_\neyman = \diag(\gq ^\T)_{\qiq  } , \qquad \Gamma_\fisher = \diag\{ (\gq  - \gp )^\T\}_{\qiq  }, \qquad \Gamma_\lin = 0_{Q\times JQ},
\enda 
 and hence $\hyl \succi \hyn , \hyf$; 
\item\label{item:wald_Q}
$N\hv'_* - V_* =  S_*  + \op$ with  $ S_*   \geq 0$.
\ende
\end{lemma}

Lemma \ref{lem:cre_Q}\eqref{item:normal_Q} ensures the consistency and asymptotic normality of $\hts = G\hys$ for estimating $\tau = GY$ under complete randomization.
Lemma \ref{lem:cre_Q}\eqref{item:wald_Q} ensures the asymptotic appropriateness of the \ehws covariance for estimating the true sampling covariance, and thereby justifies the Wald-type inference of $\tau$ based on $(\hts, G\hv_*' G^\T)$ and normal approximation.

The theory under the treatment-control experiment then follows as a special case with $\hts = \hys(1) - \hys(0)$ by the invariance of \ols to non-degenerate transformation of the regressor vector. 
Let $\hse_*$ be the {\ehw} standard error of $\hts$ from the same \ols fit. 

\begin{corollary}\label{cor:cre} 
Assume a completely randomized treatment-control experiment and Condition \ref{asym}. 
For $\nfl$, 
\begine[(i)]
\item\label{item:normal} 
\begina
\rtn   \left(\begin{array}{cc}
\hts  - \tau \\
\htx
\end{array}
\right)
\ \rightsquigarrow \   
\mN \left\{ 
0_{J+1},  
  \left(\begin{array}{cc}
v_*  &  \csi^\T \\
 \csi & \vxi \end{array}\right) \right\}, 
 \enda
with $
v_* = (-1, 1) V_* (-1, 1)^\T$,  $\vx = \ppinv\sxx$, and   
\begina
\cni =\sxx(e_0^{-1}\gamma_0 + e_1^{-1} \gamma_1),\qquad  \cfi =   \sxx(e_1^{-1}-e_0^{-1})(\gamma_1-\gamma_0 ),\qquad \cli = 0_J
\enda
satisfying $v_* - v_\lin = \cs^\T \vx^{-1} \cs \geq 0$; 
\item\label{item:wald}
$N(\hse_*)^2 - v_* = (-1, 1)S_* (-1, 1)^\T + \op$ with  $(-1, 1)S_* (-1, 1)^\T \geq 0$.
\ende
\end{corollary}

Parallel to the comments after Lemma \ref{lem:cre_Q}, Corollary \ref{cor:cre}\eqref{item:normal} states the consistency and asymptotic normality of $\hts$ for estimating $\tau = \by(1) - \by(0)$ under complete randomization, and ensures the asymptotic efficiency of $\htl$ over $\htn$ and $\htf$. 
Corollary \ref{cor:cre}\eqref{item:wald} justifies the Wald-type inference based on $(\hts, \hse_*)$ and normal approximation.

\subsection{Peakedness}
Lemma \ref{lem:gci} below states the celebrated Gaussian correlation inequality, with the recent breakthrough proof due to \cite{royen}; see also \cite{inbook}.

\begin{lemma}[Gaussian correlation inequality]\label{lem:gci}
Let $\mu$ be an $m$-dimensional Gaussian probability measure on $\mr ^m$, that is, $\mu$ is a multivariate normal distribution, centered at the origin. Then $\mu(\mc_1 \cap \mc_2) \geq \mu(\mc_1) \mu(\mc_2)$ for all convex sets $\mc_1, \mc_2 \subset \mr ^m$ that are symmetric about the origin.
\end{lemma}

Lemma \ref{lem:gci} immediately implies Corollary \ref{cor:peak} below, which states that a mean-zero Gaussian measure restricted to a  symmetric convex set is more peaked than the original unrestricted measure.

\begin{corollary}\label{cor:peak}
Let $\ep \sim \mn(0_m, \Sigma)$. Then 
$
\ep \mid \{\ep \in \mc\}  \ \succeq\ \ep
$
for arbitrary convex set $\mc \subset \mathbb R^m$ that is symmetric about the origin. 
\end{corollary}

\begin{proof}[Proof of Corollary \ref{cor:peak}]
Let $\mt \sim \ep \mid \{\ep\in\mc\}$. 
The result follows from 
\begina
\pr(\mt \in \mc_1) = \pr(\ep \in \mc_1 \mid  \ep \in \mc ) = \frac{\pr(\ep \in \mc_1, \ \ep  \in \mc )}{\pr( \ep \in \mc )}  
  \geq    \pr(\ep \in \mc_1 )
\enda 
for arbitrary symmetric convex set $\mc_1 \subset \mr^m$ by Lemma \ref{lem:gci}. 
\end{proof}

Recall that 
\begina 
\wa 
\begin{array}{clllcll}
\mt_\ts   & \ \sim \ & \ep_\ts    \mid   \{ |\ep_\ts   |  \leq  a\}, &\quad& \mt'_\ts    & \ \sim \ &   \ep_\ts     \mid  \{ |\ep_\ts    |  \leq  a , \  \|\ep_\ts   \|_\mm  \leq a_0 \}, \\
%
\mtm  & \ \sim \ &  \epm   \mid   \{ |\epm  |  \leq  a\},&\quad&
\mtm'    & \ \sim \ &  \epm    \mid  \{ |\epm|  \leq  a , \  \|\epm  \|_\mm  \leq a_0 \},\\
%
\mt_\textrm{logit}  & \ \sim \ &  \ep_\textrm{logit}  \mid   \{ |\ep_\textrm{logit} |  \leq  a\}, & \quad &\mt'_\textrm{logit}  & \ \sim \ &  \ep_\textrm{logit}   \mid  \{ |\ep_\textrm{logit} |  \leq  a , \  \|\ep_\textrm{logit}\|_\mm  \leq a_0 \},\\
%
%
\ml & \ \sim \ & \ep_0  \mid \{\|\ep_0\|_2^2\leq a_0\}, & \quad& \mtf & \ \sim \ &   \epf  \mid \{ \sumq   e_q   \ep_{\ff ,qj} ^2 \leq a'_j  \sxj \ \fall   \jej \},  
\end{array}
%
\enda
with $\ep_\ts      \sim \mn\{0_J, D( \vx)\}$, $ \epm    \sim \mn\{0_J, D(\vxinv)\}$, $\ep_\textrm{logit}  \sim \mn\{0_{J(Q-1)}, D (\vsi  )\}$, $\ep_0 \sim \mN(0_{J(Q-1)},I_{J(Q-1)})$, and 
$\epf \sim \mn(0_{JQ},\Vx)$.

\begin{lemma}\label{lem:peak}
$\mt_*, \mt'_* \succeq \ep_*$ for $* = \ts, \lm, \logit$, 
$\ml \succeq \ep_0$, and $\mtf \succeq \epf$. 
\end{lemma}

\begin{proof}[Proof of Lemma \ref{lem:peak}]
The results follow from Corollary \ref{cor:peak} and the convexity of 
$\{u \in \mr ^m: |u| \leq a\}$, $\{u\in \mr ^m: |u|\leq a, \ \|u\|_\mm \leq a_0\}$, $\{u\in \mr ^m: \|u\|_2^2 \leq a_0\}$, and $\{u = (u_{qj})_{\qiq; \ \jej} \in \mr ^{JQ}: \sum_{\qiq} e_q u_{qj}^2 \leq a_j' S_{x,j}^2 \ \fall \jej\}$.

\end{proof}


Lemma \ref{lem:peak_sum} below reviews two classical results in probability for comparing peakedness between random vectors.
\cite{AOS} used them before. 
The proofs follow from \citet[][Lemma 7.2 and Theorem 7.5]{dharmadhikari1988}.

\begin{lemma}\label{lem:peak_sum}
\begine[(i)]
\item 	If two $m\times 1$ symmetric random vectors $A$ and $B$ satisfy $A \succeq B$, then $CA \succeq CB$ for any matrix $C$ with compatible dimensions.
\item 
	Let $A$, $B_1$, and $B_2$ be three independent $m\times 1$   symmetric random vectors. If $A$ is  normal and $B_1 \succeq B_2$, then $A+B_1 \succeq A+B_2$. 
\ende
\end{lemma}

\subsection{Design-based properties of \mlr}
Theorem \ref{thm:mle_Q} below is a novel technical result, and clarifies the design-based properties of the {\mle} outputs from the logistic and {\mlr}s, respectively. The result ensures the asymptotic equivalence between the \lrts and the Wald test in terms of both the test statistics and the corresponding $p$-values. 
We relegate the proof to Section \ref{sec:Q_app}.

Recall that $\hxm = (\hx(1)^\T, \dots, \hx(Q-1)^\T)^\T$, with 
$\vxm = N\cov(\hxm)$ denoting its scaled sampling covariance under complete randomization. 
Recall that  $ V_{\Psi  }  = N\cov(\Psi \hxm) = \Psi \vxp \Psi^\T$, with $\Psi    = \{\Phi ^{-1}\diag(e_\pluss )\} \otimes  \sxxinv$, $e_\pluss  = (e_1, \dots, e_{Q-1})^\T$, and $
\Phi = \diag(e_\pluss )   - e_\pluss  e_\pluss ^\T$. 

\begin{theorem}\label{thm:mle_Q}
Consider a completely randomized experiment with $Q \geq 2$ treatment arms. 
Under Conditions \ref{asym} and \ref{cond:glm_Q}, we have 
 \begina
 \wa 
\begin{array}{lll}
\sqrtn (\tb  -   \Psi   \hx_\pluss ) = \op, &\quad \rtn \tb  \rightsquigarrow \mN ( 0_{J(Q-1)}, \vsi   ), & \quad  N\tv = \vsi   + \op,\\
\llrt  - N\hxp^\T \vxp^{-1} \hxp = \op, &\quad W_\textup{logit} - N\hxp^\T \vxp^{-1} \hxp = \op, & \quad \llrt - W_\textup{logit} = \op, \\
 \llrt  \rs \chi^2_{J(Q-1)}, & \quad \wmlr   \rs \chi^2_{J(Q-1)}. 
\end{array}
\enda
For a treatment-control experiment with $Q = 2$ and the reference level $q=2$ relabeled as 0, the results simplify to
\begina
&&\sqrtn \{\tb  -   \sxxinv\htx\} = \op, \qquad \rtn \tb  \rightsquigarrow \mN\{0_J, (\pp )^{-1}\sxxinv\},\\
&&N\tv = (\pp)^{-1}\sxxinv + \op.
\enda
\end{theorem}

\subsection{Weak convergence}\label{sec:wc_app}
Lemma \ref{lem:equiv} below underpins the asymptotic equivalence between balance criteria based on asymptotically equivalent thresholds or test statistics. 
The proof follows from standard probability calculation and is thus omitted.


\begin{lemma}\label{lem:equiv}
Let $(A_N)_{N=1}^\infty$ be a sequence of $m\times 1$ random vectors.
Let $(B_N)_{N=1}^\infty$ and $(B'_N)_{N=1}^\infty$ be two sequences of random variables with $B_N - B_N' = \op$ and $(A_N, B_N)_{N=1}^\infty$ having a continuous limiting distribution, represented by $(A, B)$. 
Let $(b_N)_{N=1}^\infty$ be a sequence of constants with a finite limit, $b_\infty = \lim_{N\to\infty} b_N  < \infty$. Then 
\begine[(i)]
\item\label{item:equiv1} 
$A_N \mid \{B_N \leq b_N\} \asim A_N \mid \{B_N \leq b_\infty\}$ provided $\pr(B \leq b_\infty) > 0$; 
\item\label{item:equiv2}
 $A_N \mid \{B_N \leq b \} \asim A_N \mid \{B'_N \leq b \}$   for arbitrary fixed $b \in\mr$ that satisfies $\pr(B \leq b) > 0$.
 \ende
\end{lemma}

Definition \ref{def:reg} below extends the notion of rerandomization with general covariate balance criterion (ReG) from \cite{LD2018}.

\begin{definition}\label{def:reg}
Let $\phi(B, C)$ be a binary {\it covariate balance indicator function}, where $\phi(\cdot, \cdot)$ is a binary indicator function and $(B, C)$ are two statistics computed from the data.  
An ReG accepts a randomization if $\phi(B, C)= 1$. 
\end{definition}

The definition of ReG is general and includes all nine criteria in Table \ref{tb:criteria_stat} as special cases. 
Table \ref{tb:reg} below summarizes the covariate balance indicator functions for ReM and the nine \repschs  in Table \ref{tb:criteria_stat} under the treatment-control experiment, respectively. 
As an illustration, $(B, C) = (\sqrtn \htx, N\ho)$ under the two-sample $t$-test model option, with $\phi(\cdot,\cdot)$ equaling
\begina
\phi(u, v) = 
\left\{
\begin{array}{ll}
1\{ |\sigma ( v)^{-1}  u |  \leq  a_\ts\}& \quad \text{under the marginal rule;}\\
1(u^\T v^{-1} u   \leq a_0)&\quad \text{under the joint rule;}\\
1\{ |\sigma ( v)^{-1}  u |  \leq  a_\ts, \ u^\T v^{-1} u  \leq a_0 \}&\quad \text{under the consensus rule.}
\end{array}
\right.
\enda
The resulting covariate balance indicator functions equal
\begina
\phi(B, C) = 
\left\{
\begin{array}{ll}
1\{ |   \shoinv   \htx|   \leq  a_\ts\}&\quad \text{under the marginal rule;}\\
1(    \htxt \ho^{-1}   \htx    \leq  a_0)&\quad \text{under the joint rule;}\\
1\{ |   \shoinv   \htx|   \leq  a_\ts, \   \htxt \ho^{-1}   \htx    \leq  a_0 \}& \quad \text{under the consensus rule}
\end{array}
\right.
\enda
given $\phi(u, v) = \phi( u / \sqrtn,v/N)$ in all three cases. 

\begin{table}[ht]\caption{\label{tb:reg}The covariate balance indicator functions and the corresponding $(B, C)$ for R\lowercase{e}M and the nine R\lowercase{e}P schemes in Table \ref{tb:criteria_stat} under the treatment-control experiment. The joint criterion under the ``\logit'' model option is given in terms of the asymptotically equivalent Wald statistic to  highlight the  analogy across different model options.}
\begin{center}
\begin{tabular}{ccc}\hline
  & \rem & t   \\ \hline
$(B, C)$ & $( \sqrtn \htx,  N \cov(\htx))$ &  $ (\sqrtn \htx,   N \ho)$  \\
marginal     & n.a. & $1\{ |   \shoinv   \htx|    \leq   a_\ts\}$ 
  \\
joint   & $  1\{\htx^\T  \cov(\htx) ^{-1} \htx \leq a_0\}$ & $  1(\htx^\T \ho^{-1} \htx \leq a_0)$   \\
consensus    & n.a. & $1\{ |   \shoinv   \htx|    \leq   a_\ts,\   \htx^\T \ho^{-1} \htx \leq a_0\}$  \\
\hline\hline
& \lm & \logit \\ \hline
$(B, C)$  & $( \sqrtn \hb,   N\hv) $ & $( \sqrtn \tb,  N\tv)$\\
marginal     & $1\{ |   \shvinv   \hb|    \leq   a_\lm\}$ & $1\{|   \stvinv   \tb|    \leq   a \}$   \\
joint   & $  1(\hb^\T \hv^{-1} \hb \leq a_\ff)$   & $  1(\tb^\T \tv^{-1} \tb \leq a_0)$   \\
consensus     & $1\{|   \shvinv   \hb|    \leq   a_\lm,\   \hb^\T \hv^{-1} \hb \leq a_\ff\}$
 & $1\{ |   \stvinv   \tb|    \leq   a,\   \tb^\T \tv^{-1} \tb \leq a_0\}$  \\
\hline
\end{tabular}
\end{center}
\end{table}

Lemma \ref{lem:weak_convergence} below is a generalization of \citet[][Proposition A1]{LD2018},  and gives the asymptotic distribution of arbitrary random elements under ReG. 
To this end, Condition \ref{cond:A1} below imposes some smoothness constraints on the associated $\phi$ to prevent the acceptance region from being a set of measure zero.
All covariate balance criteria in Table \ref{tb:reg} satisfy Condition \ref{cond:A1}.

\begin{condition}\label{cond:A1}
The binary indicator function $\phi(\cdot , \cdot )$ satisfies:
(i) $\phi(\cdot , \cdot )$ is almost surely continuous;
(ii) for $u \sim \mN(0_J, v_0 )$,  we have $\pr\{\phi(u,  v_0 ) = 1\} > 0$ for all $  v_0 > 0$, and $\cov\{u \mid  \phi (u,  v_0 ) = 1\}$ is a continuous function of
$v_0$. 
\end{condition}

\begin{lemma}[Weak convergence under ReG]\label{lem:weak_convergence}
Assume $(\phi_N)_{N=1}^\infty$ as a sequence of binary indicator functions under Condition \ref{cond:A1} that converges to $\phi$. 
For a sequence of random elements $(A_N, B_N, C_N)_{N=1}^\infty$ that satisfies $(A_N, B_N, C_N) \rightsquigarrow (A, B  ,C )$ as $N\to \infty$, we have
\begina
(A_N, B_N) 
 \mid   \{\phi_N(B_N , C_N) = 1 \} &\ \rs \ &
(
A , B )
 \mid    \{\phi(B , C) = 1 \}
\enda
in the sense that, for any continuity set $\mathcal{S}$ of $(A , B ) \mid \{\phi(B , C) = 1\}$, 
\begina
\pr\{ (A_N, B_N) \in \mathcal{S} \mid \phi_N(B_N , C_N) = 1\} = 
\pr\{ (A , B ) \in \mathcal{S} \mid \phi(B , C) = 1  \} + o(1).
\enda
\end{lemma}

Lemma \ref{lem:joint} below gives the asymptotic joint distributions of $\hts$ and the elements that determine the covariate balance measures in Table \ref{tb:reg}. 
The result provides the basis for verifying Propositions \ref{prop:t} and \ref{prop:lm_glm} in a unified way. 

\begin{lemma}\label{lem:joint}
Assume a completely randomized treatment-control experiment and Conditions \ref{asym} and \ref{cond:glm}. 
For $\nfl$, we have 
\begina
\begin{array}{lll}
\big(\sqrtn (\hts-\tau), \sqrtn \htx,  N\cov(\htx) \big) &\ \rs \ &(A_* , B ,  \vxi ),\\
\big(\sqrtn (\hts-\tau), \sqrtn \htx,  N\ho \big) & \ \rs\  & (A_* , B ,  \vxi ),\\
\big(\sqrtn (\hts-\tau), \sqrtn \hb,  N\hv \big) 
& \ \rs\  & (A_* , \vxiinv B ,  \vxiinv),\\
\big(\sqrtn (\hts-\tau), \sqrtn \tb,  N\tv \big) 
& \ \rs\  & \big(A_* ,  \sxxinv B ,  (\pp )^{-1}\sxxinv \big)
\end{array}
\enda
with  
\begina
\beginp
A_*\\
B
\endp \  \sim \ 
\mn\left\{ 0_{J+1}, \beginp
v_*  &  \csi^\T \\
 \csi & \vxi \endp \right\}.
\enda
\end{lemma}

\begin{proof}[Proof of Lemma \ref{lem:joint}]
The result on ReM follows from Corollary \ref{cor:cre}. 
The result on the ``\logit'' model option follows from Theorem \ref{thm:mle_Q}. 
We verify below the results on the two-sample $t$-test (``t'') and linear regression (``lm'') model options, respectively.  

\prg{Result on the two-sample $t$-test model option.}
The joint distribution of $\sqrtn(  \hts - \tau , \htx)$ follows from Corollary \ref{cor:cre}. 
The probability limit of $N\ho$ follows from 
$
N \ho =  (e_0e_1)^{-1} \sxx + \op = \vx + \op$
by \eqref{eq:ho} and $\hsx(q) = \sxx + \op$ under complete randomization and Condition \ref{asym}. 

\prg{Result on the linear regression model option.}
{\fwlc} ensures that
\beginy\label{eq:hb}
\hat\beta = \frac{N_1}{N-1} \sxxinv  \hx(1) =  \frac{N_0N_1}{(N-1)N} \sxxinv  \htx = \frac{N}{N-1}\vx^{-1}\htx
\endy
is a non-degenerate linear transformation of $\htx$. 
The joint distribution of $ \sqrtn(\hts-\tau,\hb)$ then follows from Corollary \ref{cor:cre}. 

Further let
\begina
\szz = \frac{1}{N-1}\sumi (Z_i - \bar Z)^2 = \frac{N}{N-1} \pp, \quad \sxz =  \frac{1}{N-1}\sumi x_i  (Z_i-\bar Z) = \frac{N}{N-1}e_1\hx (1),
\enda
with $\bar Z = N^{-1}\sumi Z_i = e_1$. 
The probability limit of $\hv$ follows from 
\begina
N\hv = \hat\sigma^2 \left(\sumi x_i x_i^\T\right)^{-1}
=\frac{N}{N-1} \hat\sigma^2 \sxxinv 
\enda
with  
$$
 \hat\sigma^2 = \frac{1}{N-1-J}\sumi (Z_i - \bar Z -  x_i^\T\hb)^2
 =  \frac{N-1}{N-1-J}( \szz + \hb^\T  \sxx \hb - 2  \hb^\T \sxz ) = \pp  + \op. 
$$
\end{proof}

Recall that $\Gamma_*' =  \Gamma_* \kappa$,  with $\kappa = (I_{Q-1}, - e_{Q}^{-1} e_\pluss )^\T \otimes I_J $,  
$\Gamma_\neyman = \diag(\gq ^\T)_{\qiq  }$, 
$\Gamma_\fisher = \diag\{ (\gq  - \gp )^\T\}_{\qiq  } $, and 
$\Gamma_\lin = 0_{Q\times JQ}$. 
It follows from  $0_J = \bar x = \sumq   e_q  \hx(q)$ that 
\beginy\label{eq:gamma x}
\hx = \kappa \hxm , \qquad \Gamma_*\hx = \Gamma_*'\hxp \qquad(\nfl).
\endy
This gives the intuition behind the definition of $\Gamma_*'$.
Lemma \ref{lem:joint_Q} gives the asymptotic joint distributions of $(\hys, \hxp)$ and $(\hys, \tb, \tv)$ under the general experiment, respectively,  analogous to Lemma \ref{lem:joint}.

\begin{lemma}\label{lem:joint_Q}
Assume a completely randomized general experiment and Conditions \ref{asym} and \ref{cond:glm_Q}. 
Let $A \sim \mn(0_Q, \vr)$, $B \sim \mn(0_{JQ}, V_x)$, and $B'  \sim \mn(0_{J(Q-1)},\vxm )$ be independent normal random vectors. 
For $\nfl$, we have
\begine[(i)]
\item $\sqrtn (\hys-\by, \hx) \rs (A+\Gamma_* B, B), \ \ \sqrtn (\hys-\by,\hxp) \rs (A+\Gamma'
_* B', B')$; 
\item $
(\sqrtn (\hys-\by), \sqrtn \tb,  N\tv)  \rs  (A+\Gamma'_*  B' ,   \Psi   B' ,  \vsi   )$; 
\item $
\hys \mid \{ \llrt\leq a_0 \} \  \asim \ \hys \mid  \{ \wlogit\leq a_0 \}
$. 

\ende

\end{lemma}

\begin{proof}[Proof of Lemma \ref{lem:joint_Q}]
The result on $ \sqrtn (\hys-\by, \hx) $ follows from Lemma \ref{lem:cre_Q}. This, together with \eqref{eq:gamma x}, further ensures   
$
\sqrtn ( \hys-\by - \Gamma'_* \hxp, \hxp   ) \rightsquigarrow (A, B')$, 
and hence 
$
 \sqrtn (\hys-\by, \hxp  ) \rightsquigarrow (A+\Gamma'_*  B'  ,    B'  )
$
for $\nfl$.

The result on $(\sqrtn (\hys-\by), \sqrtn \tb,  N\tv)$ then follows from $\sqrtn (\tb - \Psi \hxp) = \op$ and $N\tv = \vsi  + \op$ by Theorem \ref{thm:mle_Q}. 

The asymptotic  equivalence between $\hys \mid  \{ \llrt\leq a_0 \}$ and $\hys \mid  \{ \wlogit\leq a_0 \}$  follows from $\llrt - N\hxp^\T \vxp^{-1} \hxp = \op$ and $\wlogit  - N\hxp^\T \vxp^{-1} \hxp = \op$ by Theorem \ref{thm:mle_Q} and Lemma \ref{lem:equiv}.
\end{proof}

\begin{remark}\label{rmk:proof} Lemma \ref{lem:joint} and its proof also imply some of the comments we made in the main text.

First, Lemma \ref{lem:joint}, together with \eqref{eq:hb} and $\sqrtn\{\tb - \sxxinv \htx\} = \op$ from Theorem \ref{thm:mle_Q}, ensures the asymptotic equivalence between $\|\htx\|_\mm = \htxt \cov(\htx)^{-1}\htx$, $\wt = \htxt \ho^{-1}\htx$, $\wlm = \hbt \hv^{-1}\hb$, and $\wlogit = \tb^\T\tv^{-1}\tb$ in the sense that 
$W_\dg -\|\htx\|_\mm = \op$ for $\mos$.
This elucidates the asymptotic equivalence between ReM and the joint criteria under the treatment-control experiment by Lemma \ref{lem:equiv}.

Next, recall $\hse_j$ and $\hse'_j$ as the classic and \ehws standard errors of $\htxj$, respectively, from Remark \ref{rmk:1}. 
It follows from {\fwl} and \citet[][Lemma S1]{ZDfrt} that 
\begina
\hsesq_j = \frac{1}{N-2}\left(\frac{\sxxj}{\szz} - \htau_{x,j}^2 \right), 
\qquad 
\tsesqj =  \frac{(Z- 1_N\bar Z)^\T \diag(\ep_{ij}^2)_{i=1}^N (Z- 1_N\bar Z)}{\| Z-  1_N\bar Z\|_2^2} 
\enda
with $Z = (Z_1, \dots, Z_N)^\T$, $\sxxj = (N-1)^{-1}\sumi x_{ij}^2$, $\ep_{ij} = x_{ij} - Z_i \htxj$, and  
\begina
N \hsesq_j = (\pp )^{-1}\sxxj + \op, 
\qquad 
N \tsesqj = (\pp )^{-1}\sxxj + \op.
\enda
This ensures $t'_{j,\ts}  = \htxj / \hse'_j = t_{j,\ts} + \op$, and hence the asymptotic equivalence of the classic and {\ehw} standard errors for constructing the marginal criterion under the two-sample $t$-tests. 
The results for other criteria are similar and thus omitted.

 Importantly, the asymptotic equivalence between the classic and \ehws standard errors does not hold for $\hses' \ (\nfl)$ and their classic counterparts based on the default outputs of 
$\lmt(Y_i \sim 1+Z_i)$, $\lmt(Y_i \sim 1+Z_i +x_i)$, and $ \lmt(Y_i \sim 1+Z_i +x_i + Z_i x_i)$
  in general.
Specifically, the classic standard errors of $\hts \ (\nfl)$ are not necessarily asymptotically conservative for estimating the true sampling variances, and can thus lead to invalid inferences. 
As a result, the use of \ehws standard errors is immaterial for rerandomization yet crucial for analysis.

\end{remark}

\section{Proofs of the main results}\label{sec:proof_main}
\subsection{Asymptotic distributions in Propositions \ref{prop:t}--\ref{prop:glm_Q}}
\begin{proof}[Proof of Propositions \ref{prop:t} and \ref{prop:lm_glm}]
Let
\begina
\beginp
A_*\\
B
\endp
\ \sim \  
\mn\left\{  0_{J+1},\beginp v_* & c_*^\T\\ c_* & v_x \endp  \right\} 
\enda
with $\sqrt N (\hts - \tau, \htx^\T)^\T \rs (A_*, B^\T)^\T$ for $\nfl$. 
Recall the definitions of $\ma_{\dg,\dr}$ for $\mos$ and $\drs$ in terms of the test statistics from Table \ref{tb:criteria_stat}. 

\prg{Two-sample $t$-tests.}
For $\nfl$, let $(A_N, B_N, C_N) = (\sqrt N(\hts-\tau), \sqrtn \htx, N\ho)$ with $(A_N, B_N, C_N) \rs (A_*, B, v_x)$ by Lemma \ref{lem:joint}. 
\begini 
\item Recall that	$\matjt =\{\pzt \geq \alpha_0\} = \{\wt \leq a_0\}$ under the joint rule. We have 
\begina
\sqrt N(\hts-\tau) \mid \matjt &=& 
\sqrt N(\hts-\tau) \mid \{\htxt \ho^{-1} \htx \leq a_0\}\\
&\ \rs \ & A_* \mid \{B^\T \vx^{-1} B  \leq a_0\}\\
&\ \sim\ & A_* \mid \{\|B\|_\mm  \leq a_0\}
\enda 
by applying Lemma \ref{lem:weak_convergence} to $(A_N, B_N, C_N)$ and $\phi_N(u, v) = \phi(u, v) = 1(u^\T v^{-1} u \leq a_0)$. 
%
\item Recall that $\matmg =\{\pjt \geq \alpha_j, \ \jej \} = \{|T_\ts|  \leq  a_\ts\}$ under the marginal rule. We have 
\begina
\sqrt N(\hts-\tau) \mid \matmg &=& \sqrt N(\hts-\tau) \mid \{ |  \shoinv  \htx|   \leq  a_\ts \} \\
&\ \rs \ & A_* \mid  \{ | \svxinv B|   \leq  a \}
\enda
by applying Lemma \ref{lem:weak_convergence} to $(A_N, B_N, C_N)$, $\phi_N(u, v) = 1\{ |\sigma(v)^{-1} u |  \leq  a_\ts\}$, and $\phi(u, v) = 1\{ |\sigma(v)^{-1} u |  \leq  a\}$.
\item Recall that $\matcs =\{\pjt \geq \alpha_j, \ j = \zt{J} \} = \{\wt\leq a_0, \ |T_\ts|  \leq  a_\ts\}$ under the consensus rule. We have 
\begina
\sqrt N(\hts-\tau) \mid \matcs &=& \sqrt N(\hts-\tau) \mid  \{  \htxt \ho ^{-1} \htx \leq a_0, \ | \shoinv   \htx|   \leq  a_\ts \} \\
 &\ \rs \ & 
 A_* \mid \{  \|B\|_\mm  \leq a_0, \ | \svxinv B|   \leq  a \} 
\enda
by applying Lemma \ref{lem:weak_convergence} to $(A_N, B_N, C_N)$, $\phi_N(u, v) = 1(u^\T v^{-1} u \leq a_0) \cdot 1\{ |\sigma(v)^{-1} u |  \leq  a_\ts\} $, and $\phi(u, v) = 1(u^\T v^{-1} u \leq a_0)\cdot 1\{ |\sigma(v)^{-1} u |  \leq  a\}$.
\endi
It thus suffices to compute 
\begina
A_* \mid \{\|B\|_\mm  \leq a_0\}, \quad A_* \mid  \{ | \svxinv B|   \leq  a\}, \quad A_* \mid  \{ \|B\|_\mm  \leq a_0, \ | \svxinv B|   \leq  a\},
\enda respectively. 

To this end, write 
\beginy\label{eq:decompose_A}
A_*=( A_* - \cs^\T \vxinv B ) + \cs^\T \vxinv B,
\endy
with $A_* - \cs^\T \vxinv B  \sim \mn(0, \vl)$ and independent of $B $. 
This ensures
\beginy\label{eq:astar_joint}
A_*\mid \{\|B\|_\mm \leq a_0\} 
&\ \sim\ & ( A_* - \cs^\T \vxinv B ) + \cs^\T \vx^{-1} \left[ B \mid \{\|B\|_\mm \leq a_0\} \right] \nonumber\\
&\ \sim\ &  v_\lin^{1/2} \epsilon + \cs^\T \vxinv (\vx^{1/2}\ml).
\endy

Likewise for the results under the marginal  and consensus rules. In particular,  let 
$
\ep_\ts    =\svxinv  B   \sim \mn\{0, D(\vx)\}
$ to write 
\begina
B =  \sigma(\vx)  \ep_\ts , \qquad \{ | \svxinv   B|  \leq  a \}=\{ |  \ep_\ts   |  \leq  a \}.
\enda
This ensures
\begina
 B \mid \{ | \svxinv   B|  \leq  a \}
 &=&  \sigma(\vx)   \ep_\ts     \mid   \{ |  \ep_\ts   |  \leq  a \} \\
 &\ \sim\ & \sigma(\vx)   \mt_\ts,\\
  B \mid \{ \|B\|_\mm  \leq a_0, \ | \svxinv B|   \leq  a \} &=&  \sigma(\vx)   \ep_\ts     \mid  \{ \|\ep_\ts   \|_\mm  \leq a_0, \  |\ep_\ts|  \leq  a \} \\
 &\ \sim\ & \sigma(\vx)   \mt'_{\ts},
 \enda
and thus 
\begina
A_* \mid \{ | \svxinv   B|  \leq  a \} &\ \sim\ &  \vl^{1/2} \epsilon +  \cs^\T \vxinv \sigma(\vx)  \mt_\ts, \\
A_* \mid \{ \|B\|_\mm  \leq a_0, \ | \svxinv B|   \leq  a \} &\ \sim\ &  \vl^{1/2} \epsilon +   \cs^\T \vxinv \sigma(\vx)  \mt'_{\ts}
\enda by \eqref{eq:decompose_A}.

\prg{Linear regression.}
For $\nfl$, let $(A_N, B_N, C_N) = (\sqrt N(\hts-\tau), \sqrtn \hb, N\hv)$ with $(A_N, B_N, C_N) \rs (A_*, v_x^{-1} B, v_x^{-1})$ by Lemma \ref{lem:joint}.
\begini
\item Recall that $\malmjt =\{\pzlm \geq \alpha_0\} = \{\wlm \leq a_\ff\}$ under the joint rule by Lemma \ref{lem:equiv_lm}. We have
\begina
\sqrt N(\hts-\tau) \mid \malmjt  &=& 
\sqrt N(\hts-\tau) \mid \{\hbt \hv^{-1} \hb \leq a_\ff\}\\
&\ \rs \ & A_* \mid \{ (\vx^{-1}  B)^\T \vx (\vx^{-1}  B) \leq a_0\}\\
& \sim& A_* \mid \{\|B\|_\mm  \leq a_0\} 
\enda 
by applying Lemma \ref{lem:weak_convergence} to $(A_N, B_N, C_N)$, $\phi_N(u, v) =   1(u^\T v^{-1} u \leq a_\ff)$, and $ \phi(u, v) = 1(u^\T v^{-1} u \leq a_0)$. 
\item Recall that $\malmmg = \{\pjlm \geq \alpha_j, \ \jej \} = \{|T_\lm| \leq  a_\lm\}$  under the marginal rule. We have
\begina
\sqrt N(\hts-\tau) \mid \malmmg  &=& \sqrt N(\hts-\tau) \mid \{ |  \shvinv  \hb|   \leq  a_\lm \}\\
&\ \rs \ & A_* \mid  \{ | \svxinvinv \vx^{-1}B|   \leq  a \}
\enda
by applying Lemma \ref{lem:weak_convergence} to $(A_N, B_N, C_N)$, $\phi_N(u, v) = 1\{ |\sigma(v)^{-1} u |  \leq  a_\lm\}$, and $\phi(u, v) = 1\{ |\sigma(v)^{-1} u |  \leq  a\}$.
\item Recall  that $\malmcs =  \{\pjlm \geq \alpha_j, \ j = \zt{J}\} = \{\wlm \leq a_\ff, \ |T_\lm| \leq  a_\lm\}$  under the consensus rule. We have \begina
\sqrt N(\hts-\tau) \mid \malmcs &=& \sqrt N(\hts-\tau) \mid  \{\hbt \hv ^{-1} \hb \leq a_\ff, \ | \shvinv   \hb|   \leq  a_\lm \}\\
 &\ \rs \ & 
 A_* \mid  \{ (\vx^{-1}  B)^\T \vx (\vx^{-1}  B) \leq a_0, \ | \svxinvinv \vx^{-1}B|   \leq  a \}\\
 & \sim& A_* \mid \{ \|B\|_\mm  \leq a_0, \ | \svxinvinv \vx^{-1}B|   \leq  a \}
\enda
by applying Lemma \ref{lem:weak_convergence} to $(A_N, B_N, C_N)$, $\phi_N(u, v) =  1(u^\T v^{-1} u \leq a_\ff) \cdot 1\{ |\sigma(v)^{-1} u |  \leq  a_\lm\}$, and $\phi(u, v) = 1(u^\T v^{-1} u \leq a_0)\cdot 1\{ |\sigma(v)^{-1} u |  \leq  a\}$.
\endi
It thus suffices to compute $ A_* \mid  \{\|B\|_\mm  \leq a_0 \}$, 
$A_* \mid   \{ | \svxinvinv \vx^{-1}B|   \leq  a\}$, and 
$A_* \mid  \{ \|B\|_\mm  \leq a_0, \ | \svxinvinv \vx^{-1}B|   \leq  a\}$, respectively.

The distribution of $ A_* \mid  \{\|B\|_\mm  \leq a_0 \}$ is given by \eqref{eq:astar_joint}. 
For $A_* \mid   \{ | \svxinvinv \vx^{-1}B|   \leq  a\}$  and 
$A_* \mid  \{ \|B\|_\mm  \leq a_0, \ | \svxinvinv \vx^{-1}B|   \leq  a\}$, let 
 $
 \epm   =   \dd  ^{-1} \vxinv B \sim \mn\{0, D(\vxinv)\}
$
  to write   
\begina
B = \vx \dd  \epm, \qquad \big \{ | \dd  ^{-1} \vxinv B|  \leq  a \} = \{| \epm | \leq  a \}.
\enda
This ensures
 \begina
B \mid  \{ | \svxinvinv \vx^{-1}B|   \leq  a \} &=&   \vx\dd    \epm   \mid  \{ |\epm|  \leq  a\}\\
&\ \sim\ & \vx\dd  \mtm,\\
B \mid \{ \|B\|_\mm  \leq a_0, \ | \svxinvinv \vx^{-1}B|   \leq  a \} &=&  \vx\dd   \epm   \mid   \{\|\epm \|_\mm \leq a_0, \  |\epm|  \leq  a\}\\
 &\ \sim\ & \vx\dd   \mtm', 
 \enda 
and thus
\begina
A_* \mid \{ | \svxinvinv \vx^{-1}B|   \leq  a \} &\ \sim\ & \vl^{1/2} \epsilon + \cs^\T \dd   \mtm,\\
A_* \mid \{ \|B\|_\mm  \leq a_0, \ | \svxinvinv \vx^{-1}B|   \leq  a \}&\ \sim\ & \vl^{1/2} \epsilon + \cs^\T \dd   \mtm'
\enda 
by \eqref{eq:decompose_A}. 

\prg{Logistic regression.} 
For $\nfl$, let $(A_N, B_N, C_N) = (\sqrt N(\hts-\tau), \sqrtn \tb, N\tv)$ with $(A_N, B_N, C_N) \rs (A_* ,  \sxxinv B ,  (\pp )^{-1}\sxxinv )$ by Lemma \ref{lem:joint}.
We verify below that $(\hts \mid \ma_{\logit,\dr}) \asim (\hts \mid \ma_{\lm,\dr})$ for $\drs$. 
\begini
\item Recall that $\malogitjt =  \{\pzmlr\geq \alpha_0 \} = \{ \llrt \leq a_0\}$  under the joint rule. 
By $\hys \mid \{ \llrt\leq a_0\} \asim \hys \mid \{ \wlogit\leq a_0\}$ from Lemma \ref{lem:joint_Q}, 
we have
\beginy\label{eq:lrt_wald_2}
  \hts   \mid \malogitjt
& \ = \ &  
 \hts    \mid\{ \llrt \leq a_0\}\\
 & \  \asim\ & 
  \hts  \mid\{ \wmlr \leq a_0\} =\hts  \mid\{\tb^\T \tv^{-1} \tb \leq a_0\},\nonumber
 \endy
  with 
\begina
 \sqrtn(\hts - \tau)   \mid\{\tb^\T \tv^{-1} \tb \leq a_0\} &\ \rs \ & 
A_* \mid \{   B^\T\sxxinv  (\pp\sxx) \sxxinv B  \leq a_0\} \\
&\ \sim\ & A_*  \mid \{\|B\|_\mm  \leq a_0\}
\enda
by applying Lemma \ref{lem:weak_convergence} to $(A_N, B_N, C_N) $ and $\phi_N(u, v) = \phi(u, v) = 1(u^\T v^{-1} u \leq a_0)$. 
This ensures
\begina
\sqrt N(\hts - \tau) \mid\malogitjt
&\ \rs \ & A_*  \mid \{\|B\|_\mm  \leq a_0\},
\enda 
identical to the limiting distribution of $\sqrt N(\hts - \tau) \mid\malmjt$. 
\item Recall that $\malogitmg = \{\pjglm \geq \alpha_j, \ \jej \} = \{|T_\logit|  \leq  a\}$  under the marginal rule. We have
\begina
\sqrt N(\hts-\tau) \mid \malogitmg &=& \sqrt N(\hts-\tau) \mid \{ |  \stvinv  \tb|   \leq  a \}\\
&\ \rs \ & A_* \mid  \{ | \svxinvinv \vx^{-1}B|   \leq  a \} 
\enda
by applying Lemma \ref{lem:weak_convergence} to $(A_N, B_N, C_N)$ and $\phi_N(u, v) =\phi(u, v) = 1\{ |\sigma(v)^{-1} u |  \leq  a\}$; in particular, the limit of $\{ |  \stvinv  \tb|   \leq  a  \}$ follows from
\begina
 \stvinv  \tb &\ \rs \ & \sigma\left( (\pp )^{-1}\sxxinv \right)  \sxxinv B\\
&= &
\sigma\left( (\pp )^{-2}\vxinv \right)  (\pp) \vxinv B 
= 
\sigma( \vxinv)   \vxinv B
\enda
given  $\vx = (\pp )^{-1}\sxx$ and $\sigma( (\pp )^{-2}\vxinv) = (\pp )^{-1} \sigma( \vxinv) $. 
This is identical to the limiting distribution of $\sqrt N(\hts - \tau) \mid\malmmg$. 
\item Recall that $\malogitcs =\{\pjglm \geq \alpha_j, \ j = \zt{J}\} = \{\llrt \leq a_0, \ |T_\logit|  \leq  a\}$ under the consensus rule. The same reasoning as in  \eqref{eq:lrt_wald_2} ensures
\begina
\hts-\tau \mid \malogitcs &\ = \ & \hts-\tau \mid  \{\llrt \leq a_0, \ | \stvinv   \tb|   \leq  a \}\\
 &\ \asim\ & 
\hts-\tau\mid  \{ \tbt \tv^{-1}\tb \leq a_0, \ | \stvinv   \tb|   \leq  a \},
\enda
 with 
\begina
&&\sqrt N(\hts-\tau) \mid  \{ \tbt \tv^{-1}\tb \leq a_0, \ | \stvinv   \tb|   \leq  a \}\\
 &\ \rs \ & 
 A_* \mid \{ \|B\|_\mm  \leq a_0, \ | \svxinvinv \vx^{-1}B|   \leq  a \}
\enda
by applying Lemma \ref{lem:weak_convergence} to $(A_N, B_N, C_N)$ and $\phi_N(u, v) = \phi(u, v) = 1(u^\T v^{-1} u \leq a_0) \cdot 1\{ |\sigma(v)^{-1} u |  \leq  a\}$.
This ensures 
\begina
\sqrt N(\hts-\tau) \mid \malogitcs
  &\ \rs \ &
 A_* \mid \{ \|B\|_\mm  \leq a_0, \ | \svxinvinv \vx^{-1}B|   \leq  a \}, 
\enda
identical to the limiting distribution of $\sqrt N(\hts - \tau) \mid\malmcs$. 
\endi
\end{proof}

\begin{proof}[Proof of Proposition \ref{prop:F}]
Recall $\mafmg = \{ F_j \leq \ajf \ \fall \jej \}$ from \eqref{eq:afmg} with the explicit forms of $F_j$ in \eqref{eq:fj}. 
With
$
(N-1)\sxj = \sumi x_{ij}^2 =   \sumqi \{x_{ij} - \hxj(q)\}^2 + \sumq   N_q    \hxj^2(q) 
$
by direct algebra, the expression of $\mafmg$ can be simplified to 
\begina
\mafmg = \{ F_j \leq \ajf \ \ \fall j \}  =  \left\{  \sumq   e_q  \{\sqrtn \hx_j(q)\}^2 \leq a''_j \ \ \fall j \right\} , 
\enda
where 
\begina
a''_j = \frac{(N-1)\sxj}{ (N-Q)/\{(Q-1)\ajf\}+1} = a'_j  \sxj + o(1)
\enda
by $\ajf = (Q-1)^{-1} a'_j  + o(1)$. 

Observe that  
 $\sqrtn\hx \rs \ep_\ff$ and $\sqrtn \hat x_j(q) \rs \ep_{\ff,qj}$ by Lemma \ref{lem:cre_Q}.  
The result follows from Lemmas \ref{lem:weak_convergence} and \ref{lem:joint_Q} as
\begina
&& \sqrtn(\hys - \by)\mid \mafmg\\
&=& \sqrtn(\hys - \by - \Gamma_*\hx) +\Gamma_* \sqrtn\hx ~ \left |~ \left\{  \sumq   e_q  \{\sqrtn \hx_j(q)\}^2 \leq a''_j\ \ \fall j \right\}\right. \\
&\rs &    \vr^{1/2}  \epsilon + \Gamma_*    \epf ~ \left |~ \left\{ \sumq   e_q  \ep_{\ff,qj}^2 \leq a_j' \sxj \ \ \fall j \right\}\right.. 
\enda

\end{proof}

\begin{proof}[Proof of Proposition \ref{prop:glm_Q}]
Renew $a_0$ as the $(1-\alpha_0)$th quantile of the $\chi^2_{J(Q-1)}$ distribution.
Renew $a = (a_{qj})_\qj$, where $a_{qj}$ denotes the $(1-\alpha_{qj}/2)$th quantile of the  standard normal distribution.
The marginal, joint, and consensus criteria based on the \mlrs equal 
\beginy\label{eq:criteria_mlr}
\begin{array}{ll}
\malogitmg = \{|T_\logit|  \leq  a\}, &\quad 
\malogitjt = \{\llrt \leq a_0\},\\
\malogitcs = \{\llrt \leq a_0, \ |T_\logit|  \leq  a\},
\end{array}
\endy
respectively, with $
\tmlogit = \diag(\tv_{qj,qj}^{-1/2}) \tb = \stvinv  \tb$ and $\llrt- \tb^\T \tv^{-1} \tb = \op$.

Let $
A \sim \mn(0_Q, \vr)$ and $ B  \sim \mn(0_{J(Q-1)},\vxm )$ be two independent normal random vectors with $\cov(\Psi B) = \vsi $.  
For $\nfl$, let $(A_N, B_N, C_N) = (\sqrt N(\hys-\by), \sqrtn \tb, N\tv)$ with $ (A_N, B_N, C_N) \rs (A+\Gamma'_*  B  ,   \Psi   B ,  \vsi   ) $ by Lemma \ref{lem:joint_Q}. 
\begini 
\item Under the joint rule, Lemma \ref{lem:joint_Q} ensures 
\beginy\label{eq:lrt_wald}\hys   \mid \malogitjt \ = \ \hys   \mid \{\llrt \leq a_0\} \ \asim \ \hys     \mid\{\tb^\T \tv^{-1} \tb \leq a_0\}
\endy
with 
\begina
 \sqrtn(\hys - \by)   \mid\{\tb^\T \tv^{-1} \tb \leq a_0\} &\ \rs \ & 
A+\Gamma'_*  B  \mid \{ (\Psi   B)^\T \vsi ^{-1} (\Psi   B)  \leq a_0\} \\
&\ \sim\ & A+\Gamma'_*  B  \mid \{\|B\|_\mm  \leq a_0\}
\enda
by applying Lemma \ref{lem:weak_convergence} to $(A_N, B_N, C_N) $ and $\phi_N(u, v) = \phi(u, v) = 1(u^\T v^{-1} u \leq a_0)$. This ensures
\begina
\sqrt N(\hys - \by ) \mid \malogitjt
&\ \rs \ & A+\Gamma'_*  B  \mid \{\|B\|_\mm  \leq a_0\}. 
\enda 
\item Under the marginal rule, we have
\begina
\sqrt N(\hys - \by ) \mid\malogitmg
&=& \sqrt N(\hys - \by ) \mid \{ |  \stvinv  \tb|   \leq  a \} \\
&\ \rs \ & A+\Gamma'_*  B  \mid  \{ | \sigma(\vsi )^{-1} \Psi B|   \leq  a \}
\enda
by applying Lemma \ref{lem:weak_convergence} to $(A_N, B_N, C_N) $ and $\phi_N(u, v)  = \phi(u, v) = 1\{ |\sigma(v)^{-1} u |  \leq  a\}$.
\item Under the consensus rule, we have
\begina
\hys  \mid \malogitcs
&\ = \ & \hys   \mid  \{\llrt \leq a_0, \ |  \stvinv  \tb|   \leq  a \} \\
 &\ \asim\ &  \hys  \mid  \{\tbt\tv^{-1}\tb\leq a_0, \ |  \stvinv  \tb|   \leq  a \}
\enda
by \eqref{eq:lrt_wald} with 
\begina
&& \sqrt N(\hys - \by ) \mid  \{\tbt\tv^{-1}\tb\leq a_0, \ |  \stvinv  \tb|   \leq  a \}\\
&\ \rs \ &  A+\Gamma'_*  B  \mid \{  \|B\|_\mm  \leq a_0, \  | \sigma(\vsi )^{-1} \Psi B|   \leq  a \} 
\enda
by applying Lemma \ref{lem:weak_convergence} to $(A_N, B_N, C_N) $ and  $\phi_N(u, v) = \phi(u, v) = 1(u^\T v^{-1} u \leq a_0)\cdot 1\{ |\sigma(v)^{-1} u |  \leq  a\}$.
This ensures
\begina
 \sqrt N(\hys - \by )  \mid \malogitcs &\ \rs \ &  A+\Gamma'_*  B  \mid \{  \|B\|_\mm  \leq a_0, \  | \sigma(\vsi )^{-1} \Psi B|   \leq  a \}.
\enda
\endi
With $A \sim V_\lin^{1/2}\ep$ and independent of $B$, it suffices to compute $B  \mid \{  \|B\|_\mm  \leq a_0  \} $,  $B  \mid \{   | \sigma(\vsi )^{-1} \Psi B|   \leq  a\} $, and $B  \mid \{  \|B\|_\mm  \leq a_0, \  | \sigma(\vsi )^{-1} \Psi B|   \leq  a\} $, respectively.

For the joint criterion, we have 
$
B \mid  \{\|B\|_\mm \leq a_0\} \sim \vxm^{1/2} \ml
$, and thus 
\begina
 A+\Gamma'_*  B  \mid \{  \|B\|_\mm  \leq a_0  \} &\ \sim\ & V_\lin^{1/2}\ep  + \Gamma'_*  \vxm^{1/2} \ml.
\enda
For the marginal and consensus criteria, let 
$
 \ep_\logit   =    \sigma(\vsi  ) ^{-1} \Psi  B \sim \mn\{0, D(\vsi  )\}$
  to write  
\begina
 B = \Psi ^{-1}   \sigma(\vsi  )   \ep_\logit , \qquad \{ |  \sigma(\vsi  ) ^{-1} \Psi  B|  \leq  a \} = \big \{ | \ep_\logit |  \leq  a \}. 
\enda
This ensures
 \begina
B \mid  \{ |  \sigma(\vsi  ) ^{-1} \Psi  B|  \leq  a \} &=&   \Psi ^{-1}     \sigma(\vsi  )  \ep_\logit   \mid \{ | \ep_\logit | \leq a\}\\
&\ \sim\ &\Psi ^{-1}   \sigma( \vsi  )   \mt_\logit,\\
B  \mid \{  \|B\|_\mm  \leq a_0, \  | \sigma(\vsi )^{-1} \Psi B|   \leq  a \} &= &\Psi ^{-1}    \sigma(\vsi  )   \ep_\logit   \mid \{ |\ep_\logit  | \leq a, \ \|\ep_\logit \|_\mm \leq a_0\} \\
 &\ \sim\ & \Psi ^{-1}   \sigma(\vsi  )  \mt'_\logit, 
 \enda
and thus 
\begina
 A+\Gamma'_*  B  \mid \{    | \sigma(\vsi )^{-1} \Psi B|   \leq  a \} &\ \sim\ & V_\lin^{1/2}\ep  + \Gamma'_*  \Psi ^{-1}   \sigma(\vsi  )  \mt_\logit, \\
 A+\Gamma'_*  B  \mid \{  \|B\|_\mm  \leq a_0, \  | \sigma(\vsi )^{-1} \Psi B|   \leq  a \} &\ \sim\ & V_\lin^{1/2}\ep  + \Gamma'_*  \Psi ^{-1}   \sigma(\vsi  )  \mt'_\logit.
\enda
\end{proof}

\subsection{Covariate balance and asymptotic relative efficiency in Theorems \ref{thm:t}--\ref{thm:glm_Q}}

%
%

\begin{proof}[Proof of Theorems  \ref{thm:t}--\ref{thm:glm_Q}] 
The asymptotic relative efficiency of $\hts \ (\nfl)$ follows immediately from Propositions \ref{prop:t}--\ref{prop:glm_Q} and Lemmas \ref{lem:peak}--\ref{lem:peak_sum}. 
We verify below the improved covariate balance under the \repschs based on the \mlr. 
The results under the two-sample $t$-test-based, linear or logistic regression-based, and marginal $F$-test-based criteria are analogous and thus omitted. 
Note that the asymptotic conditional bias and the difference between different estimators are all functions of $\htx$. 
The results associated with the covariance reduction factor $\rho(J, a_0)$ under the joint rules then follow from the results on ReM from \cite{morgan2012rerandomization} and \cite{LD2018}. 

Recall the acceptance criteria by test statistics from \eqref{eq:criteria_mlr}. 
Recall from \eqref{eq:gamma x} that  $\hx = \kappa \hxm$ and hence $\htx = G_x \hx = G_x\kappa \hxm$. 
Let $A \sim \mn(0_{J(Q-1)}, \vxp)$ with 
$\sqrtn \hxp \rs A$ and $\sqrtn \htx \rs G_x\kappa A$  by Lemma \ref{lem:joint_Q}.  
The result follows from 
\begina
\sqrtn \htx \mid \malogitjt
&\ =\ & 
G_x\kappa (\sqrtn \hxm) \mid \{\llrt \leq a_0 \}\\
&\ \asim\ &  G_x\kappa (\sqrtn \hxm) \mid \{  N\hxp^\T \vxp^{-1} \hxp \leq a_0 \} \\
&\ \rs\ &  G_x\kappa A \mid \{ A^\T \vxp^{-1} A \leq a_0 \}\\
&\succeq& G_x\kappa A, \\
\sqrtn \htx \mid \malogitmg
&\ =\ & 
G_x\kappa (\sqrtn \hxm) \mid \{ |\stvinv \tb|  \leq  a \}\\
&\ \asim\ & 
G_x\kappa (\sqrtn \hxm) \mid \{ |\sigma(\vsi)^{-1} \sqrtn  \Psi \hxm|  \leq  a \}\\
&\ \rs\ &  G_x\kappa A \mid \{ |\sigma(\vsi)^{-1} \Psi A|  \leq  a \}\\
&\succeq& G_x\kappa A, \\
\sqrtn \htx \mid \malogitcs
&\ =\ & 
G_x\kappa (\sqrtn \hxm) \mid \{\llrt \leq a_0 , \ |\stvinv \tb|  \leq  a \}\\
&\ \asim\ & 
G_x\kappa (\sqrtn \hxm) \mid \{ N\hxp^\T \vxp^{-1} \hxp \leq a_0 , \ |\sigma(\vsi)^{-1} \sqrtn  \Psi \hxm|  \leq  a \}\\
&\ \rs\ & G_x\kappa A \mid \{A^\T \vxp^{-1} A \leq a_0 , \ |\sigma(\vsi)^{-1} \Psi A|  \leq  a \}\\
&\succeq& G_x\kappa  A. 
\enda
In particular, the three ``$\asim$'' follow from 
$\llrt - N\hxp^\T \vxp^{-1} \hxp = \op$,  $\sqrtn (\tb -  \Psi\hxp) = \op$, and $N \tv - \vsi = \op$; the three ``$\rs$'' follow from Lemma \ref{lem:weak_convergence}; the three ``$\succeq$'' follow from Corollary \ref{cor:peak}. 
\end{proof}

\section{Proof of Theorem \ref{thm:mle_Q}}\label{sec:Q_app}
We verify in this section the properties of the logistic and {\mlr}s in Theorem \ref{thm:mle_Q}.
The results ensure the asymptotic equivalence of the \lrts and the Wald test for \lmlrs from the design-based perspective.  
To this end, we first review some useful numeric facts about the \mlrs in Section \ref{sec:mle_Q_numerical}, and  then give the proof in Section \ref{sec:mle_Q_proof}. 

Recall that $e_\pluss  = (e_1, \dots, e_{Q-1})^\T$, $\diag(e_\pluss ) = \diag(e_q )_{\qiqp}$, and $\Psi    = \{\Phi ^{-1}\diag(e_\pluss )\} \otimes  \sxxinv$, 
with $
\Phi = \diag(e_\pluss )   - e_\pluss  e_\pluss ^\T$.
Let $\rp   =\diag(e_\pluss )$ be a shorthand for $\diag(e_\pluss)$ to write $\Psi    =  (\Phi ^{-1} \rp  ) \otimes  \sxxinv$ with
\begina
\Phi   =  \rp   - e_\pluss  e_\pluss ^\T = 
\begin{pmatrix}
e_1(1-e_1) & -e_1 e_2 & \dots & -e_1 e_{Q-1}\\
- e_2e_1 &  e_2(1- e_2) & \dots & -e_2 e_{Q-1}\\
\vdots& \vdots& & \vdots\\
- e_{Q-1}e_1 & -e_{Q-1} e_2 & \dots &  e_{Q-1} (1-e_{Q-1})
\end{pmatrix}.
\enda
Recall that $\hxm = (\hx(1)^\T, \dots, \hx(Q-1)^\T)^\T$, with 
$\vxm = N\cov(\hxm)$. 
Then $\vxm  = ( \rp^{-1}   -1_{(Q-1)\times (Q-1)} )\otimes \sxx $ equals the upper $J(Q-1) \times J(Q-1)$ submatrix of $\Vx=\{ \diag(e_q^{-1})_{\qiq}   -1_{Q\times Q}\}\otimes \sxx $. 
This ensures 
\beginy\label{eq:vsi}
\qquad \quad\vsi  & =& N\cov(\Psi\hxp)= \Psi \vxp \Psi^\T \\
& =&   \left\{(\Phi^{-1}\rp)\otimes \sxxinv\right\}  \left\{ (\rp  ^{-1} -1_{(Q-1)\times(Q-1)}) \otimes \sxx\right\}\left\{(\Phi^{-1}\rp)^\T\otimes \sxxinv\right\} \nonumber\\
&=&  \left\{\Phi ^{-1} \rp   (\rp  ^{-1} - 1_{(Q-1)\times (Q-1)})\rp   \Phi ^{-1}\right\} \otimes \sxxinv\nonumber \\
&=& \linv \otimes \sxxinv   \nonumber 
\endy 
given $
\Phi = \rp( \rp^{-1}   -1_{(Q-1)\times (Q-1)} ) \rp$. 
As a result, we have $
\Psi = \vsi  (\rp  \otimes I_J)$ and hence
\beginy\label{eq:rsr}
(\rp \otimes I_J) \vsi  (\rp \otimes I_J) = \Psi^\T \vsi^{-1} \Psi = \vxp^{-1}. 
\endy 

\subsection{Numeric facts about the \mlrs}\label{sec:mle_Q_numerical}
Recall the multinomial logistic model from \eqref{eq:mlr}.  
We have
\beginy\label{eq:pq}
\pi_q (x_i) = \pi_q (\txi , \theta) = \frac{\exp(\txbi_q )}{1+\sum_{q'\in\mqp} \exp(\txbi_{q'}) } \qquad \text{for} \ \  q=\ot{Q}, 
\endy
with $\txi  = (1, x_i^\T)^\T$, $\theta_Q = 0_{J+1}$, $\theta_q  = (\bqz , \bq ^\T)^\T$ for $\qiqp$, and $\theta = (\theta^\T_1, \dots, \theta^\T_{Q-1})^\T $. 
The scaled log-likelihood function of $(x_i, Z_i)_{i=1}^N$ equals 
\beginy\label{eq:l}
\bar L (\theta) &\ =\ & \meani \log\{\pi_{Z_i} (x_i)\} \\
&=& \meani \left[ \sumqmo  \wiq  \txit  \theta_q - \log\left\{ 1+\sumqmo \exp(\txit  \theta_q )\right\}  \right]. \nonumber
\endy
The score function of $\bar L (\theta)$ equals 
\beginy\label{eq:u}
U(\theta) =  \frac{\partial  \bar L (\theta)}{\partial \theta} = 
\begin{pmatrix}
U_1(\theta)\\
 \vdots\\
 U_{Q-1}(\theta)
\end{pmatrix} 
\endy
with
\begina
U_q (\theta) =  \frac{\partial  \bar L (\theta)}{\partial \theta_q } = N^{-1} \sumi \txi \{\wiq  - \pi_q (\txi , \theta)\} \qquad (\qiqp  ).
\enda
The Hessian matrix of $\bar L(\theta)$ equals
\beginy\label{eq:hth}
H(\theta)  
  =   \frac{\partial^2  \bar L (\theta)}{\partial \theta \partial \theta^\T}, 
\endy
 with the explicit form given by Condition \ref{cond:glm_Q}.

\subsection{The proof}\label{sec:mle_Q_proof}
Lemmas \ref{lem:interchange} and \ref{lem:McFadden} provide the basis for proving Theorem \ref{thm:mle_Q}. 
\begin{lemma}\label{lem:interchange}\cite[][Theorem 7.17]{rudin1976}
Suppose $\{f_N(x)\}_{N=1}^\infty$ is a sequence of functions, differentiable on $[a, b]$
and such that $\{f_N(x_0)\}_{N=1}^\infty$ converges for some point $x_0$ on $[a, b]$. 
If the sequence of derivatives, $\{   f'_N(x)\}_{N=1}^\infty$,  converges
uniformly on $[a, b ]$, then $\{f_N(x)\}_{N=1}^\infty$ converges uniformly on $[a, b ]$, to a function $f(x)$, and
\begina
\lim_{N\to\infty}   f'_N(x) =  f'(x)  \qquad ( a\leq x \leq b).
\enda
\end{lemma}

\begin{lemma}\label{lem:McFadden}\cite[][Theorem 2.7]{Newey1994}
If there is a function $Q_0(\theta)$ such that (i) $Q_0(\theta)$ is uniquely maximized at $\theta_0$; (ii) $\theta_0$ is an element of the interior of a convex set $\Theta$ and $Q_N(\theta)$ is concave; and (iii) $ Q_N(\theta) = Q_0(\theta) +   \op $ for all $\theta \in \Theta$, then the maximizer of $Q_N(\theta)$, denoted by $\hth_N$, exists with probability approaching one and $\hth_N = \theta_0+ \op$. 
\end{lemma}

\begin{proof}[Proof of Theorem \ref{thm:mle_Q}]
\citet[][Lemma S5]{ZDfrt} ensures that 
\beginy\label{eq:as}
\hx(q) = o(1)\qquad \text{for all $\qiq $}
\endy  almost surely under Condition \ref{asym}. 
For notational simplicity, we assume that  \eqref{eq:as} is true in the following proof. The simplification does not affect the validity of the proof given all results concern either convergence in probability or convergence in distribution.

Let $\tbq $ and $\tth_q = (\tbqz, \tbq ^\T)^\T$  be the {\mle}s  of $\bq $ and $\theta_q  = (\bqz , \bq ^\T)^\T$ for $\qiqp$ in \eqref{eq:mlr}, respectively,  concatenated as $\tb=(\tbb_1^\T, \dots, \tbb_{Q-1}^\T)^\T$ and $\tth=(\tth_1^\T, \dots, \tth_{Q-1}^\T)^\T$.

\prg{Convergence of $\tth$.}
As a key intermediate result, we first verify
\beginy\label{eq:lim}
\tth = \theta^* + \op
\endy
for 
 \begina
\theta^* =  ( (\theta_1^*)^\T, \ldots, (\theta_\qmo^*)^\T )^\T, \qquad \text{where} \ \  \theta_q ^* = (\bqzs , 0_J^\T)^\T \ \ \text{with}  \ \ \bqzs  = \log(e_q /e_Q). 
\enda
By Lemma \ref{lem:McFadden}, 
it suffices to show that there exists a function $\bar L_\infty(\theta)$ such that (i) $\bar L_\infty(\theta)$ is uniquely maximized at $\theta^*$; (ii) $\bar L(\theta)$ is concave on $\mrjq$; and (iii) $\bar L(\theta) =\bar L_\infty(\theta) + \oo$ for all $\theta\in \mrjq $.
We verify below these three sufficient conditions in the order of (iii) to (i) to (ii).

First, it follows from $\txit \theta^*_q  = \bqzs  =\log(e_q /e_Q)$   that $\pi_q (\txi , \theta^*) = e_q $ for all $q\in\mq$ and $i = \ot{N}$ by \eqref{eq:pq}. Plug this in the expressions of $\bar L(\theta)$, $U(\theta)$, and $H(\theta)$ from \eqref{eq:l}, \eqref{eq:u}, and Condition \ref{cond:glm_Q} to see
\beginy\label{eq:Uq}
\bar L (\theta^*) & \ = \ & \sumq   e_q \log(e_q),\nonumber\\
U(\theta^*)  & \ = \ & \beginp
U_1 (\theta^*)\\
\vdots\\
U_{Q-1} (\theta^*)
\endp, \quad \text{where} \ \ U_q (\theta^*) = \beginp 0\\ e_q  \hx(q) \endp,\\
H(\theta^*)  & \ = \ & -\Phi\otimes (\delta \stx) \quad \text{with} \ \ 
 H_{qq'}(\theta^*) =  e_q \{ e_{q'} - 1(q=q')\} ( \temp  \stx), \nonumber
\endy
where  $\temp  = 1-N^{-1}$ and $\stx  = (N-1)^{-1}\sumi \txi \txit  =\diag(\temp ^{-1}, \sxx)$. 
Under \eqref{eq:as},  this ensures that 
$\bar L(\theta)$, $U(\theta)$, and $H(\theta)$ all converge pointwise at $\theta = \theta^*$, with  
\beginy\label{eq:h}
\hst = H_\infty(\theta^*)  = - \Phi \otimes\diag(1, \sxx) < 0. 
\endy 

In addition, Condition \ref{cond:glm_Q} ensures that $\partial U(\theta)/ \partial \theta = H(\theta)$ converges uniformly to $H_\infty(\theta)$ on any compact set in $\mrjq $.
Let 
$
U_{qj}(\theta)= \pd \bar L(\theta)/\pd \beta_{qj}$ 
be the $(qj)$th element of $U(\theta)$ for $ \qiqp$ and $j = \zt{J}$. 
Applying Lemma \ref{lem:interchange} component-wise to  $f_N = U_{qj}$ ensures that 
there exists a function, denoted by $U_\infty(\theta)$, such that 
\begina
U(\theta) = U_\infty(\theta) + \oo, \qquad \pd U_\infty(\theta) / \pd \theta = H_\infty(\theta)
\enda
 for all $\theta \in \mrjq $, and the convergence is uniform  on any compact set in $\mrjq $. 

Sufficient condition (iii) then follows from applying Lemma \ref{lem:interchange} component-wise to  $f_N = \bar L$, which ensures that there exists a function, denoted by $\bar L_\infty(\theta)$, such that 
\begina 
\bar L(\theta) = \bar L_\infty(\theta) + \oo, \qquad \pd \bar L_\infty(\theta)/ \pd \theta = U_\infty(\theta)
\enda
 for all $\theta \in \mrjq $.

Sufficient condition (i) then follows from $\pd^2 \bar L_\infty(\theta)/ \pd \theta \pd\theta^\T = H_\infty(\theta) < 0$ by Condition \ref{cond:glm_Q} and $ U_\infty(\theta^*) = \lim_{N\to \infty} U(\theta^*) = 0$  
 by \eqref{eq:Uq}. 
 
For sufficient condition (ii), 
let 
$
H_i(\theta) =  (H_{i, qq'}(\theta)  )_{q,q' \in\mqp} 
$
with
\begina
H_{i, qq'}(\theta) =\pi_q (\txi , \theta)\{\pi_{q'}(\txi , \theta) -1(q=q')\}  \txi \txit.
\enda 
Then $H(\theta) = N^{-1}\sumi H_i(\theta)$ by the explicit form of $H(\theta)$ in Condition \ref{cond:glm_Q}.
Observe that $H_i(\theta) = - \Phi _i \otimes (\txi \txit)$, with $ \Phi _i = (\Phi_{i, qq'})_{q, q' \in \mqp}$ where $\Phi_{i,qq'} = \pi_q (\txi , \theta)\{1(q=q')-\pi_{q'} (\txi , \theta)\}$. 
It follows from $\Phi_i \geq 0$ that $H_i(\theta) \leq 0$ and hence $H(\theta)\leq 0$.

\prg{Asymptotic equivalence of $\tb$ and $\Psi\hxm$.}
We next verify $\sqrtn (\tb - \Psi\hxm) = \op$. 
The proof follows from the same reasoning as that of \citet[][Theorem 3.1]{Newey1994} for independent and identically distributed samples.

Recall 
$
U_{qj}(\theta)= \pd \bar L(\theta)/\pd \beta_{qj}$ 
as the $(qj)$th element of $U(\theta)$ for $ \qiqp$ and $j = \zt{J}$. 
Let $
H_{qj}(\theta) =   \pd U_{qj}(\theta)/ \pd \theta\in \mathbb R^{(J+1)(Q-1)}$, with 
\begina
H(\theta) = \pd U(\theta) / \pd \theta^\T = \left(H_{1,0}(\theta) , H_{1,1}(\theta) , \ldots, H_{Q-1, J}(\theta) \right)^\T. 
\enda
Expanding $U_{qj}(\theta)$ at $\theta^*$ yields
\beginy\label{eq:taylor_component}
0 = U_{qj}(\tth) = U_{qj}(\theta^*) + \{H_{qj}(\theta'_{qj})\}^\T  (\tth- \theta^*),
\endy
where $\theta'_{qj} \in \mathbb{R}^{(J+1)(Q-1)}$ is a point on the line segment between $\tth$ and $\theta^*$.  
That $\tth= \theta^*+\op$ from \eqref{eq:lim} ensures $\theta_{qj}' = \theta^* + \op$ and hence $H_{qj}(\theta_{qj}') = H_{qj}(\theta^*) + \op$ for all $\qiqp$ and $j = \zt{J}$.

Let $H'$ be the matrix with rows $H_{qj}(\theta_{qj}')$ in lexicographical order of  $(qj)$. 
Then
$
H' = H^* + \op
$ 
with $H^* < 0$ and hence 
\begina
\oh = 1(\text{$H'$ is nonsingular})  = 1+\op
\enda by Condition \ref{cond:glm_Q}.
Stacking \eqref{eq:taylor_component} in lexicographical order of $(qj)$ yields $0 = U(\ths) + H'(\tth -\theta^*)$ and hence
\begina
\oh\sqrt N (\tth -\theta^*) =  -\oh (H')^{-1}\sqrt N  U (\theta^*).
\enda
This, together
with $\sqrtn U(\theta^*)$ being asymptotically normal by \eqref{eq:Uq} and Lemma \ref{lem:cre_Q}, ensures 
\beginy
\sqrt N (\tth -\theta^*)  = O_P(1)\label{eq:bigo}
\endy and hence 
\beginy
\sqrt N (\tth -\theta^*) 
 &\ =\ &   \oh\sqrt N (\tth -\theta^*)  
+ (1 - \oh) \sqrt N (\tth -\theta^*)\\
&= &  - \oh (H')^{-1}\sqrt N  U (\theta^*) 
+ (1 - \oh) \sqrt N (\tth -\theta^*) \nonumber\\
\label{eq:theta}
  & =& - \hsti   \sqrtn U (\theta ^*) +\op \nonumber
\endy
by Slutsky's theorem.  
Observe that $
(\hst)^{-1}   = - \Phi^{-1} \otimes\diag\{1, \sxxinv\} $
from \eqref{eq:h}. 
Removing the dimensions corresponding to $\{\bqz:\qiqp  \}$ in \eqref{eq:theta} yields
\begina
\sqrtn \tb  = \{\Phi^{-1}   \otimes  \sxxinv\}  \sqrtn(\rp  \otimes I_J) \hx_\pluss   +\op =\sqrtn \Psi \hxm + \op
\enda
by the explicit form of $U(\theta^*)$ from \eqref{eq:Uq}.


\prg{Asymptotic normality of $\tb$.} That $\rtn \tb  \rightsquigarrow \mN ( 0_{J(Q-1)}, \vsi)$ then follows from the asymptotic normality of $\hxm$  by Lemma \ref{lem:joint_Q} and Slutsky's theorem.


\prg{Convergence of $N \tv$.} 
Let $\tilde H = H(\tth)$ be the value of $H(\theta)$ evaluated at the \mle. 
Then $\tv$ equals the submatrix of  $(-N\tilde H)^{-1}$ after removing the rows and columns corresponding to the $Q-1$ intercepts, namely $\{\tbqz:\qiqp  \}$. 
That $ N\tv = \vsi   + \op$ follows from 
\begina
 \tilde H^{-1} =  \hsti + \op = -\Phi ^{-1} \otimes \diag\{1, \sxxinv\} + \op
 \enda by \eqref{eq:lim} and \eqref{eq:h}, and the definition of the Kronecker product.

\prg{Asymptotic equivalence of $\lambda_\text{LRT}$ and $N\hxp^\T \vxp^{-1} \hxp$.}
The \lrts tests $\logitt(Z_i \sim 1+x_i)$ against  $H_0: \logitt(Z_i \sim 1)$. 
Let $\Theta_0 = \{\theta = (\theta_1^\T, \ldots, \thqmo^\T)^\T: \theta_q = (\bqz,0_J^\T)^\T\} $ be the restricted parameter space under $H_0$, with $\tthn \in \Theta_0$ as the \mle.
The test statistic equals 
\begina
\llrt  = -2  N \left\{  \sup_{\theta\in\Theta_0} \bar L(\theta) - \sup_{\theta \in \mathbb R^{(J+1)(Q-1)}}\bar L(\theta)\right\}  = 2N \{\bar L(\tth)  - \bar L(\tthn)\}.
\enda
For $\theta \in \Theta_0$,  we have $\txit \thq = \bqz$ such that \eqref{eq:l} reduces to
$
\bar L(\theta) =  \sumq e_q \log(\pi_q)
$,
with   
\begina
\pi_q = \frac{\exp(\bqz)}{1+\sum_{q'\in\mqp} \exp(\beta_{q' 0}) } \qquad \text{for} \ \  \qiq
\enda 
denoting the identical value of $\pi_q(x_i)$ across $i = \ot{N}$; see \eqref{eq:pq}. 
The invariance of \mles to non-degenerate transformation of the parameters ensures that the {\mle}s of $\pi_q$ and $\bqz$ equal $e_q$ and $\log(\eq / \eqq) = \bqzs$, respectively, for $\qiqp$.   This ensures $\tthn = \ths$ and hence 
\beginy\label{eq:lmd}
\llrt =   2N \{\bar L(\tth)  - \bar L(\ths)\}. 
\endy
We verify below $\llrt  - N\hxp^\T \vxp^{-1} \hxp = \op$.

First, 
$\bar L(\ths)
=  \bar L(\tth)  + 2^{-1} (\ths - \tth)^\T H(\theta') (\ths -\tth)$ 
for some $\theta'$ on the line segment of $\tth$ and $\ths$. 
This, together with \eqref{eq:lmd}, ensures
\begina
\llrt = - N(\tth - \ths)^\T H(\theta') (\tth - \ths)= - N(\tth - \ths)^\T \hst (\tth - \ths) + \op
\enda
given $H(\theta') = \hst+ \op$ by \eqref{eq:lim} and $\sqrtn (\tth - \ths) = O_P(1)$ by \eqref{eq:bigo}. 

Next, it follows from \eqref{eq:theta} that 
\begina
 - N(\tth - \ths)^\T \hst (\tth - \ths)   
= - N\{U (\theta ^*)\}^\T  \hsti  U (\theta ^*) + \op.
\enda
The result then follows from
\begina
&& -N \{U(\theta^*)\}^\T \hsti  U(\theta^*) \\
&=&
N\big(\eo \hx(1)^\T, \ldots, \emo \hx(Q-1)^\T \big)\{ \Phi ^{-1} \otimes \sxxinv \}  \beginp
\eo\hx(1)\\
\vdots\\
\emo\hx(Q-1)
\endp \\
&=& 
N \{ \hxp^\T (\rp \otimes I_J) \} \vsi \{ (\rp \otimes I_J) \hxp \} \\
&=& N\hxp^\T \vxp^{-1} \hxp
\enda
by \eqref{eq:Uq}--\eqref{eq:h} and \eqref{eq:vsi}--\eqref{eq:rsr}.

\prg{Asymptotic equivalence of $\wmlr$ and $N\hxp^\T \vxp^{-1} \hxp$.}
The result follows from 
\begina
\wmlr = \tb^\T \tv^{-1} \tb = N (\Psi \hxp)^\T \vsi ^{-1} (\Psi\hxp) + \op =  N\hxp^\T \vxp^{-1} \hxp+ \op 
\enda
with $\sqrtn(\tb- \Psi\hxm) = \op$, $N\tv  - \vsi = \op$ as we just proved and  
$
\Psi^\T \vsi ^{-1} \Psi = \vxp^{-1} $ by \eqref{eq:rsr}. 

\prg{Asymptotic distributions of $\lambda_\text{LRT}$ and $\wmlr$.} The result follows from $N\hxp^\T \vxp^{-1} \hxp \rs \chisq_{J(Q-1)}$ by Lemma \ref{lem:joint_Q} and Slutsky's theorem. 

\prg{Simplification under the treatment-control experiment.}
The result follows from $\hx_\pluss  = \hx(1) =  e_{0}\htx$ and $\Psi = e_0^{-1}\sxxinv$ such that $ \Psi \hxm = e_0^{-1} \sxxinv \hx(1) = \sxxinv \htx$ and $\vsi  = (\pp)^{-1}\sxxinv$. 
\end{proof}

\end{document}